\documentclass[12pt, hyper]{article}
\usepackage{amsmath,amsopn,amssymb,amsfonts}
\numberwithin{equation}{section}
\usepackage{amsthm}
\usepackage{hyperref}
\usepackage[pdftex]{graphicx}
\usepackage{epsfig}
\usepackage{cleveref}
\usepackage{subfigure}
\usepackage{mathrsfs}
\DeclareMathAlphabet{\mathpzc}{OT1}{pzc}{m}{it}
\usepackage{verbatim}
\usepackage{multirow}
\usepackage{tikz}
\usepackage{pgfplots}
\usepackage{makecell}
\usepackage{arydshln}
\usepackage{enumitem}
\setlist{leftmargin=2mm}
\usepackage{cite}
\usepackage{amsthm}
\usepackage[margin=1cm]{caption}
\usepackage[font={small}]{caption}

\usepackage{braket}
\usetikzlibrary{arrows,calc,positioning,shadows,shapes}

\crefname{equation}{}{}

\def\CN{{\cal N}}\def\CO{{\cal O}}

\def\CT{{\cal T}}\def\CV{{\cal V}}
\def\CW{{\cal W}}

\def\a{\alpha}\def\g{\gamma}
\def\d{\delta}\def\e{\epsilon}

\def\l{\lambda}
\def\m{\mu}
\def\r{\rho}\def\s{\sigma}
\def\t{\tau}
\def\G{\Gamma}
\def\D{\Delta}

\def\half{\frac{1}{2}}

\usepackage{cleveref}

\begin{document}

\begin{titlepage}
\vfill
\begin{flushright}
{\tt\normalsize KIAS-P19045}\\
{\tt\normalsize LCTP-19-29}\\
\end{flushright}
\vfill
\begin{center}
{\LARGE\bf Notes on Superconformal Representations in Two Dimensions}

\vfill

Siyul Lee$^{\sharp,\flat}$ and Sungjay Lee$^\dagger$

\vskip 5mm
$^\sharp${\it Department of Physics and Leinweber Center for Theoretical Physics, \\
University of Michigan, Ann Arbor, MI 48109, USA}
\vskip 3mm
$^\flat${\it Department of Physics and Astronomy, \\
Seoul National University, Seoul 08826, Korea}
\vskip 3mm
$^\dagger${\it Korea Institute for Advanced Study \\
85 Hoegiro, Dongdaemun-Gu, Seoul 02455, Korea}

\end{center}
\vfill

\begin{abstract}
\noindent We study global subalgebras of superconformal algebras in
two dimensions and their unitary representations.
Global superconformal multiplets are decomposed into conformal multiplets
using Racah-Speiser algorithm, revealing many essential aspects of
superconformal theories such as stress-energy tensor, conserved current,
supersymmetric deformation and supersymmetry enhancement.
Character formulae for the representations are presented.
We further find a collection of conserved charges that are $k$-forms under
the R-symmetry, which must be
part of the super Virasoro algebra with $\CN \geq 3$ supersymmetries.
\end{abstract}

\vfill
\end{titlepage}

\parskip 0.1 cm
\tableofcontents
\renewcommand{\thefootnote}{\#\arabic{footnote}}
\setcounter{footnote}{0}

\parskip 0.2 cm

\section{Introduction and Conclusion}

Conformal field theory(CFT) is one of the key ingredients of theoretical physics,
with its far-reaching applications from string theory to condensed matter theory.
Power of the conformal field theories lies on the abundance of symmetries,
which constrain the structure and contents of the theory to a large degree.

The power of symmetries becomes particularly pronounced with introduction
of supersymmetries. The superconformal symmetry, considered to be the
most general symmetry group in four spacetime dimensions \cite{Haag:1974qh},
enables a plethora of progress based on `kinematics' of the theory alone,
let aside the `dynamics'.

The conformal symmetry becomes significantly extended to have an
infinite number of generators in two space-time dimensions,
known as the Virasoro algebra \cite{Virasoro:1969zu}.
In addition to the Virasoro algebra, many extensions of two-dimensional
conformal algebra such as Ka\v{c}-Moody algebra \cite{Kac_1968,Moody:1968zz}
or $\CW$-algebra \cite{Zamolodchikov:1985wn} have been studied,
see also \cite{DiFrancesco:1997nk,Blumenhagen:2009zz}.

However, the supersymmetric extension of the Virasoro algebra,
namely super Virasoro algebra, is increasingly complicated. It has been
fully studied only for a relatively fewer number of supersymmetries $\CN \leq 4$
\cite{Ademollo:1975an,Ademollo:1976pp,Schoutens:1987uu,Schoutens:1988ig,
Sevrin:1988ew,Eguchi:1987sm,Eguchi:1987wf,Odake:1988bh,Odake:1989dm}.
For $\CN \geq 5$, the full super Virasoro algebra has not been clearly
constructed to the best of our knowledge.

Numerous authors have attempted to construct super Virasoro algebra with
a large number $\CN$ of supercurrents. The results are yet incomplete
and do have certain unconventional features. For instances, see
\cite{Englert:1987fm,Gastmans:1987up,Sevrin:1988ps,Schoutens:1988tg,Gates:1998ss,
deBoer:1998kjm,Ito:1998vd,Henneaux:1999ib,Kraus:2007vu}.
It is the existence of certain generators, other than what are expected by
a straightforward generalization of cases with fewer $\CN$, that
makes the construction nontrivial.

Under the circumstances,
we first study in the present work the structure of global subalgebra of the
super Virasoro algebra in two dimensions and its unitary representations.
We find that the study of the global subalgebra provides certain concrete
implications on the poorly understood super Virasoro algebra.

Recently, the unitary representations
of superconformal algebras with any number of supersymmetries $\CN$ in
dimensions $3 \leq d \leq 6$ were systematically organized in \cite{Cordova:2016emh}.
The key was to decompose each representation of a superconformal
algebra into those of a conformal algebra,
utilizing the Racah-Speiser algorithm to organize
the conformal primaries into representations of Lorentz and R-symmetry groups.
Since the global conformal algebra $\mathfrak{so}(2,2)$ in two dimensions is highly
analogous to its higher-dimensional counterpart $\mathfrak{so}(d,2)$,
we can maximize utilization of the methodology
of \cite{Cordova:2016emh}.

The problem further simplifies in two dimensions because
the two-dimensional conformal algebra splits into two copies of Virasoro algebra.
The two copies are often referred to as \emph{left-moving} and \emph{right-moving}
sectors, or also as \emph{holomorphic} and \emph{anti-holomorphic} sectors.
Each copy of Virasoro algebra can be extended to accommodate any number
$\CN$ of supersymmetries that results in super Virasoro algebra.
Therefore, one can study multiplets of individual sectors, then simply take
direct product of multiplets in each sector to form a full conformal multiplet.

Despite our limitation to the global subalgebra, we find that it contains
many essential features of the larger super Virasoro algebra,
particularly involving the shortening, or unitary, conditions.
Thus, much can be inferred about the full superconformal theory
in two dimensions by studying the global subalgebra.

We list some of the most notable results below.

\begin{itemize}

\item In two-dimensional global subalgebra of super Virasoro algebra
with any number $\CN$ of supersymmetries,
a multiplet that is constant on the conformal manifold,
a.k.a. absolutely protected multiplet
does not exist.
In other words, every short multiplet that saturates the unitarity bound
may \emph{recombine} with another short multiplet to form a long multiplet
in the limit of its saturation of unitarity bound.
See section \ref{subsecrecomb}.

\item In all global superconformal theory with any numbers $(\CN, \bar{\CN})$
of supersymmetries, holomorphic and anti-holomorphic conserved currents
that are supersymmetric, R-neutral, and have spins $s=1,\frac32,2,\cdots$ are allowed.
In particular, a conserved current with $s=2$ identified as the stress-energy
tensor and a supersymmetric higher-spin current $s=\frac{\CN}{2}-1$
for $\CN \geq 7$ are universal in all theories.  See section \ref{subsecconcur}.

\item In all global superconformal theory with $(\CN \neq 4 , \bar{\CN} \neq 4)$,
or with large $\CN=4$ or $\bar{\CN}=4$ with equal Ka\v{c} level for
two copies of $\mathfrak{su}(2) \subset \mathfrak{so}(4)$,
relevant supersymmetric deformations with conformal dimension
$\D=\frac32$ that are Lorentz scalars and R-spinors are allowed.
However, their existence is not guaranteed, and in two dimensions the
\emph{universal mass} as defined in \cite{Cordova:2016xhm} does not exist
in any superconformal theory. See section \ref{subsecdeform}.

\item In all global superconformal theory with any numbers $(\CN, \bar{\CN})$
of supersymmetries, a marginal supersymmetric deformation that is a
Lorentz scalar and an R-singlet is allowed.
In particular, we rediscover that a universal marginal
deformation is guaranteed to exist in the large $(4,4)$
superconformal theory. See \cite{Gukov:2004ym} and section \ref{subsecdeform}.

\item A global superconformal theory with $\CN=5$ (or $\bar{\CN}=5$) is
automatically enhanced to an $\CN=6$ (or $\bar{\CN}=6$) theory.
See section \ref{subsecenhance}.

\item A super Virasoro algebra with number of supersymmetries $\CN \geq 5$
must contain conserved current operators that are $k$-forms under the
R-symmetry group $SO(\CN)$ and have scaling dimensions ($L_0$-eigenvalue)
$\frac{k}{2}-1$, where $k=3,4,\cdots,\CN$. The current is bosonic when
$k$ is even and fermionic when $k$ is odd.
(Anti-)commutation relation between the supercurrent and the $k$-form current
must yield, among others, the $(k+1)$-form current and the $(k-1)$-form current.
See section \ref{subsecimplyalgebra}.

\end{itemize}

This article is organized as follows.
We begin with preliminaries in section \ref{prelim},
where the global subalgebra, and decomposition principle and unitarity condition
for its multiplets are explained.
Presented in section \ref{multipletsection} are lists of unitary multiplets,
for global subalgebras with every number of supersymmetries $\CN$.
Then, we discuss various implications of the results in order.
First, we discuss more or less straightforward results in section \ref{propertysection},
namely the recombination rules and character formulae for the multiplets.
Then, section \ref{applicationsection} contains more physically
significant applications such as conserved currents, deformations, and
supersymmetry enhancements, highlighted by the stress-energy tensor.
Finally in section \ref{virasorosection}, we take a step further to discuss what we
can infer about the super Virasoro algebra from our results on the global subalgebra.

\section{Preliminaries}\label{prelim}

\subsection{Global Superconformal Algebras}

We start with a brief review on the global superconformal algebra with
an arbitrary number of supercharges in two dimensions. Generic cases
are first discussed, followed by a special case of the $\CN=4$
superconformal algebra. We limit ourselves to the left-moving sector,
as the algebra of the right-moving sector is identical.

\subsubsection{Generic Global Subalgebras}

A superconformal algebra with $\CN$ supercharges is an $\mathfrak{so}(\CN)$
Ka\v{c}-Moody algebra \cite{Knizhnik:1986wc,Bershadsky:1986ms}.
Its global subalgebra is generated by
three Laurent modes $L_{-1},L_0,L_1$ of the Virasoro operator,
two modes $G^a_{-\frac{1}{2}},G^a_{\frac{1}{2}}$ of $\CN$ supercharges,
and one mode $T_0^{ab}=-T_0^{ba}$ of R-symmetry generators,
where $a,b$ are $\mathfrak{so}(\CN)$ vector indices.
Let us present the non-trivial (anti-)commutation relations below,

\begin{subequations}\label{genericalgebra}
\begin{align}
[L_m,L_n] &= (m-n) L_{m+n},
\\
[L_m,G_r^a] &= (\frac{m}{2}-r) G_{m+r}^a,
\\
[L_m,T_0^{ab}] &= 0,
\\
[T_0^{ab},T_0^{cd}] &= i ( \d^{ac} T_0^{bd} - \d^{bc} T_0^{ad} - \d^{ad} T_0^{bc} + \d^{bd} T_0^{ac}) , \\
[T_0^{cd},G_r^a] &= -(V^{cd})_{ab} G_{r}^b, \\
\{ G_r^a, G_s^b \} &= 2 L_{r+s} \delta^{ab} - (r-s) (V^{cd})_{ab}T_0^{cd} ,
\end{align}
\end{subequations}
where
\begin{equation} \label{defV}
(V^{cd})_{ab}=-i (\d^c_a \d^d_b - \d^d_a \d^c_b )
\end{equation}
are $\mathfrak{so}(\CN)$ generators $T_0^{cd}$ in the vector representation.
We work in Neveu-Schwarz sector, so $r,s=-\half,\half $ and $m,n=-1,0,1$.

\subsubsection{$\CN=4$ Global Subalgebras}
The case $\CN=4$ where $\mathfrak{so}(4) \simeq \mathfrak{su}(2) \oplus \mathfrak{su}(2)$
calls for a special treatment. As found in \cite{Sevrin:1988ew}, there exists a one-parameter
family of $\CN=4$ superconformal algebra, where we use $\alpha$ to parametrize
relative levels of the two $\mathfrak{su}(2)$'s. Six $\mathfrak{so}(4)$ generators are
arranged into two mutually commuting sets of $\mathfrak{su}(2)$ generators $T^{\pm i}_0$.
For the global subalgebra of the superconformal algebra, last three subequations
of (\ref{genericalgebra}) are modified as follows:
\begin{subequations}\label{N4largealgebra}
\begin{align}
[T^{\pm i}_0,T^{\pm j}_0] &= i \e^{ijk} T_{0}^{\pm k}, \qquad [T^{\pm i}_0,T^{\mp j}_0] = 0, \qquad (i,j,k=1,2,3)\\
[T_0^{\pm i},G_r^a] &= i \eta^{\pm i}_{ab} G_{r}^b, \\
\label{N4largealgebraGG} \{ G_r^a, G_s^b \} &= 2 L_{r+s} \d^{ab} +
4(r-s) \big( \frac{\a}{1+\a} i \eta^{+i}_{ab}T^{+i}_0 + \frac{1}{1+\a} i \eta^{-i}_{ab}T_0^{-i} \big),
\end{align}
\end{subequations}
where
\begin{equation}
\eta^{\pm i}_{ab} = \pm \d^i_{[a} \d^4_{b]} + \frac{1}{2} \e_{iab},
\end{equation}
and the parameter $\a$ is related to the $\mathfrak{su}(2)$ levels by
\begin{equation}
\a= \frac{k_-}{k_+}.
\end{equation}

This global subalgebra is named D(2,1;$\a$). When
$\a=1$, D(2,1;$\a$) is $\mathfrak{osp}(4 \vert 2)$, which is
precisely what one obtains by putting in $\CN=4$ in the generic algebra
(\ref{genericalgebra}). In doing so, one must be careful with a numerical factor in
the relation between generators of $\mathfrak{so}(4)$ and $\mathfrak{su}(2) \oplus
\mathfrak{su}(2)$. That is, $(T^i_{SO(4)})^2=2((T^i_{SU(2)_+})^2+(T^i_{SU(2)_-})^2)$,
which accounts for the extra factor of 2 in the second term of (\ref{N4largealgebraGG}).

In subsequent sections, we will usually leave $\a$ as a free parameter and refer to
D(2,1;$\a$) as \emph{large $\CN=4$} global subalgebra,
although we will frequently give it the value 1.

Meanwhile, another subalgebra can be obtained from the above by, e.g., taking a
limit $\a \rightarrow \infty$. This leaves us with only one $\mathfrak{su}(2)$ generated
by $T^i_0$, under which four supercharges transform as two independent sets of spinors
$G_r^\pm,\bar{G}_s^\pm$ where $r,s=\pm \frac{1}{2}$ as before.
 The algebra is summarized by, in addition to the first three subequations of
(\ref{genericalgebra}), (see \cite{Eguchi:1987sm,Eguchi:1987wf})
\begin{subequations}\label{N4smallalgebra}
\begin{align}
[T^{i}_0,T^{j}_0] &= i \epsilon^{ijk} T_{0}^{k}, \quad \qquad (i,j,k=1,2,3)\\
[T_0^{i},G_r^a] &= -\frac{1}{2} \sigma^i_{ab} G_{r}^b, \qquad (a,b=1,2) \\
[T_0^{i},\bar{G}_s^a] &= \frac{1}{2} (\sigma^i_{ab})^* \bar{G}_{s}^b, \\
\{ G_r^a, G_s^b \} &= \{ \bar{G}_r^a, \bar{G}_s^b \} = 0, \\
\{ G_r^a, \bar{G}_s^b \} &= 2 L_{r+s} \delta^{ab} -2(r-s) \sigma^i_{ab}T^i_{0},
\end{align}
\end{subequations}
where $\sigma^i_{ab}$ are the usual Pauli matrices. This global subalgebra is
$\mathfrak{su}(2 \vert 1,1)$ (see \cite{Sevrin:1988ew}), to which we refer as
\emph{small $\CN=4$} global subalgebra.

\subsubsection{Comment on Super Virasoro Algebras}

In two dimensions, super Virasoro algebras with $\CN$ supercharges are constructed
by central charge extension from the corresponding global subalgebras \cite{Virasoro:1969zu}.
As a result, the full algebra contains an infinite number of Laurent modes for each of
the operators $L$, $G^a$, $T^{ab}$, with non-trivial commutation relations such as
\begin{equation}
  \big[ L_m, L_n \big] =
  (m-n) L_{m+n} + \frac{c}{12}(m^3-m) \delta_{m+n,0}, \qquad (m,n \in \mathbb{Z})
\end{equation}

However, the super Virasoro algebra in general is not completed by the infinite modes of
$L$, $G^a$, and $T^{ab}$. For example, the super Virasoro algebra for the large $\CN =4$ contains
an additional $\mathfrak{u}(1)$ generator among others \cite{Sevrin:1988ew}, and higher-$\CN$
algebras are expected to contain more extra generators. See \cite{Gates:1998ss} for an example.

While the super Virasoro algebras for $2 \leq \CN \leq 4$ have been thoroughly studied
\cite{Ademollo:1975an,Ademollo:1976pp,Schoutens:1987uu,Schoutens:1988ig,
Sevrin:1988ew,Eguchi:1987sm,Eguchi:1987wf,Odake:1988bh,Odake:1989dm},
those for generic $\CN \geq 5$ are not fully understood
to the best of our knowledge. It is one of the main goals of this article to find extra generators
that must enter the algebra by studying their global subalgebras, in particular their stress-energy
tensor multiplets. See section \ref{subsecimplyalgebra} for this account.

\subsection{Superconformal Multiplets}\label{subsecmultiplets}

We now turn to global superconformal multiplets allowed by each global
subalgebra with an arbitrary number $\CN$ of supercharges. Let us first discuss
the unitary multiplets of the left-movers $G_{1/2}^a$ below.
Same argument applies to the right-movers as well.

An irreducible superconformal multiplet is fully determined by its superconformal primary $\CV$
that is annihilated by $\{ G^a_{1/2}, L_{1} \} $.
A superconformal primary furnishes an irreducible representation
under the maximally compact bosonic subalgebra
$\mathfrak{sl}(2) \oplus \mathfrak{so}(\CN)$.\footnote{There is an exception of
small $\CN=4$ superconformal algebra where $\mathfrak{so}(4)$ reduces
to $\mathfrak{su}(2)$.}
The left-moving multiplet then consists of the primary $\mathcal{V}$ and
its superconformal descendants,
obtained by consecutive actions of $L_{-1}$ and $G^a_{-1/2}$ on the primary.

Given any superconformal primary, it is straightforward to build a multiplet with
its descendants. However, unitarity conditions impose a bound on the allowed
$L_0$-eigenvalue $h_0$ of the superconformal primary.
Let us examine this bound in any $\CN$ in two dimensions.
This argument closely follows that of \cite{Minwalla:1997ka}, and we shall thus be brief.

Let us denote the state corresponding to the superconformal primary $[R]_{h_0}$ by $\ket{[R]_{h_0}}_\alpha$,
where $[R]=[R_1 \, R_2 \, \cdots \,R_r]$ collectively denotes the highest weight $\mathfrak{so}(\CN)$ Dynkin labels
with $r=\lfloor \CN/2 \rfloor$  and $\alpha$ is an index for the representation $[R]$.
The unitarity bound can be obtained by
enforcing all first-level components in the multiplet to have non-negative norms.
To be more explicit, let us consider a matrix element\footnote{We actually need conjugate
indices for the bra, but it is irrelevant as we are only interested in the eigenvalues.}
\begin{equation}
A_{b\beta ; a\alpha} = \bra{[R]_{h_0}}_\beta G^b_{1/2} G^a_{-1/2} \ket{[R]_{h_0}}_\alpha.
\end{equation}
Since $(G^a_{1/2})^\dagger = G^a_{-1/2}$, all its eigenvalues are required to be non-negative.

We can proceed with the (anti-)commutation relation (\ref{genericalgebra}) since the primary
ket is annihilated by $G_{1/2}$:
\begin{align}
  A_{b\beta ; a\alpha} &= \bra{[R]_{h_0}}_\beta (2L_0 \delta_{ba} - \frac{1}{2} (V^{cd})_{ba} T^{cd}_0 ) \ket{[R]_{h_0}}_\alpha
  \nonumber \\ &=
  2h_0 (\mathbb{I})_{b\beta ; a\alpha} - \frac{1}{2} (V^{cd})_{ba} \otimes
  (\rho(T_0)^{cd})_{\beta \alpha},  \label{unicon1}
\end{align}
where $(V^{cd})_{ba}= -i (\d^c_b\d^d_a- \d^d_b\d^c_a)$ as in (\ref{defV}) and
$(\rho(T_0)^{cd})_{\beta \alpha}$ represents matrix elements of $\mathfrak{so}(\CN)$
generators in the vector and $[R]$ representations. Eigenvalues of the matrix (\ref{unicon1})
can be obtained as in a well known quantum mechanics problem.
The result is:
\begin{equation}\label{unicon2}
h_0 \geq \frac{1}{4} ( c_2([R']) - c_2([R]) - c_2(V) ),
\end{equation}
where $c_2([R])$ denotes the quadratic Casimir of the representation $[R]$,
$V$ denotes the vector representation, and $[R']$ is any irreducible representation that
composes the tensor product $[R] \otimes V$.
For the generic case where the R-symmetry group is $\mathfrak{so}(\CN)$,
under which the primary has Dynkin labels $[R_1 \cdots R_r]$ and the vector has $[1 \, 0 \cdots 0]$,
(\ref{unicon2}) becomes
\begin{equation}\label{unicon3}
  2 h_0 \geq h_1=\begin{cases} R_1 + \cdots + R_{r-2} + \frac{R_{r-1}+R_r}{2}
  & \CN \text{ is even}, \\ R_1 + \cdots + R_{r-1} + \frac{R_r}{2} &
  \CN \text{ is odd}, \end{cases}
\end{equation}
where $h_1$ is the first orthogonal weight.
The orthogonal basis will be used exclusively in section \ref{subsecchar}.
For later convenience, we present the relation between the Dynkin labels $R_i$
and the orthogonal weights $h_i$ below:
\begin{itemize}
    \item[~] When $\CN$ is even,
    \begin{gather}
      h_i = R_i + \cdots + R_{r-2} + \frac{R_{r-1}+R_r}{2} \quad (i=1,2,\cdots, r-2)\ ,
      \nonumber \\
      h_{r-1} = \frac{R_{r-1}+R_r}{2}\ , \quad
      h_r =  \frac{R_{r-1}-R_r}{2}\ .
    \label{defineorthogonaleven}
    \end{gather}
    \item[~] When $\CN$ is odd,
    \begin{gather}
      h_i = R_i + \cdots + R_{r-1} + \frac{R_r}{2} \quad (i=1,2, \cdots ,r-1)\ , \quad
      h_{r} = \frac{R_r}{2}\ .
    \label{defineorthogonalodd}
    \end{gather}
\end{itemize}

When the BPS condition (\ref{unicon3}) is saturated, $A_{b\beta ; a\alpha}$
acquires a zero eigenvalue. Since (\ref{unicon3}) corresponds to (\ref{unicon2})
with $[R']$ that yields the strongest bound, this indicates that among many states
$G^a_{-1/2} \ket{[R]_{h_0}}_\alpha$ that transform in $[R_1 \cdots R_r] \otimes [1 \, 0 \cdots 0]$
representation of the R-symmetry group, those belonging to the irreducible representation
$[R_1+1 \, R_2 \cdots R_r]$ are null. Then, not only the component
$[R_1+1 \, R_2 \cdots R_r]_{h_0+\frac{1}{2}}$ but also its conformal descendants must be
removed from the superconformal multiplet. A systematic procedure of such a removal is
discussed in \cite{Cordova:2016emh}. Such superconformal multiplets with null states are referred
to as \textit{short} multiplets, as opposed to \textit{long} multiplets. Following the convention
of \cite{Cordova:2016emh}, a long multiplet will be denoted as
$$L[R]_{h}\qquad h>h_1(R),$$
where $[R]$ and $h$ are quantum numbers of its superconformal primary. Similarly
a short multiplet will be denoted as
$$A[R]_h\qquad h=h_1(R).$$
In particular, a vacuum multiplet will be denoted as
$$V[0]_0,$$
although it is also a short multiplet. The vacuum multiplet is defined as a multiplet
whose primary is annihilated by all operators so there are no descendants. All of its
quantum numbers vanish.

So far we have discussed the unitarity condition of first-level states only.
We conjecture that it alone suffices, because the first-level states impose the strongest
bound.

A complication arises in higher dimensions where the Lorentz group is also non-abelian.
The strongest unitarity bound for a given primary arises from the descendant whose Dynkin
labels for the Lorentz group are small and that for the R-symmetry are large.
As a result, when the primary is a Lorentz singlet, second-level unitarity
bound can be stronger than that of the first-level, where the states are
necessarily Lorentz spinors rather than of smaller Dynkin labels that would have been present
(and given stronger bound) for generic primaries.
This non-generic phenomenon leads to a diversity of short multiplets,
including those with higher-level null states and more interestingly,
\emph{isolated} short multiplets. See \cite{Cordova:2016emh} for more details.
In contrast, all short multiplets in two dimensions are limiting cases of long multiplets, which
results in the absence of absolutely protected multiplets.
We will discuss these features further in section \ref{subsecrecomb}.

A superconformal multiplet is decomposed into conformal multiplets \cite{Cordova:2016emh}
that consist of conformal primaries (annihilated by $L_1$) and their descendants
(obtained by consecutive actions of $L_{-1}$ on the primaries).
Thus it is useful to express the superconformal multiplet as a collection of conformal
primaries whose conformal multiplets make up the superconformal multiplet.
We refer to these conformal multiplets as \emph{components} of the superconformal
multiplet. The collection consists of the superconformal primary $\mathcal{V}$ and
the operators that are obtained by (repeatedly) acting only $G^a_{-1/2}$'s on
$\mathcal{V}$. Note that we can effectively set
\begin{equation}
\{ G^a_{-1/2}, G^b_{-1/2} \} = 2 L_{-1} \delta^{ab}  \sim 0,
\end{equation}
because the $L_{-1}$ action does not generate a new conformal
primary but generates a descendant.
Due to this Fermi-Dirac statistic, decomposition into conformal multiplets is finite.

Since conformal primaries are also in the representations of
$\mathfrak{sl}(2) \times \mathfrak{so}(\CN)$ \cite{Cordova:2016emh}, it proves
convenient to specify each conformal multiplet by
the $L_0$ eigenvalue $h$ and the
highest weight $\mathfrak{so}(\CN)$ Dynkin label $[R_1 \cdots R_r]$
of the corresponding conformal primary. Thus,
$$ [R]_h \text{ : a conformal multiplet with corresponding primary.} $$
Note that we are using the same notation to refer to superconformal multiplets
as to conformal multiplets, except that a letter $L$, $A$, or $V$ to indicate the
presence of null states is omitted for the latter.

Conformal multiplet decomposition of a superconformal multiplet can be performed by
consecutive actions of $G^a_{-1/2}$ on the superconformal primary, and
organizing into irreducible representations of the bosonic subalgebra
$\mathfrak{sl}(2) \oplus \mathfrak{so}(\CN)$.
This process is best done via Racah-Speiser algorithm, inspired by \cite{Dolan:2002zh}
and thoroughly explained in \cite{Cordova:2016emh}. In our case of two dimensions this is
particularly simple, because $G^a_{-1/2}$ simply act as raising operators for the
$\mathfrak{sl}(2)$ and the only non-trivial part is the R-symmetry $\mathfrak{so}(\CN)$.

Number of operations of $G^a_{-1/2}$ required on the superconformal primary to
obtain a particular conformal primary is referred to as \emph{level} of the component.
Thanks to the Fermi-Dirac statistics, the level is bounded from above by $\CN$ in
any superconformal multiplet. In fact, long multiplets are always terminated at the
level $\CN$ while short multiplets are terminated earlier. In two dimensions where
$G^a_{-1/2}$ simply raises the $L_0$-eigenvalue by $\frac{1}{2}$, any component
at level $l$ has an $L_0$-eigenvalue $h_0 + \frac{l}{2}$ where $h_0$ is that of the
superconformal primary.

Note that in all of the discussions so far, and in most of the discussions that will
follow, conformal and superconformal multiplets refer to representations of the global
subalgebras. In general, multiplets of the full super Virasoro algebras include many
global multiplets, as they include the action of all negative modes such as $L_{n<-1}$,
$G^a_{r<-1/2}$, $T^{[ab]}_{n<0}$ on the primary. For example, the vacuum multiplet
and the stress-energy tensor multiplet for the global subalgebra belong to the same
super Virasoro multiplet.

\subsection{Supersymmetric Deformations and Conserved Currents}\label{subsecdeformation}

One of the applications of decomposition of superconformal multiplets is to
look for possible deformations of CFTs. Following \cite{Cordova:2016xhm},
we seek possible deformations of SCFTs in two dimensions in vicinity of
RG fixed points by relevant or marginal local operators $\mathcal{O}$.
That is, given a superconformal theory, we aim to find a local operator that
i) is a Lorentz singlet, ii) has scaling dimension less than or equal to the
dimension, which is $2$ throughout this paper, iii) is not a total derivative,
and iv) is supersymmetric. Note that the operator must reside in an allowed
superconformal multiplet of the theory.

In two dimensions, superconformal algebras separate themselves into left-
and right-moving sectors. The $L_0$- (in the left) and $\bar{L}_0$- (in the right)
eigenvalues $h_0$ and $\bar{h}_0$ of an operator sum up to the scaling
dimension, and their difference represents the spin of the operator.
Thus, to satisfy the condition i) we require $h_0=\bar{h}_0$, and further for ii),
it suffices to look for operators with $h_0 = \bar{h}_0 \leq 1$.
The condition iii) suggests to consider only the conformal primaries,
as conformal descendants are obtained by applying $L_{-1} \sim \partial_z $
to another local operator.

The condition iv) is a little trickier to satisfy. One obvious way for a conformal
primary belonging to a superconformal multiplet to be supersymmetric
(i.e. to be annihilated by all supersymmetries $G_{-1/2}$) is to be a
\emph{generic top} component of the superconformal multiplet:
to reside at the highest level of the multiplet since an application of $G_{-1/2}$
raises the level by unity. As discussed in the last subsection,
a long multiplet always has its generic top component at the level $\CN$
while short multiplets have them at lower levels, and possibly more than
one of them. Every superconformal multiplet possesses at least
one generic top component.

Nevertheless, there are components of short superconformal multiplets
at the level lower than the generic top component, that however are
annihilated by all supersymmetries. Following \cite{Cordova:2016xhm},
we refer to them as \textit{sporadic} top components, and they prove to be
very fruitful in discussion of conserved currents in two dimensions.

A sporadic top component is easily identified when, as in \cite{Cordova:2016xhm},
there exists a conformal primary whose Dynkin labels match none of those
at the next level when added by any of supercharges. In such case,
we infer that all conformal primaries at the next level must be produced by
acting on other components at the previous level by supercharges.
However, note that this is sufficient but not necessary a condition to be a
top component. We shall see counterexamples in sections
\ref{N3 multiplets}, \ref{N5 multiplets}, and \ref{N7 multiplets}.

Also of our interest are conserved currents. When an operator satisfies
the conditions iii) and iv) above but has either $h_0=0$ or $\bar{h}_0=0$,
the supersymmetric operator is annihilated by $L_0 \sim \partial_z$ or
$\bar{L}_0 \sim \partial_{\bar{z}}$, respectively. In other words, it is conserved.
In particular, if such an operator is an R-singlet and has $(h_0,\bar{h}_0)=(2,0)$
it could be the holomorphic part of the stress-energy tensor and if it has
$(h_0,\bar{h}_0)=(0,2)$ it could be the anti-holomorphic part,
expected to exist in all physical theories.
Further, to the same multiplet as the stress-energy tensor but at the previous
level must belong the supercurrents (in the vector representation of the
R-symmetry) and yet at the level below the R-symmetry currents
(in the adjoint representation). Meanwhile, a supersymmetric and R-singlet
operator with $(h_0,\bar{h}_0)=(0,1)$ or $(1,0)$ would similarly indicate a
flavor current, and those with $(h_0,\bar{h}_0)=(0,s>1)$ or $(s>1,0)$
the higher-spin currents.

Supersymmetric deformations and conserved currents will be discussed
in sections \ref{subsecconcur} and \ref{subsecdeform}, respectively.

\section{List of Multiplets}\label{multipletsection}

We tabulate in this section all superconformal multiplets, long or short as
explained in section \ref{subsecmultiplets}, for each number of supersymmetries.
In this section, we restrict only to the left-moving sector,
as the right-moving sector may have a same list of multiplets with its own $\bar{\CN}$.
In doing so, we will explicitly list top components, generic or sporadic as explained
in section \ref{subsecdeformation}, to be discussed in detail in the following sections.

\subsection{$\CN=1$}
\begin{table}[t] \centering
\begin{tabular}{| c | c | c | c | c | c |} \hline
& Primary & Unitarity bound & Null component & Sporadic top & Generic top \\ \hline \hline
L & $[0]_{h_0}$ & $h_0 > 0$ & - & - & $[0]_{h_0+\frac{1}{2}}$ \\ \hline
V &$[0]_{0}$ & - & $[0]_\frac{1}{2}$ & - & $[0]_{0}$ \\
\hline \end{tabular}
\caption{\label{N1table} List of $\CN=1$ multiplets.}
\end{table}
Let us begin with the simplest case, $\CN=1$ that contains no R-symmetry.
Representation with respect to the R-symmetry group is always trivial: $[0]$.
For any superconformal primary $[0]_{h_0}$, the unitarity bound simply becomes
\begin{equation}\label{N1 bound}
h_0 \geq 0.
\end{equation}
When the bound is saturated, the only superconformal descendant
$G_{-1/2} \ket{[0]_{h_0}} = \ket{[0]_{{h_0}+\frac{1}{2}}}$ is null
and the superconformal multiplet consists of only one conformal multiplet.
Then a complete list of superconformal multiplets and their top components
is as simple as Table \ref{N1table}.

\subsection{$\CN=2$}

\begin{table}[t] \centering
\begin{tabular}{| c | c | c | c | c | c |} \hline
& Primary & Unitarity bound & Null component & Sporadic top & Generic top \\ \hline \hline
L & $[j_0]_{h_0}$ & $h_0 > \frac{| j_0 |}{2}$ & - & - & $[j_0]_{h_0+1}$ \\ \hline
A &$[j_0>0]_{ \frac{ j_0 }{2}}$ & - & $[j_0+1]_{\frac{ j_0 }{2}+\frac{1}{2}}$ & - & $[j_0-1]_{ \frac{ j_0 }{2}+\frac{1}{2} }$ \\
A &$[j_0<0]_{ -\frac{ j_0 }{2}}$ & - & $[j_0-1]_{-\frac{ j_0 }{2}+\frac{1}{2}}$ & - & $[j_0+1]_{ -\frac{ j_0 }{2}+\frac{1}{2} }$ \\
V &$[0]_{0}$ & - & $[\pm 1]_{\frac{ 1}{2}}$ & - & $[0]_0$ \\
\hline \end{tabular}
\caption{\label{N2table} List of $\CN=2$ multiplets.}
\end{table}
The $\CN=2$ supersymmetry has an abelian R-symmetry $SO(2) \sim U(1)$
under which the two supercharges $G^\pm_{-1/2}$
are charged by $\pm 1$. A superconformal primary $[j_0]_{h_0}$ has to
satisfy the unitarity bound below:
\begin{align}
h_0 \geq \frac{| j_0 |}{2} ,
\end{align}
where $j_0$ denotes a $U(1)$ R-charge rather than a Dynkin label.

Note that this bound is weaker than the unitarity bounds
put forward by the super Virasoro case \cite{Boucher:1986bh}.
A list of superconformal multiplets, along with their top components,
is summarized in Table \ref{N2table}.

\subsection{Small $\CN=4$}

Let us examine the $\CN=4$ superconformal algebra
before $\CN=3$ for the reason that will soon be clear.
The small $\CN=4$ superconformal algebra has two independent
sets of supercharges $G^\pm_{-1/2}$ and $\bar{G}^\pm_{-1/2}$
in the fundamental representation of $SU(2)$ R-symmetry.\footnote{One can think of an extra
$U(1)$ that distinguishes the two, which we choose not to because it plays no role other
than labelling the two and justifying some of (\ref{N4smallalgebra}).}
The superconformal primary is labelled by the $SU(2)$ Dynkin label as $[R_1]_{h_0}$,
where $R_1$ is a non-negative integer. For instance, the fundamental representation
is labelled by $[1]$. The small $\CN=4$ superconformal
unitarity bound is
\begin{align}
  h_0 \geq \frac{R_1}{2},
\end{align}
and as the bound is saturated, two copies of raising operators
$G_{-1/2}^+$ and $\bar G_{-1/2}^+$  simultaneously annihilate the primary.
We summarize the list of small $\CN=4$ multiplets in Table \ref{N4stable}.
\begin{table}[t] \centering
\begin{tabular}{| c | c | c | c | c | c |} \hline
& Primary & Unitarity & Null & Sporadic & Generic \\
& & bound & component & top & top \\ \hline \hline
L & $[R_1]_{h_0}$ & $h_0 > \frac{R_1}{2}$ & - & - & $[R_1]_{h_0+2}$ \\ \hline
A &$[R_1 \geq 2]_{ \frac{ R_1 }{2}}$ & - &
$[R_1+1]_{\frac{R_1}{2}+\frac{1}{2}} \oplus [R_1+1]_{\frac{R_1}{2}+\frac{1}{2}}$
& - & $[R_1-2]_{ \frac{R_1}{2}+1}$ \\
A &$[1]_{\frac{1}{2}}$ & - & $[2]_{1} \oplus [2]_{1}$ & - & $[0]_{1} \oplus [0]_{1}$ \\
V &$[0]_{0}$ & - & $[1]_{\frac{ 1}{2}} \oplus [1]_{\frac{ 1}{2}}$ & - & $[0]_0$ \\
\hline \end{tabular}
\caption{\label{N4stable} List of small $\CN=4$ multiplets.}
\end{table}
Here, $A[1]_{\frac12}$ is an example where
there exist two degenerate top components $[0]_{1}$.

\subsection{Large $\CN=4$} \label{subsec N4}

The large global $\CN=4$ superconformal algebra $D(2,1;\a)$
contains two copies of $\mathfrak{su}(2)$ algebra. When the free
parameter $\a$ is set to unity, the R-symmetry group
becomes $SO(4)$.\footnote{The free parameter $\a$ is related to
the levels $k_+ = c(1+\a)/(6\a)$ and $k_-=c(1+\a)/6$ of two $\mathfrak{su}(2)$ current
algebras when the global superconformal algebra is promoted to
the large $\CN=4$ super Virasoro algebra. See section
\ref{subsecimplyalgebra} for more details.}
The Dynkin labels $R_1$ and $R_2$ now correspond
to those of two $\mathfrak{su}(2)$'s such that the
four supercharges are in the representation $[1;1]$.
Note that $[1;1]$ is the highest weight of a vector
representation when the R-symmetry group becomes $SO(4)$.

The unitarity bound on the conformal weight $h$ of
the large $\CN=4$ superconformal primary is given by
\begin{align}
h_0 \geq \frac{h_1}{2} = \frac{\a R_1+  R_2}{2(1+\a)}.
\end{align}
One can find the list of superconformal multiplets
in Table \ref{N4ltable}.
\begin{table}[t] \centering
\begin{tabular}{| c | c | c | c | c | c |} \hline
& Primary & Unitarity & Null & Sporadic & Generic \\
& & bound & component & top & top \\ \hline \hline
L & $[R_1;R_2]_{h_0}$ & $h_0 > \frac{h_1}{2}$ & - & - & $[R_1;R_2]_{h_0+2}$ \\ \hline
A & $ \begin{matrix} [R_1 \geq 1 ;R_2 \geq 1]_{h_1/2} \\ \text{(but not }R_1=R_2=1) \end{matrix} $ & - & $[R_1+1;R_2+1]_{\frac{h_1}{2}+ \frac{1}{2}}$ & - & $[R_1-1;R_2-1]_{\frac{h_1}{2} + \frac{3}{2}}$ \\
A & $[1;1]_\frac{1}{2} $ & - & $[2;2]_1$ & $[0;0]_1$ & $[0;0]_2$ \\
A &$[R_1\geq 2 ;0]_{ \frac{ \a R_1}{2(1+\a)}}$ & - & $[R_1+1; 1]_{ \frac{ \a R_1}{2(1+\a)}+\frac{1}{2}}$ & - & $[R_1-2;0]_{  \frac{ \a R_1}{2(1+\a)} +1}$ \\
A &$[0;R_2\geq 2]_{ \frac{R_2}{2(1+\a)}}$ & - & $[1;R_2+1]_{ \frac{R_2}{2(1+\a)}+\frac{1}{2}}$ & - & $[0;R_2-2]_{ \frac{R_2}{2(1+\a)} +1}$ \\
A &$[1 ;0]_{ \frac{ \a}{2(1+\a)}}$ & - & $[2; 1]_{ \frac{ \a}{2(1+\a)}+\frac{1}{2}}$ & - & $[0;1]_{\frac{\a}{2(1+\a)} +\frac{1}{2}}$ \\
A &$[0;1]_{ \frac{1}{2(1+\a)}}$ & - & $[1;2]_{ \frac{1}{2(1+\a)}+\frac{1}{2}}$ & - & $[1;0]_{  \frac{1}{2(1+\a)} +\frac{1}{2}}$ \\
V &$[0;0]_{0}$ & - & $[1;1]_{\frac{ 1}{2}} $ & - & $[0;0]_0$ \\
\hline \end{tabular}
\caption{\label{N4ltable} List of large $\CN=4$ multiplets.}
\end{table}
Since the multiplet $A[1,1]_\frac12$ is the first example where a sporadic top component appears,
let us pause to examine this. The superconformal multiplet $A[1;1]_\frac12$
can be decomposed into various conformal multiplets as in Figure \ref{N4 sporadic}.
\tikzstyle{line} = [draw, -latex']

\tikzset{block/.style={draw,text width=1.5cm, align=center,  inner sep=2ex, anchor=east, minimum height=4em}}

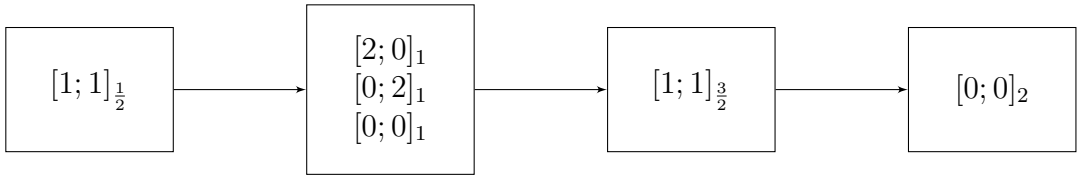
\begin{figure}[!htb]
\centering
\begin{tikzpicture}[node distance=4cm,auto,>=latex']
\node [block] (node1) {$[1;1]_\frac12$};
\node [block, right of=node1] (node2) {$\begin{matrix} [2;0]_1 \\ [0;2]_1 \\ [0;0]_1 \end{matrix}$};
\node [block, right of=node2] (node3) {$[1;1]_\frac{3}{2}$};
\node [block, right of=node3] (node4) {$[0;0]_2$};

\path [line] (node1) -- (node2);
\path [line] (node2) -- (node3);
\path [line] (node3) -- (node4);

\end{tikzpicture}

\caption{An $\CN=4$ multiplet with a sporadic top component.}
\label{N4 sporadic}
\end{figure}
We see that the null state $[2;2]_1$ and
the Racah-Speiser trial state constructed by acting the supercharge of weight $[1;1]$ on
the primary have the same quantum numbers.
As explained in \cite{Cordova:2016emh}, any RS states involving
the supercharge of weight $[1;1]_\frac12$ have to be removed from the
superconformal multiplet. Then, none of the three other supercharges
can act on $[0;0]_1$ at the first level to produce $[1;1]_\frac{3}{2}$ at the
second level. In other words, the conformal primary $[0;0]_1$ is
annihilated by all supercharges and becomes a sporadic top component.
As we will see in sections \ref{subsecdeform} and \ref{subsecimplyalgebra},
the sporadic top component $[0;0]_1$ leads to two important
features that any large $\CN=4$ superconformal theories share.
One of them is related to the universal marginal operator and the other to
an extra $U(1)$ symmetry.

\subsection{$\CN=3$} \label{N3 multiplets}

The $\CN=3$ superconformal algebra is $OSp(3|2)$ with R-symmetry $SO(3)$.
Three supercharges transform as $[2]$, namely the vector representation under $SO(3)$.
One can label a superconformal primary as $[R_1]_{h_0}$ where the Dynkin label $R_1$
is a non-negative integer. Note that we use the $SO(3)$ Dynkin label rather than
SU(2), which differ by a factor of 2.
The unitarity condition (\ref{unicon3}) is then
\begin{equation}
h_0 \geq \frac{R_1}{4}.
\end{equation}

For any short superconformal multiplet $A[R_1]_{ h=\frac{ R_1 }{4}}$, the
$Q$-descendant $[R_1+2]_{\frac{R_1}{4}+{\frac{1}{2}}}$ becomes null.
However, since top components differ when $R_1$ is small, we list them
separately in Table \ref{N3table}. In particular, for the
primary $[0]_{\e>0}$, $[2]_{\frac12+\e}$ consists the first level alone,
and as $\epsilon \rightarrow 0$ it becomes null and the multiplet terminates already
at the level zero. This phenomenon is universal: for all $\CN$ the superconformal
multiplet $V[0]_0$ consists of a superconformal primary only, which corresponds to the identity
operator ${\bf 1}$. This is because $[0]_0$ can be annihilated by all lowering operators of $SO(3)$:
once the supercharge of the highest $\mathfrak{so}(3)$ weight annihilates it, every
supercharge annihilates it. We refer to the multiplet $V[0]_0$ as the vacuum multiplet.
\begin{table}[t] \centering
\begin{tabular}{| c | c | c | c | c | c |} \hline
& Primary & Unitarity bound & Null component & Sporadic top & Generic top \\ \hline \hline
L & $[R_1]_{h_0}$ & $h_0 > \frac{R_1}{4}$ & - & - & $[R_1]_{h_0+\frac{3}{2}}$ \\ \hline
A &$[R_1 \geq 3]_{ \frac{ R_1 }{4}}$ & - & $[R_1+2]_{\frac{R_1}{4}+\frac{1}{2}}$ & - &
$[R_1-2]_{ \frac{R_1}{4}+1}$ \\
A &$[2]_{ \frac{1}{2}}$ & - & $[4]_{1}$ & $[0]_1$ & $[0]_{\frac{3}{2}}$ \\
A &$[1]_{\frac{1}{4}}$ & - & $[3]_{\frac{3}{4}}$ & - & $[1]_{\frac{3}{4} }$ \\
V &$[0]_{0}$ & - & $[2]_{\frac{ 1}{2}}$ & - & $[0]_0$ \\
\hline \end{tabular}
\caption{\label{N3table} List of $\CN=3$ multiplets.}
\end{table}

The $A[2]_\frac{1}{2}$ multiplet that contains a sporadic top component calls for special
attention. Its decomposition into conformal multiplets is given in Figure \ref{N3 sporadic}.
It is not obvious if the $ [0]_1$ at the first level is indeed a top component, because the
supercharge of weight [0] might produce the $[0]_\frac32$ at the second level when acted
on the $[0]_1$ in the sense of Racah-Speiser algorithm. However,
this is not always the case because the actual state represented by $[0]_1$ is some linear
combination of elements of $[2]_\frac{1}{2}$ acted by an appropriate supercharge, and
thus the null condition is highly non-trivial. For details, see \cite{Cordova:2016emh}.

\tikzset{block/.style={draw,text width=1.5cm, align=center,  inner sep=2ex, anchor=east, minimum height=4em}}

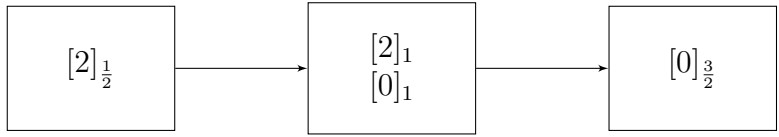
\begin{figure}[!htb]
\centering
\begin{tikzpicture}[node distance=4cm,auto,>=latex']
\node [block] (node1) {$[2]_\frac{1}{2}$};
\node [block, right of=node1] (node2) {$\begin{matrix} [2]_1 \\ [0]_1 \end{matrix}$};
\node [block, right of=node2] (node3) {$ [0]_\frac{3}{2} $};
\path [line] (node1) -- (node2);
\path [line] (node2) -- (node3);
\end{tikzpicture}
\caption{An $\CN=3$ multiplet with a sporadic top component.}
\label{N3 sporadic}
\end{figure}

In order to argue that $[0]_1$ is indeed a sporadic top component, let us
consider the large $\CN=4$ superconformal theory that can be viewed as
a special case of $\CN=3$ superconformal theory.
As will be explained in section \ref{subsecconcur},
any $\CN=4$ superconformal theory must have the stress tensor multiplet
$A[1;1]_\frac12$ that we have analyzed in Figure \ref{N4 sporadic}.
Decomposition of each conformal primary in $A[1;1]_\frac12$
into $SO(3)$ representations, summarized in Figure \ref{N4 sporadic decomposed},
shows how the $\CN=3$ short multiplet $A[2]_\frac{1}{2}$ can be
embedded in the $\CN=4$ short $A[1; 1]_\frac{1}{2}$.
In particular, we see that the sporadic top component $[0; 0]_1$ in $A[1;1]_\frac12$,
annihilated by four supercharges, can be identified as $[0]_1$ in $A[2]_\frac12$.
Since three supercharges of $\CN=3$ superconformal algebra
are a subset of the four supercharges of $\CN=4$ one,
this is sufficient to argue that $[0]_1$ in Figure \ref{N3 sporadic} is
indeed a sporadic top component as well.

\tikzstyle{line} = [draw, -latex']
\tikzset{block/.style={draw,text width=1.5cm, align=center,  inner sep=2ex, anchor=east, minimum height=4em}}

\begin{figure}[!htb]
\centering
\begin{tikzpicture}[node distance=4cm,auto,>=latex']

\node [block] (node1) {$\begin{matrix} [2]_\frac{1}{2} \\ [0]_\frac{1}{2} \end{matrix}$};
\node [block, right of=node1] (node2) {$\begin{matrix} [2]_1 \\ [2]_1 \\ [0]_1 \end{matrix}$};
\node [block, right of=node2] (node3) {$\begin{matrix} [2]_\frac{3}{2} \\ [0]_\frac{3}{2} \end{matrix}$};
\node [block, right of=node3] (node4) {$[0]_2$};

\path [line] (node1) -- (node2);
\path [line] (node2) -- (node3);
\path [line] (node3) -- (node4);

\end{tikzpicture}

\caption{An $\CN=4$ multiplet with a sporadic top component,
decomposed into $SO(3)$ representations.}
\label{N4 sporadic decomposed}
\end{figure}
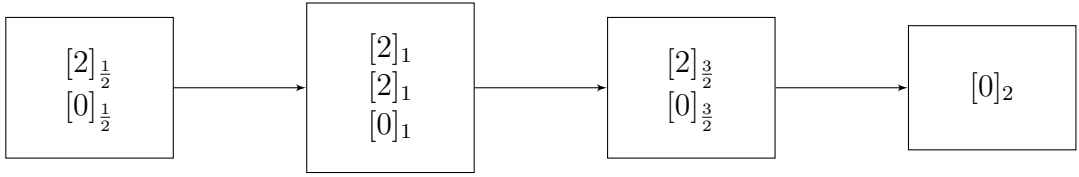

\subsection{$\CN=6$}

Let us examine the $\CN=6$ superconformal algebra before the $\CN=5$
superconformal algebra for the same reason as for the $\CN=4$ and $\CN=3$ algebras.
The R-symmetry is now $SO(6)\simeq SU(4)$\footnote{$SO(6)$ and
$SU(4)$ Dynkin labels are related by exchange of the first two. We choose to use $SO(6)$
labels to be coherent with different values of $\CN$.}, under which supercharges transform
as a vector $[1 \, 0 \, 0]$. We label a superconformal primary as
$[R_1 \, R_2 \, R_3]_{h_0}$. The unitarity bound for $\CN \geq 5$ shall always be given
by (\ref{unicon3}). In this case, it is
\begin{equation}
h_0 \geq \frac{h_1}{2} = \frac{R_1}{2} + \frac{R_2 + R_3}{4}.
\end{equation} We tabulate the list in Table \ref{N6table}.

From this subsection, we do not attempt to give a complete list of short multiplets,
but skip many of those that are irrelevant to the subsequent sections.
Those short multiplets however can be easily reproduced from long multiplets
via the procedure described in \cite{Cordova:2016emh}.

\begin{table} \centering
\begin{tabular}{| c | c | c | c | c | c |} \hline
& Primary & Unitarity bound & Null component & Sporadic top & Generic top \\ \hline \hline
L & $[R_1 \, R_2 \, R_3]_{h_0}$ & $h_0 > \frac{h_1}{2}$ & - & - & $[R_1 \, R_2 \, R_3]_{h_0+3}$ \\ \hline
A & $[1 \, 0 \, 0]_\frac{1}{2}$ & - & $[2 \, 0 \, 0]_1$ & $[0 \, 0 \, 0]_1$ & $[0 \, 0 \, 0]_3$ \\
A & $[0 \, 1 \, 1]_\frac{1}{2}$ & - & $[1 \, 1 \, 1]_1 $ & $[0 \, 0 \, 0]_\frac{3}{2}$ & $[0 \, 0 \, 0]_\frac{5}{2}$ \\
A & $[0 \, 0 \, 2]_\frac{1}{2}$ & - & $[1 \, 0 \, 2]_1$ & - & $[0 \, 0 \, 0]_2$ \\
A & $[0 \, 2 \, 0]_\frac{1}{2}$ & - & $[1 \, 2 \, 0]_1$ & - & $[0 \, 0 \, 0]_2$ \\
A & $[0 \, 0 \, 1]_\frac{1}{4}$ & - & $[1 \, 0 \, 1]_\frac{3}{4}$ & - & $[0 \, 1 \, 0]_\frac{3}{4}$ \\
A & $[0 \, 1 \, 0]_\frac{1}{4}$ & - & $[1 \, 1 \, 0]_\frac{3}{4}$ & - & $[0 \, 0 \, 1]_\frac{3}{4}$ \\
V & $[0 \, 0 \, 0]_0$ & - & $[1 \, 0 \, 0]_\frac{1}{2}$ & - & $[0 \, 0 \, 0]_0$ \\
\hline \end{tabular}
\caption{\label{N6table} List of $\CN=6$ multiplets.}
\end{table}

It proves useful to examine the short multiplet $A[0 \, 1 \, 1]_\frac{1}{2}$ in detail.
Its decomposition into conformal multiplets is given by Figure \ref{N6 sporadic}.
$[0 \, 0 \, 0]_\frac{3}{2}$ at the second level is a sporadic top component because
none of the supercharges, except the one of the highest weight $[1 \, 0 \, 0]$
that annihilates the primary in short multiplets,
can act on the R-singlet to produce the R-vector component at the next level.

\tikzset{block/.style={draw,text width=1.5cm, align=center,  inner sep=2ex, anchor=east, minimum height=4em}}

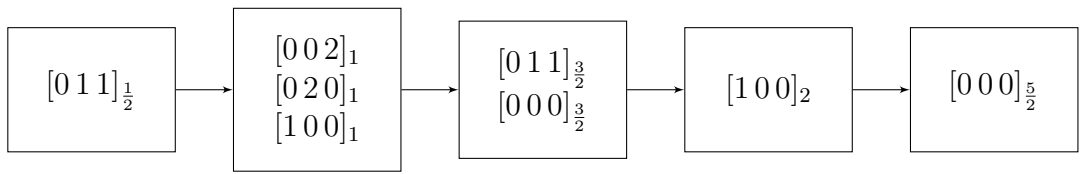
\begin{figure}[!htb]
\centering
\begin{tikzpicture}[node distance=3cm,auto,>=latex']
\node [block] (node1) {$[0 \, 1 \, 1]_\frac{1}{2}$};
\node [block, right of=node1] (node2) {$ [0 \, 0 \, 2]_1$ \\ $[0 \, 2 \, 0]_1$ \\ $[1 \, 0 \, 0]_1$};
\node [block, right of=node2] (node3) {$ [0 \, 1 \, 1]_\frac{3}{2}$ \\ $[0 \, 0 \, 0]_\frac{3}{2}$};
\node [block, right of=node3] (node4) {$ [1 \, 0 \, 0]_2 $};
\node [block, right of=node4] (node5) {$[0 \, 0 \, 0]_\frac{5}{2} $};
\path [line] (node1) -- (node2);
\path [line] (node2) -- (node3);
\path [line] (node3) -- (node4);
\path [line] (node4) -- (node5);
\end{tikzpicture}

\caption{An $\CN=6$ multiplet with a sporadic top component.}
\label{N6 sporadic}
\end{figure}

\subsection{$\CN=5$}\label{N5 multiplets}
\begin{table}[t] \centering
\begin{tabular}{| c | c | c | c | c | c |} \hline
& Primary & Unitarity bound & Null component & Sporadic top & Generic top \\ \hline \hline
L & $[R_1 \, R_2]_{h_0}$ & $h_0 > \frac{h_1}{2}$ & - & - & $[R_1 \, R_2]_{h_0+\frac{5}{2}}$ \\ \hline
A & $[1 \, 0]_\frac{1}{2}$ & - & $[2 \, 0]_1 $ & $[0 \, 0]_1$ & $[0 \, 0]_\frac{5}{2}$ \\
A & $[0 \, 2]_\frac{1}{2}$ & - & $[1 \, 2]_1$ & $[0 \, 0]_\frac{3}{2}$ & $[0 \, 0]_2$ \\
A & $[0 \, 1]_\frac{1}{4}$ & - & $[1 \, 1]_\frac{3}{4}$ & - & $[0 \, 1]_\frac{3}{4}$ \\
V & $[0 \, 0]_0$ & - & $[1 \, 0]_\frac{1}{2}$ & - & $[0 \, 0]_0$ \\
\hline \end{tabular}
\caption{\label{N5table} List of $\CN=5$ multiplets.}
\end{table}
The $\CN=5$ superconformal algebra has the $SO(5)\simeq Sp(4)$ R-symmetry\footnote{$SO(5)$
and $Sp(4)$ Dynkin labels are related by exchange of the two labels.
We choose to use $SO(5)$ labels to be coherent with
different values of $\CN$.} of which the supercharges are
in the vector representation $[1\,0]$. The unitarity
provides a bound on the conformal weight $h_0$ of
a superconformal primary $[R_1 \, R_2]_{h_0}$,
\begin{equation}\label{aaa}
h_0 \geq \frac{h_1}{2} = \frac{R_1}{2} + \frac{R_2}{4}.
\end{equation}
We tabulate a partial list of $\CN=5$ superconformal multiplets in Table \ref{N5table}.

The $A[0 \, 2]_\frac{1}{2}$ multiplet that contains a sporadic top component calls for
special attention. Its decomposition into conformal multiplets is given in Figure
\ref{N5 sporadic}. It is not obvious if the $[0 \, 0]_\frac{3}{2}$ at the second level is
indeed a top component, because the supercharge of weight $[0 \, 0]$ might produce
the $[0 \, 0]_2$ at the third level when acted on the $[0 \, 0]_\frac32$ in the
sense of Racah-Speiser algorithm. However, this is not always the case because the actual
state represented by $[0 \, 0]_\frac{3}{2}$ is some linear combination of elements of
$[0 \, 2]_\frac{1}{2}$ acted by an appropriate combination of two supercharges, and
thus the null condition is highly non-trivial. For details, see \cite{Cordova:2016emh}.
\tikzset{block/.style={draw,text width=1.5cm, align=center,  inner sep=2ex, anchor=east, minimum height=4em}}

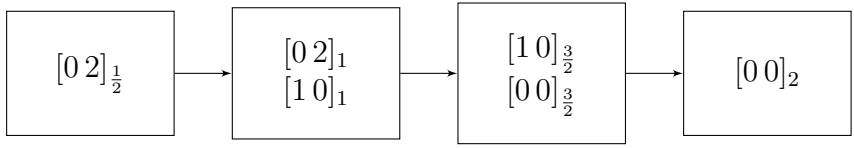
\begin{figure}[!htb]
\centering
\begin{tikzpicture}[node distance=3cm,auto,>=latex']
\node [block] (node1) {$[0 \, 2]_\frac{1}{2}$};
\node [block, right of=node1] (node2) {$\begin{matrix} [0 \, 2]_1 \\ [1 \, 0]_1 \end{matrix}$};
\node [block, right of=node2] (node3) {$\begin{matrix} [1 \, 0]_\frac{3}{2} \\ [0 \, 0]_\frac{3}{2} \end{matrix}$};
\node [block, right of=node3] (node4) {$ [0 \, 0]_2 $};
\path [line] (node1) -- (node2);
\path [line] (node2) -- (node3);
\path [line] (node3) -- (node4);
\end{tikzpicture}

\caption{An $\CN=5$ multiplet with a sporadic top component.}
\label{N5 sporadic}
\end{figure}
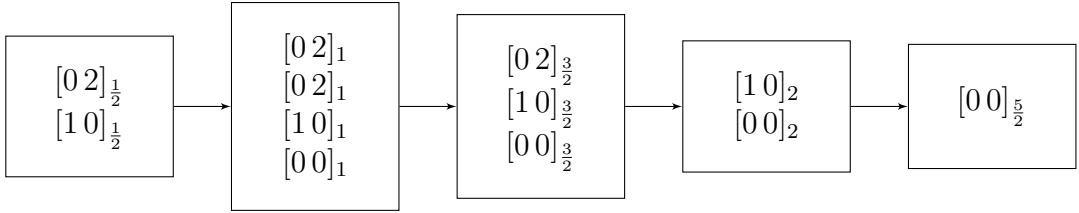
\begin{figure}[!htb]
\centering
\begin{tikzpicture}[node distance=3cm,auto,>=latex']
\node [block] (node1) {$\begin{matrix} [0 \, 2]_\frac{1}{2} \\ [1 \, 0]_\frac{1}{2} \end{matrix}$};
\node [block, right of=node1] (node2) {$\begin{matrix} [0 \, 2]_1 \\ [0 \, 2]_1  \\ [1 \, 0]_1 \\ [0 \, 0]_1 \end{matrix}$};
\node [block, right of=node2] (node3) {$\begin{matrix} [0 \, 2]_\frac{3}{2}  \\ [1 \, 0]_\frac{3}{2} \\ [0 \, 0]_\frac{3}{2} \end{matrix}$};
\node [block, right of=node3] (node4) {$\begin{matrix} [1 \, 0]_2 \\ [0 \, 0]_2 \end{matrix} $};
\node [block, right of=node4] (node5) {$[0 \, 0]_\frac{5}{2} $};
\path [line] (node1) -- (node2);
\path [line] (node2) -- (node3);
\path [line] (node3) -- (node4);
\path [line] (node4) -- (node5);
\end{tikzpicture}

\caption{An $\CN=6$ multiplet with a sporadic top component,
decomposed into $SO(5)$ representations.}
\label{N6 sporadic decomposed}
\end{figure}
In order to argue that $[0 \, 0]_\frac{3}{2}$ is indeed a sporadic top component, let us
consider the $\CN=6$ superconformal theory that can be viewed as a special case of
$\CN=5$ superconformal theory. In this perspective, the $A[0 \, 2]_\frac{1}{2}$ multiplet
in the $\CN=5$ theory is a part of the $A[0 \, 1 \, 1]_\frac{1}{2}$ multiplet in the $\CN=6$
theory described in Figure \ref{N6 sporadic}. As shown in Figure \ref{N6 sporadic decomposed},
decomposition of each conformal primary in $A[0 \, 1 \, 1]_\frac{1}{2}$ into $SO(5)$ representations
shows how the $\CN=5$ short multiplet $A[0 \, 2]_\frac{1}{2}$ can be embedded in the $\CN=6$ multiplet
$A[0 \, 1 \, 1]_\frac{1}{2}$. In particular, we see that the sporadic top component
$[0 \, 0 \, 0]_\frac{3}{2}$ in $A[0 \, 1 \, 1]_\frac{1}{2}$, annihilated by $\CN=6$ supercharges,
can be identified as $[0 \, 0]_\frac{3}{2}$ in $A[0 \, 2]_\frac{1}{2}$. Since
the $\CN=5$ supercharges are a subset of the $\CN=6$ supercharges,
this is sufficient to argue that $[0 \, 0]_\frac{3}{2}$ in Figure
\ref{N5 sporadic} is indeed a sporadic top component as well.

\subsection{$\CN=8$}

Let us examine the case $\CN=8$ before $\CN=7$ for the same reason as before.
The R-symmetry is $SO(8)$ under which supercharges transform in the vector
representation $[1 \, 0 \, 0 \, 0]$, and the superconformal primary is labelled by
$[R_1 \, R_2 \, R_3 \, R_4]_{h_0}$. The unitarity bound is
\begin{equation}
h_0 \geq \frac{h_1}{2} = \frac{R_1 + R_2}{2} + \frac{R_3 + R_4}{4}.
\end{equation}
We tabulate a partial list of $\CN=8$ superconformal multiplets in Table \ref{N8table}.

\begin{table} \centering
\begin{tabular}{| c | c | c | c | c | c |} \hline
& Primary & Unitarity bound & Null component & Sporadic top & Generic top \\ \hline \hline
L & $[R_1 \, R_2 \, R_3 \, R_4]_{h_0}$ & $h_0 > \frac{h_1}{2}$ & - & - & $[R_1 \, R_2 \, R_3 \, R_4]_{h_0+4}$ \\ \hline
A & $[1 \, 0 \, 0 \, 0]_\frac{1}{2}$ & - & $[2 \, 0 \, 0 \, 0]_1$ & $[0 \, 0 \, 0 \, 0]_1$ & $[0 \, 0 \, 0 \, 0]_4$ \\
A & $[0 \, 1 \, 0 \, 0]_\frac{1}{2}$ & - & $[1 \, 1 \, 0 \, 0]_1$ & $[0 \, 0 \, 0 \, 0]_\frac{3}{2}$ & $[0 \, 0 \, 0 \, 0]_\frac{7}{2}$ \\
A & $[0 \, 0 \, 1 \, 1]_\frac{1}{2}$ & - & $[1 \, 0 \, 1 \, 1]_1 $ & $[0 \, 0 \, 0 \, 0]_2$ & $[0 \, 0 \, 0 \, 0]_3$ \\
A & $[0 \, 0 \, 0 \, 2]_\frac{1}{2}$ & - & $[1 \, 0 \, 0 \, 2]_1$ & - & $[0 \, 0 \, 0 \, 0]_\frac{5}{2}$ \\
A & $[0 \, 0 \, 0 \, 1]_\frac{1}{4}$ & - & $[1 \, 0 \, 0 \, 1]_\frac{3}{4}$ & - & $[0 \, 0 \, 1 \, 0]_\frac{3}{4}$ \\
V & $[0 \, 0 \, 0 \, 0]_0$ & - & $[1 \, 0 \, 0 \, 0]_\frac{1}{2}$ & - & $[0 \, 0 \, 0 \, 0]_0$ \\
\hline \end{tabular}
\caption{\label{N8table} List of $\CN=8$ multiplets.}
\end{table}

\tikzset{block/.style={draw,text width=1.5cm, align=center,  inner sep=1.5ex, anchor=east, minimum height=4em}}

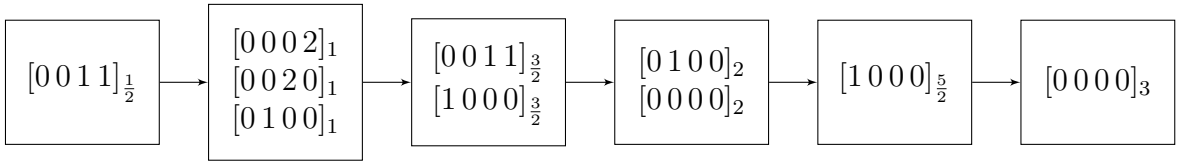
\begin{figure}[b]
\centering
\begin{tikzpicture}[node distance=2.7cm,auto,>=latex']
\node [block] (node1) {$[0 \, 0 \, 1 \, 1]_\frac{1}{2}$};
\node [block, right of=node1] (node2) {$\begin{matrix} [0 \, 0 \, 0 \, 2]_1 \\ [0 \, 0 \, 2 \, 0]_1 \\ [0 \, 1 \, 0 \, 0]_1 \end{matrix}$};
\node [block, right of=node2] (node3) {$\begin{matrix} [0 \, 0 \, 1 \, 1]_\frac{3}{2} \\ [1 \, 0 \, 0 \, 0]_\frac{3}{2} \end{matrix}$};
\node [block, right of=node3] (node4) {$\begin{matrix} [0 \, 1 \, 0 \, 0]_2 \\ [0 \, 0 \, 0 \, 0]_2 \end{matrix}$};
\node [block, right of=node4] (node5) {$[1 \, 0 \, 0 \, 0]_\frac{5}{2} $};
\node [block, right of=node5] (node6) {$[0 \, 0 \, 0 \, 0]_3 $};
\path [line] (node1) -- (node2);
\path [line] (node2) -- (node3);
\path [line] (node3) -- (node4);
\path [line] (node4) -- (node5);
\path [line] (node5) -- (node6);

\end{tikzpicture}

\caption{An $\CN=8$ multiplet with a sporadic top component.}
\label{N8 sporadic}
\end{figure}
It again proves useful to examine the stress-energy tensor multiplet $A[0 \, 0 \, 1 \, 1]_\frac{1}{2}$
in detail. Its decomposition into conformal multiplets is given in Figure \ref{N8 sporadic}.
$[0 \, 0 \, 0 \, 0]_2$ at the third level is a sporadic top component because none of
the supercharges, except the one of the highest weight $[1 \, 0 \, 0 \, 0]$ that annihilates
the primary, can act on the R-singlet $[0 \, 0 \, 0 \, 0]_2$ to
generate an R-vector $[1 \, 0 \, 0 \, 0]_\frac52$
at the next level. This pattern looks similar to other examples with sporadic top
components we have examined in Figure \ref{N4 sporadic} and Figure \ref{N6 sporadic}.
We will discuss this universal feature in section \ref{N9 multiplets}.

\subsection{$\CN=7$}\label{N7 multiplets}
\begin{table} \centering
\begin{tabular}{| c | c | c | c | c | c |} \hline
& Primary & Unitarity bound & Null component & Sporadic top & Generic top \\ \hline \hline
L & $[R_1 \, R_2 \, R_3]_{h_0}$ & $h_0 > \frac{h_1}{2}$ & - & - & $[R_1 \, R_2 \, R_3]_{h_0+\frac{7}{2}}$ \\ \hline
A & $[1 \, 0 \, 0]_\frac{1}{2}$ & - & $[2 \, 0 \, 0]_1$ & $[0 \, 0 \, 0]_1$ & $[0 \, 0 \, 0]_\frac{7}{2}$ \\
A & $[0 \, 1 \, 0]_\frac{1}{2}$ & - & $[1 \, 1 \, 0]_1 $ & $[0 \, 0 \, 0]_\frac{3}{2}$ & $[0 \, 0 \, 0]_3$ \\
A & $[0 \, 0 \, 2]_\frac{1}{2}$ & - & $[1 \, 0 \, 2]_1$ & $[0 \, 0 \, 0]_2$ & $[0 \, 0 \, 0]_\frac{5}{2}$ \\
A & $[0 \, 0 \, 1]_\frac{1}{4}$ & - & $[1 \, 0 \, 1]_\frac{3}{4}$ & - & $[0 \, 0 \, 1]_\frac{3}{4}$ \\
V & $[0 \, 0 \, 0]_0$ & - & $[1 \, 0 \, 0]_\frac{1}{2}$ & - & $[0 \, 0 \, 0]_0$ \\
\hline \end{tabular}
\caption{\label{N7table} List of $\CN=7$ multiplets.}
\end{table}
For the $\CN=7$ superconformal algebra, the R-symmetry is $SO(7)$, under which supercharges transform as
a vector $[1 \, 0 \, 0]$. Labelling the superconformal primary by
$[R_1 \, R_2 \, R_3]_{h_0}$, the unitarity bound on the conformal weight $h_0$ is
\begin{equation}
h_0 \geq \frac{h_1}{2} = \frac{R_1 + R_2}{2} + \frac{R_3}{4}.
\end{equation}
We tabulate a partial list of $\CN=7$ superconformal multiplets in Table \ref{N7table}.

The component $[0\,0\,0]_2$ in the short multiplet $A[0\,0\,2]_\frac12$
is a sporadic top component.  To see this, we use an argument
analogous to those in sections \ref{N3 multiplets} and \ref{N5 multiplets}.
Comparing Figure \ref{N7 sporadic} to Figure \ref{N8 sporadic decomposed},
we can identify $[0\,0\,0]_2$ at the third level of $A[0\,0\,2]_\frac12$
as the sporadic top component $[0\,0\,0\,0]_2$ of
the $\CN=8$ short multiplet $A[0 \, 0 \, 1 \, 1]_\frac{1}{2}$. This implies that
the component $[0\,0\,0]_2$ has to be annihilated by all $\CN=7$ supercharges.
This observation is crucial because this is the only supersymmetric
$[0 \, 0 \, 0]_2$ component one can find in the $\CN=7$ superconformal algebra,
which however is required for the existence of stress-energy tensor.

\tikzset{block/.style={draw,text width=1.5cm, align=center,  inner sep=1.5ex, anchor=east, minimum height=4em}}

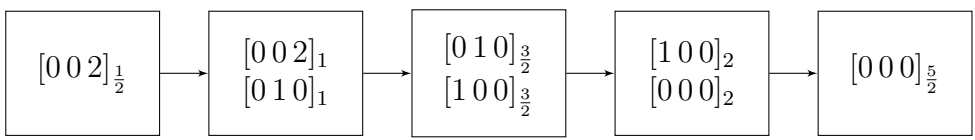
\begin{figure}[!htb]
\centering
\begin{tikzpicture}[node distance=2.7cm,auto,>=latex']
\node [block] (node1) {$[0 \, 0 \, 2]_\frac{1}{2}$};
\node [block, right of=node1] (node2) {$\begin{matrix} [0 \, 0 \, 2]_1 \\ [0 \, 1 \, 0]_1 \end{matrix}$};
\node [block, right of=node2] (node3) {$\begin{matrix} [0 \, 1 \, 0]_\frac{3}{2} \\ [1 \, 0 \, 0]_\frac{3}{2} \end{matrix}$};
\node [block, right of=node3] (node4) {$ \begin{matrix} [1 \, 0 \, 0]_2 \\ [0 \, 0 \, 0]_2 \end{matrix}$};
\node [block, right of=node4] (node5) {$[0 \, 0 \, 0]_\frac{5}{2} $};
\path [line] (node1) -- (node2);
\path [line] (node2) -- (node3);
\path [line] (node3) -- (node4);
\path [line] (node4) -- (node5);
\end{tikzpicture}

\caption{An $\CN=7$ multiplet with a sporadic top component.}
\label{N7 sporadic}
\end{figure}

\tikzset{block/.style={draw,text width=1.5cm, align=center,  inner sep=1.5ex, anchor=east, minimum height=4em}}

\begin{figure}[!htb]
\centering
\begin{tikzpicture}[node distance=2.7cm,auto,>=latex']
\node [block] (node1) {$\begin{matrix} [0 \, 1 \, 0]_\frac{1}{2} \\ [0 \, 0 \, 2]_\frac{1}{2} \end{matrix}$};
\node [block, right of=node1] (node2) {$\begin{matrix} [0 \, 0 \, 2]_1 \\ [0 \, 0 \, 2]_1 \\ [1 \, 0 \, 0]_1 \\ [0 \, 1 \, 0]_1 \end{matrix}$};
\node [block, right of=node2] (node3) {$\begin{matrix} [0 \, 1 \, 0]_\frac{3}{2} \\ [0 \, 0 \, 2]_\frac{3}{2} \\ [1 \, 0 \, 0]_\frac{3}{2} \\ [0 \, 0 \, 0]_\frac{3}{2} \end{matrix}$};
\node [block, right of=node3] (node4) {$\begin{matrix} [1 \, 0 \, 0]_2 \\  [0 \, 1 \, 0]_2 \\ [0 \, 0 \, 0]_2 \end{matrix}$};
\node [block, right of=node4] (node5) {$\begin{matrix} [1 \, 0 \, 0]_\frac{5}{2} \\ [0 \, 0 \, 0]_\frac{5}{2} \end{matrix}$};
\node [block, right of=node5] (node6) {$[0 \, 0 \, 0 \, 0]_3 $};
\path [line] (node1) -- (node2);
\path [line] (node2) -- (node3);
\path [line] (node3) -- (node4);
\path [line] (node4) -- (node5);
\path [line] (node5) -- (node6);

\end{tikzpicture}

\caption{An $\CN=8$ multiplet with a sporadic top component,
decomposed into $SO(7)$ representations.}
\label{N8 sporadic decomposed}
\end{figure}
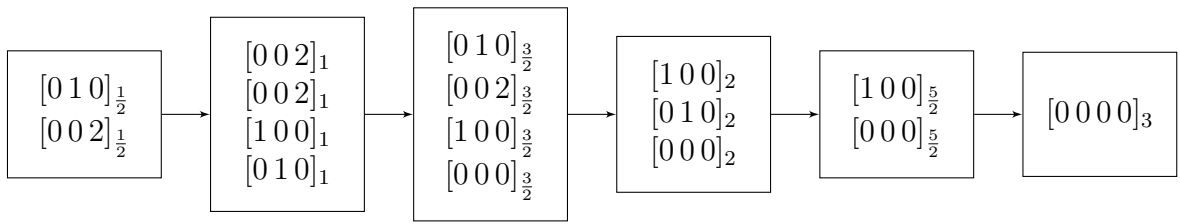

\subsection{$\CN \geq 9$}\label{N9 multiplets}

Having worked out up to $\CN=8$, which was necessary to manifest existence
of stress-energy tensors in all $\CN$, we are ready to generalize the patterns into
generic values of $\CN$. Under the R-symmetry group $SO(\CN)$, $\CN$ being even
or odd, $\CN$ supercharges transform as the vector $[1 \, 0 \cdots 0]$.
The superconformal primary is labelled as $[R_1 \cdots R_r]_{h_0}$ where
$r=[\frac{\CN}{2}]$ is the rank of the R-symmetry group. Let us repeat the
unitarity condition (\ref{unicon3}) for completeness:
\begin{equation}\label{unicon3rep}
2 h_0 \geq h_1 =
\begin{cases}
R_1 + \cdots + R_{r-2} + \frac{R_{r-1}+R_r}{2} & \CN \text{ is even}, \\
R_1 + \cdots + R_{r-1} + \frac{R_r}{2} & \CN \text{ is odd}.
\end{cases}
\end{equation}
\begin{table}[b] \centering
\begin{tabular}{| c | c | c | c | c | c |} \hline
& Primary & Unitarity bound & Null component & Sporadic top & Generic top \\ \hline \hline
L & $[R_1 \cdots R_r]_{h_0}$ & $h_0 > \frac{h_1}{2}$ & - & - & $[R_1 \cdots \, R_r]_{h_0+\frac{\CN}{2}}$ \\ \hline
A & $[1 \, 0 \cdots 0]_\frac{1}{2}$ & - & $[2 \, 0 \cdots 0]_1$ & $[0 \cdots 0]_1$ & $[0 \cdots 0]_\frac{\CN}{2}$ \\
A & $[0 \, 1 \, 0 \cdots 0]_\frac{1}{2}$ & - & $[1 \, 1 \, 0 \cdots 0]_1$ & $[0 \cdots 0]_\frac{3}{2}$ & $[0 \cdots 0]_\frac{\CN-1}{2}$ \\
\vdots & \vdots & \vdots  & \vdots & \vdots & \vdots \\
A & $ [0 \cdots 0 \, 1 \, 1]_\frac{1}{2}$ & - & $[1 \, 0 \cdots 0 \, 1 \, 1]_1$ & $ [0 \cdots 0]_\frac{r}{2}$ & $[0 \, 0 \, 0 \, 0]_\frac{r+2}{2}$ \\
A & $[0 \cdots 2]_\frac{1}{2}$ & - & $[1 \, 0 \cdots 0 \, 2]_1$ & - & $[0 \cdots 0]_\frac{r+1}{2}$ \\
A & $[0 \cdots 1]_\frac{1}{4}$ & - & $[1 \, 0 \cdots 0 \, 1]_\frac{3}{4}$ & - & $[0 \cdots 0 \, 1 \, 0]_\frac{3}{4}$ \\
V & $[0 \, 0 \, 0 \, 0]_0$ & - & $[1 \, 0 \, 0 \, 0]_\frac{1}{2}$ & - & $[0 \, 0 \, 0 \, 0]_0$ \\
\hline \end{tabular}
\caption{\label{Netable} List of $\CN = 10,12,\cdots$ multiplets.}
\end{table}
\begin{table} \centering
\begin{tabular}{| c | c | c | c | c | c |} \hline
& Primary & Unitarity bound & Null component & Sporadic top & Generic top \\ \hline \hline
L & $[R_1 \cdots R_r]_{h_0}$ & $h_0 > \frac{h_1}{2}$ & - & - & $[R_1 \cdots \, R_r]_{h_0+\frac{\CN}{2}}$ \\ \hline
A & $[1 \, 0 \cdots 0]_\frac{1}{2}$ & - & $[2 \, 0 \cdots 0]_1$ & $[0 \cdots 0]_1$ & $[0 \cdots 0]_\frac{\CN}{2}$ \\
A & $[0 \, 1 \, 0 \cdots 0]_\frac{1}{2}$ & - & $[1 \, 1 \, 0 \cdots 0]_1$ & $[0 \cdots 0]_\frac{3}{2}$ & $[0 \cdots 0]_\frac{\CN-1}{2}$ \\
\vdots & \vdots & \vdots  & \vdots & \vdots & \vdots \\
A & $[0 \cdots 0 \, 2]_\frac{1}{2} $ & - & $[1 \, 0 \cdots 0 \, 2]_1 $ & $[0 \cdots 0]_\frac{r+1}{2} $ & $[0 \cdots 0]_\frac{r+2}{2}$ \\
A & $[0 \cdots 0 \, 1]_\frac{1}{4}$ & - & $[1 \, 0 \cdots 0 \, 1]_\frac{3}{4}$ & - & $[0 \cdots 0 \, 1]_\frac{3}{4}$ \\
V & $[0 \, 0 \, 0 \, 0]_0$ & - & $[1 \, 0 \, 0 \, 0]_\frac{1}{2}$ & - & $[0 \, 0 \, 0 \, 0]_0$ \\
\hline \end{tabular}
\caption{\label{Notable} List of $\CN =9,11,\cdots$ multiplets.}
\end{table}

We tabulate partial lists of superconformal multiplets in Tables \ref{Netable} and
\ref{Notable}, for even and odd $\CN$. A generic pattern is apparent.
Figure \ref{Ngeneric} shows a generic short superconformal multiplet
$A[0 \cdots 0 \, 1 \, 0 \cdots 0]$ with $1$ being the $k^{th}$ Dynkin label,
decomposed into conformal primaries. This multiplet has two top components,
one of which is placed at the level $k$ and the other at the level $(\CN-k)$.
Both top components are R-singlets.

\tikzset{block/.style={draw,text width=3cm, align=center,  inner sep=1ex, anchor=east, minimum height=4em}}
\tikzset{check/.style={text width=2cm, align=center,  inner sep=0ex, anchor=east, minimum height=4em}}
\begin{figure}[!htb]
\centering

\begin{tikzpicture}[node distance=2cm,auto,>=latex']
\node [block] (node1) {$[\wedge^{k}V]_\frac{1}{2}$: \\ $\quad$ \\ $[ \underbrace{0 \cdots 0 }_{k-1} \, 1 \, 0 \, \cdots 0]_\frac{1}{2}$ };
\node [check, above of=node1] (node11) {};
\node [check, right of=node11] (node12) {};
\node [block, right of=node12] (node13) {$[\wedge^{k+1}V]_1$: \\ $\quad$ \\ $[ \underbrace{0 \cdots 0 }_{k} \, 1 \, 0 \, \cdots 0]_1$};
\node [check, right of=node13] (node14) {};
\node [check, right of=node14] (node15) {$\cdots$};
\node [check, right of=node15] (node17) {};
\node [block, right of=node17] (node18) {$[\wedge^{\CN}V]_\frac{\CN -k+1}{2}$: \\ $\quad$ \\ $[ 0 \cdots 0]_{\frac{\CN-k+1}{2}}$};
\node [check, below of=node1] (node21) {};
\node [check, right of=node21] (node22) {};
\node [block, right of=node22] (node23) {$[\wedge^{k-1}V]_1$: \\ $\quad$ \\ $[ \underbrace{0 \cdots 0 }_{k-2} \, 1 \, 0 \, \cdots 0]_1$};
\node [check, right of=node23] (node24) {};
\node [check, right of=node24] (node25) {$\cdots$};
\node [check, right of=node25] (node26) {};
\node [block, right of=node26] (node27) {$[\wedge^{0}V]_\frac{k+1}{2}$: \\ $\quad$ \\ $[ 0 \cdots 0]_{\frac{k+1}{2}}$};
\path [line] (node1) -- (node13);
\path [line] (node13) -- (node15);
\path [line] (node15) -- (node18);
\path [line] (node1) -- (node23);
\path [line] (node23) -- (node25);
\path [line] (node25) -- (node27);

\end{tikzpicture}

\caption{A generic short multiplet with primary $[\wedge^k V]_\half$ for generic $\CN$,
written in both Dynkin labels and anti-symmetric tensor product notation.}
\label{Ngeneric}
\end{figure}
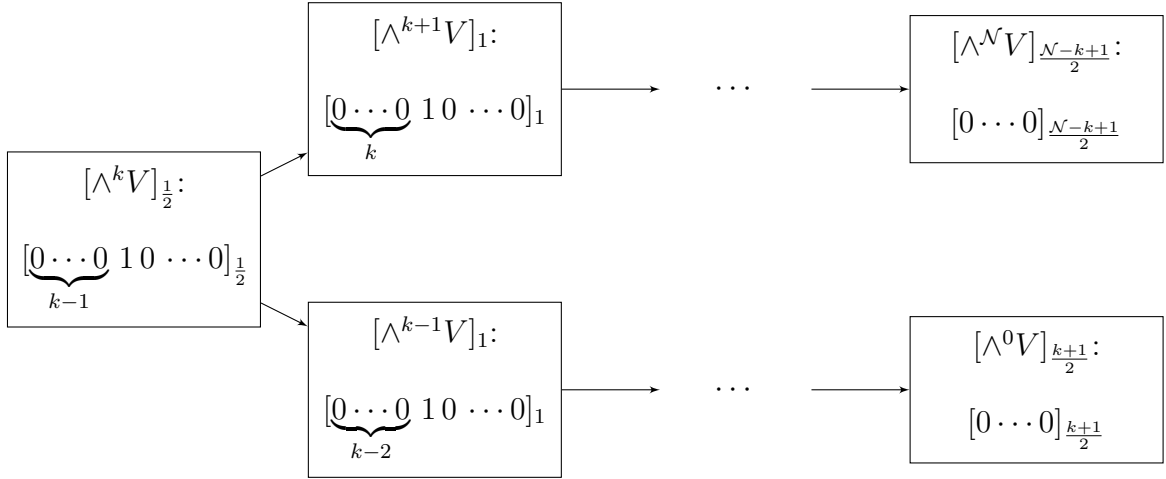

Structure of this decomposition is physically clearer if we interpret the R-representation
$[0 \cdots 0 \, 1 \, 0 \cdots 0]$ as the $k^{th}$ anti-symmetric product of vector
representations, denoted as $\wedge^k V$. This structure is further justified by
the fact that the tensor product
\begin{equation}\label{wedgeproducts}
\wedge^k V \otimes \wedge^1 V \supset \wedge^{k+1} V \oplus \wedge^{k-1} V,
\end{equation}
where the other parts vanish in short multiplets. Note that $\wedge^1 V$ here represents
the supersymmetry $G^a_{-1/2}$ that takes a component to the next level. This not only
simplifies the notation, but also manifests the fact that midway in the decomposition
\begin{align}\label{midway representation}
\begin{cases}
[\wedge^{N} V]: \quad [0 \cdots 0 \, 2 \, 0 ] \oplus [0 \cdots 0 \, 0 \, 2] & \text{for SO}(2 N) \\
[\wedge^{N} V], [\wedge^{N+1}V] : \quad [0 \cdots 0 \, 2] & \text{for SO}(2 N+1)
\end{cases}
\end{align}
appear, and also explains clearly why there are two `towers' of components that
both terminate with an R-singlet: a scalar ($\wedge^{0} V$) or a pseudoscalar
($\wedge^{\CN} V$). However, the most important role it plays will become clear
in section \ref{subsecN6stress}.

\section{Properties of the Multiplets}\label{propertysection}

Given the lists of multiplets, many implications and applications are in order.
In this section, we discuss the recombination phenomenon that happens
when the conformal weight $h$ of a long multiplet $L[R]_{h}$
hits the unitarity bound. We also present character formulae
for both long and short global superconformal multiplets.

\subsection{Recombination Rules}\label{subsecrecomb}
Decomposition of superconformal multiplets into conformal multiplets makes the
recombination rules extremely apparent. For instance, let us consider
an example of a long multiplet $L[0 \, 1]_{h}$ of the $\CN=5$ superconformal algebra,
where $h$ is bound by $h \geq \frac{1}{4}$ \eqref{aaa}.
As $h$ approaches $1/4$, the long multiplet $L[0\,1]_{h\to\frac14}$ splits into
two short multiplets, one of which is $A[0\,1]_{\frac14}$ with
the same quantum numbers and the other is $A[1\,1]_{\frac34}$
that contains the null states of $A[0\;1]_\frac14$,
\begin{equation}\label{bbb}
  L[0 \, 1]_h \, \xrightarrow{\quad  h \rightarrow \frac{1}{4} \quad} \,
  A[0 \, 1]_\frac{1}{4} \oplus A[1 \, 1]_\frac{3}{4}.
\end{equation}
The above recombination rule is apparently demonstrated in Figure \ref{N5 recomb}.
\tikzset{block/.style={draw,text width=1.5cm, align=center,  inner sep=1.5ex, anchor=east, minimum height=4em}}
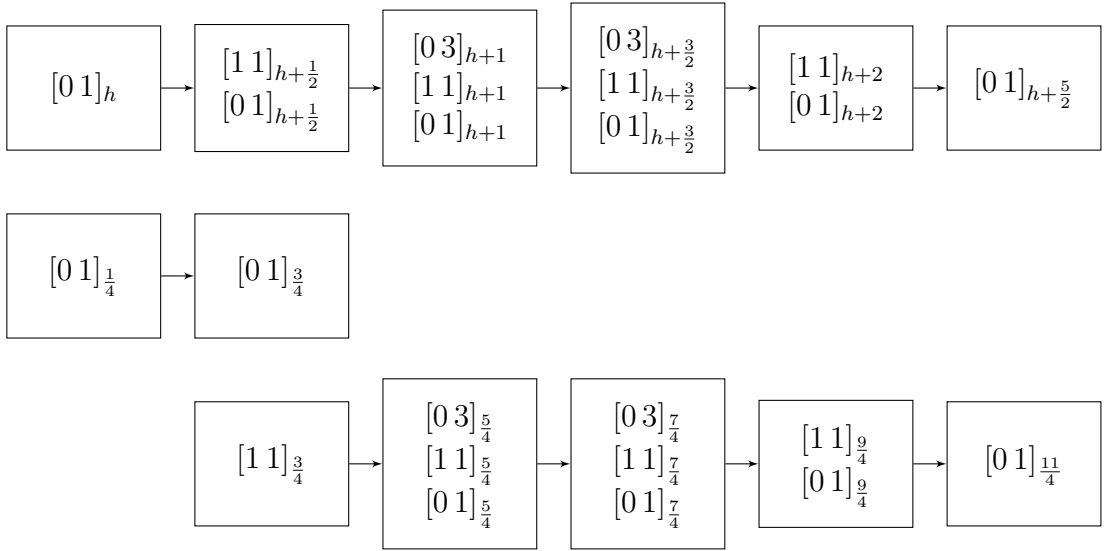
\begin{figure}[t]
\centering
\begin{tikzpicture}[node distance=2.5cm,auto,>=latex']
\node [block] (node1) {$[0 \, 1]_h$};
\node [block, right of=node1] (node2) {$\begin{matrix} [1 \, 1]_{h+\frac{1}{2}} \\ [0 \, 1]_{h+\frac{1}{2}} \end{matrix}$};
\node [block, right of=node2] (node3) {$\begin{matrix} [0 \, 3]_{h+1} \\ [1 \, 1]_{h+1} \\ [0 \, 1]_{h+1} \end{matrix}$};
\node [block, right of=node3] (node4) {$\begin{matrix} [0 \, 3]_{h+\frac{3}{2}} \\ [1 \, 1]_{h+\frac{3}{2}} \\ [0 \, 1]_{h+\frac{3}{2}} \end{matrix}$};
\node [block, right of=node4] (node5) {$\begin{matrix} [1 \, 1]_{h+2} \\ [0 \, 1]_{h+2} \end{matrix}$};
\node [block, right of=node5] (node6) {$[0 \, 1]_{h+\frac{5}{2}}$};

\node [block, below of=node1] (node11) {$[0 \, 1]_{\frac{1}{4}}$};
\node [block, right of=node11] (node12) {$[0 \, 1]_{\frac{3}{4}}$};

\node [block, below of=node12] (node21) {$[1 \, 1]_{\frac{3}{4}}$};
\node [block, right of=node21] (node22) {$\begin{matrix} [0 \, 3]_{\frac{5}{4}} \\ [1 \, 1]_{\frac{5}{4}} \\ [0 \, 1]_{\frac{5}{4}} \end{matrix}$};
\node [block, right of=node22] (node23) {$\begin{matrix} [0 \, 3]_{\frac{7}{4}} \\ [1 \, 1]_{\frac{7}{4}} \\ [0 \, 1]_{\frac{7}{4}} \end{matrix}$};
\node [block, right of=node23] (node24) {$\begin{matrix} [1 \, 1]_{\frac{9}{4}} \\ [0 \, 1]_{\frac{9}{4}} \end{matrix}$};
\node [block, right of=node24] (node25) {$\begin{matrix} [0 \, 1]_{\frac{11}{4}} \end{matrix}$};
\path [line] (node1) -- (node2);
\path [line] (node2) -- (node3);
\path [line] (node3) -- (node4);
\path [line] (node4) -- (node5);
\path [line] (node5) -- (node6);
\path [line] (node11) -- (node12);
\path [line] (node21) -- (node22);
\path [line] (node22) -- (node23);
\path [line] (node23) -- (node24);
\path [line] (node24) -- (node25);
\end{tikzpicture}

\caption{One long $L[0\,1]_h$ and two short $\CN=5$ multiplets
$A[0\,1]_{\frac14}$ and $A[1\,1]_{\frac34}$.}
\label{N5 recomb}
\end{figure}

In fact, this recombination rule generalizes to any long multiplets with generic
$\CN$.\footnote{The following argument holds for $\CN \geq 3$, except for the
small $\CN=4$.} We present a rather wordy proof for this statement.

Consider an arbitrary long multiplet $L[R_1 \cdots R_r]_h$ of a global subalgebra
with any value of $\CN$.
As $h$ approaches the unitarity bound $\frac{h_1}{2}$ in accordance with (\ref{unicon3}),
its components are classified into two: they either belong to the short multiplet
$A[R_1 \cdots R_r]_{h_1 /2}$, or are null components of the short multiplet.
According to Racah-Speiser algorithm, any components at level $l$ have
R-symmetry Dynkin labels that can be obtained by adding $l$
different weights of the vector representation to that of the primary. We refer to these
$l$ weights as a \emph{path from the primary}. Then, the null components can again
be classified by whether the path includes the highest weight $[1 \, 0 \cdots 0]$,
or $[1;1]$ for the large $\CN=4$, or not:

\begin{enumerate}
\item Null components whose path from the primary does include the highest weight
$[1 \, 0 \cdots 0]$, form a short multiplet $A[R_1+1 \cdots R_r]_{\frac{h_1}{2}+\frac{1}{2}}$.
Such components are always included in the short multiplet, because they are obtained
by adding $l-1$ different weights of the vector representation
to $[R_1+1 \cdots R_r]$. Also, it is obvious that every component of the short multiplet
$A[R_1+1 \cdots R_r]_{\frac{h_1}{2}+\frac{1}{2}}$ appears as a null component in the
original short multiplet $A[R_1 \cdots R_r]_{\frac{h_1}{2}}$.

\item Null components whose path from the primary does \emph{not} include the highest
weight $[1 \, 0 \cdots 0]$ arise only when first $k \geq 1$ Dynkin labels $R_1, \cdots , R_k$
are zero. In such a case, some of the R-symmetry group's lowering operators annihilate
the primary, thus not only the highest weight supercharge but also some lowered supercharges
annihilate the primary. See section 3.3.3. of \cite{Cordova:2016emh}.

Such components complicate the argument because they do not seem to be included
in the short multiplet $A[R_1+1 \cdots R_r]_{\frac{h_1}{2}+\frac{1}{2}}$. However,
these components are eliminated among themselves in the
Racah-Speiser algorithm, and thus do not appear in the long multiplet.
In other words, all components of the long multiplet that becomes null as it hits
the unitarity bound fall into enumeration 1 above.

To see how, consider one of such components whose path from the primary includes
$[0 \cdots 0 \, -1 \, 1 \, 0 \cdots 0]$ but not
$[1 \, 0 \cdots 0], \cdots ,[0 \cdots -1 \, 1 \, 0 \, 0 \cdots 0]$.
$R_1=\cdots = R_k =0$ with sufficiently large $k$ is required for this to be a null
component of $A[R_1 \cdots R_r]_{\frac{h_1}{2}}$. For this component to appear in the
decomposition of long multiplet, that is, to avoid a Dynkin label equal to $-1$, the path
from the primary must also include either $[0 \cdots 0 \, 1 \, -1 \, 0 \cdots 0]$ or
$[0 \cdots 1 \, -1 \, 0 \, 0 \cdots 0]$ but not both.
Therefore, there is a one-to-one correspondence between components that belong to the
enumeration 2: one whose path from the primary includes the former but not the latter and
vice versa. However, treatment of Dynkin label equal to $-2$ by the RS algorithm precisely
cancels the two.
\end{enumerate}

Therefore, all components that appear in the long multiplet $L[R_1 \cdots R_r]_h$ fall
into either the short $A[R_1 \cdots R_r]_{\frac{h_1}{2}}$ or another short
$A[R_1+1 \cdots R_r]_{\frac{h_1}{2}+\frac{1}{2}}$, and we can write the generic
recombination rule as follows:
\begin{equation}\label{genrecomb}
L[R_1 \cdots R_r]_h \, \xrightarrow{\quad  h \rightarrow \frac{h_1}{2} \quad} \, A[R_1 \cdots R_r]_{\frac{h_1}{2}} \oplus A[R_1+1 \cdots R_r]_{\frac{h_1}{2}+\frac{1}{2}}.
\end{equation}

Non-generic cases are easy to examine because the decompositions only contain
few components. We simply state the result.

\begin{align}
\label{N1recomb} &
\begin{cases}
L[0]_h \, \xrightarrow{\quad  h \rightarrow 0 \quad} \, A[0]_0 \oplus L[0]_\frac{1}{2}
\end{cases}
& \CN=1, \\
\label{N2recomb} &
\begin{cases}
L[j_0>0]_h \, & \xrightarrow{\quad  h \rightarrow \frac{j_0}{2} \quad} \, A[j_0]_\frac{j_0}{2} \oplus  A[j_0+1]_{\frac{j_0}{2} +\frac{1}{2}} \\
L[j_0<0]_h \, & \xrightarrow{\quad  h \rightarrow -\frac{j_0}{2} \quad} \, A[j_0]_{-\frac{j_0}{2}} \oplus  A[j_0-1]_{-\frac{j_0}{2} +\frac{1}{2}} \\
L[0]_h \, & \xrightarrow{\quad  h \rightarrow 0 \quad} \, A[1]_{\frac{1}{2}} \oplus A[-1]_{\frac{1}{2}}
\end{cases}
& \CN=2, \\
\label{N4recomb} &
\begin{cases}
L[j_0]_h \, \xrightarrow{\quad  h \rightarrow \frac{j_0}{2} \quad} \, & A[j_0]_\frac{j_0}{2} \oplus A[j_0+1]_{\frac{j_0}{2}+\frac{1}{2}} \\ & \oplus A[j_0+1]_{\frac{j_0}{2}+\frac{1}{2}} \oplus A[j_0+2]_{\frac{j_0}{2}+1}
\end{cases} & \text{small } \CN=4 .
\end{align}

A consistency check is to compare the number of physical states
on both sides of the equations (\ref{genrecomb})-(\ref{N4recomb}). This can be done
by adding up dimensions of all R-representations in the decomposition of each
multiplet on both sides.

To count the number of physical states, we first combine
a left-moving and a right-moving multiplet into a two-sided
superconformal multiplet. A two-sided multiplet can be labeled by
two letters with one unbarred and one barred to
indicate the left-moving and the right-moving null states. For instance,
a multiplet $$L\bar A[R_l|R_r]_{h,\bar h}$$ can be understood
as a tensor product of the left-moving multiplet $L[R_l]_h$
and the right-moving multiplet $A[R_r]_{\bar h}$. Components of the
two-sided multiplets are also given by tensor products of
components in respective sectors, which we denote as
$$[R_l]_h \otimes \overline{[R_r]}_{\bar h}$$
with a bar over the right-moving component.

When the recombination phenomenon happens for the left-moving
sector, the number of states of each physical two-sided multiplet
can be obtained by dimension of the left-moving sector multiplet multiplied by a
common factor, which is dimension of the common right-moving sector multiplet.
One exception is that the component $[0]_0$ in the left-moving sector
has to be counted as zero. This is because it combines with
the right-mover to lead to a conserved current.
Conserved current will be discussed in section
\ref{subsecconcur}, and comparing dimensions in recombination
rules has been discussed in detail in \cite{Cordova:2016emh}.

We can see that the number of physical states on both sides of the recombination
rules indeed agree level-by-level, with one exception: let us consider a
recombination phenomenon where the long multiplet has
an R-singlet primary. This long multiplet $L[0]_h$ splits into
the vacuum multiplet $V[0]_0$ and a short multiplet $A[1 \, 0 \cdots 0]_\half$
as $h\to 0$. As discussed in the last paragraph, the vacuum multiplet
has to be counted as zero. On the other hand,
the short multiplet contains a component $[0]_1$ at the level one
that does not appear as a conformal primary in the long multiplet $L[0]_{h \to 0}$,
but appears as a conformal descendant of the superconformal primary $[0]_{h \to 0}$.
This will be further discussed in the next subsection.
To summarize, the primary $[0]_{h\to 0}$ and its descendants of the long
multiplet are split into the conserved current $[0]_0$ which is the vacuum multiplet,
and $[0]_1$ and its descendants which appear in the short multiplet.

We find that unlike in higher dimensions, absolutely protected multiplets do not exist.
A multiplet is absolutely protected when it does not appear in any of the recombination
rules, so its spectrum is constant on the supersymmetric conformal manifold.
Deformations of CFT are discussed in section
\ref{subsecdeform}, see also \cite{Cordova:2016xhm}. This is a consequence of the
absence of isolated short multiplets: every short multiplet in two dimensions appears
in the unitarity limit of the long multiplet with same quantum numbers.

\subsection{Character Formulae for the Global Multiplets}\label{subsecchar}

Let us present character formulae for unitary representations,
both long and short, of a superconformal algebra. A character
of a superconformal representation can be defined as
\begin{equation}\label{definechar}
\text{Ch}^{(g)}_{h,R}(\tau,\{z_i \})
= \text{Tr}_{V(h,R)}\Big[e^{2 \pi i \tau (L_0-\frac{c}{24})} \prod_i e^{2 \pi i z_i T^i_0}\Big]
= \text{Tr}_{V(h,R)}\Big[q^{L_0-\frac{c}{24}}\prod_i \big( {y_i}^{T^i_0}\big) \Big],
\end{equation}
where $V(h,R)$ denotes a representation built on a primary
of conformal weight $h$ and R-charge Dynkin label $R$. Here
$T^i_0$ are the Cartan generators of the R-symmetry group,
and $q=e^{2 \pi i \tau}, y=e^{2 \pi i z}$.
Although the central charge $c$ is irrelevant to the global subalgebra,
we include its contribution to make connection with super Virasoro characters.
See section \ref{subsecsvc}.

To express superconformal characters, it is rather convenient to
use orthogonal basis for the special orthogonal
Lie group $SO(\CN)$ than the fundamental
basis chosen in the previous section. We can find
the linear relation between
fundamental weights $[R_1 \cdots R_r]$ and orthogonal weights
$[h_1 \cdots h_r]$ of the SO($\CN$)
in \eqref{defineorthogonaleven} and \eqref{defineorthogonalodd}.

Character formulae for long multiplets follow directly from the structure
of the multiplets. Given a superconformal primary, which is an
irreducible representation of the R-symmetry group SO($\CN$),
superconformal descendants are obtained by
applying successively the
supercharges $G_{-1/2}$ and the Virasoro generator $L_{-1}$
on the primary. The supercharges are
in the vector representation of the $SO(\CN)$ and
their orthogonal weights are
\begin{equation}
 G^a_{-1/2} \ :  \quad [\pm 1 \, 0 \cdots 0], [0 \, \pm1 \cdots 0], \cdots ,
 [0 \cdots 0 \, \pm1] \underbrace{, [0 \cdots 0]}_{\text{for odd } \CN}. \nonumber
\end{equation}
The character formula for a long multiplet $L[\{h_i\}]_h$
then becomes
\begin{align}\label{gcho}
\text{Ch}_{h,\{h_i\}}^{(g)}(\t,\{z_j\}) =
q^{h-\frac{c}{24}}
\chi_{o}(\{h_i\})
\cdot
\frac{ \displaystyle (1+q^{1/2}) \bigg( \prod_{i=1}^{r}\prod_{\e = \pm 1}
(1+y_i^{\e} q^{1/2} ) \bigg)}{1-q}
\end{align}
for odd $\CN$ with rank $r$, and
\begin{align}\label{gche}
\text{Ch}_{h,\{h_i\}}^{(g)}(\t,\{z_j\}) =
q^{h-\frac{c}{24}}
\chi_{e}(\{h_i\})
\cdot
\frac{ \displaystyle  \bigg( \prod_{i=1}^{r}\prod_{\e=\pm 1}
(1+y_i^{\e} q^{1/2} ) \bigg)}{1-q}
\end{align}
for even $\CN$ with rank $r$. Here
$\chi_o(\{h_i\})$ and $\chi_e(\{h_i\})$ denote
the Weyl character formula for $\mathfrak{so}(\CN)$
with odd and even $\CN$ respectively \cite{Fulton:1991},
\begin{align}\label{Weylchar}
  \chi_{o}(\{h_i\}) =
  \frac{ \left\lvert y_j^{h_i+r-i+\frac{1}{2}} - y_j^{-(h_i+r-i+\frac{1}{2})} \right\rvert}
  { \left\lvert y_j^{r-i+\frac{1}{2}} - y_j^{-(r-i+\frac{1}{2})} \right\rvert} ,~
  \chi_e(\{h_i\})   = \frac{\displaystyle \sum_{\e=\pm 1}  \left\lvert y_j^{h_i+r-i} + \e y_j^{-(h_i+r-i)} \right\rvert}
  { \left\lvert y_j^{r-i} + y_j^{-(r-i)} \right\rvert},
\end{align}
where $|a_{ij}|$ denotes determinant of the matrix with indices $i,j=1,2,\cdots,r$.
The structure of these characters is straightforward. The primary contributes $q^{h-\frac{c}{24}}$ multiplied by Weyl character formula
corresponding to its R weights. The other factors in \eqref{gcho}
and \eqref{gche} account for the contribution from
the superconformal descendants.

We need more elaborations to obtain the superconformal
characters for short multiplets $A[\{h_i\}]_{h}$ due to
the presence of null states. One might naively remove the factor
$(1+y_1 q^{1/2})$ that accounts for the highest weight supercharge
that produces the null states. However, although it is highly obscured in the
Racah-Speiser algorithm, it is not actually the highest weight supercharge
that produces the null states, but it is the highest weight representation obtained
from the highest weight supercharge acting on the highest weight of the primary.

The short multiplet character can be derived using the recombination rule
from section \ref{subsecrecomb}. We can rewrite the recombination rule
(\ref{genrecomb}) using the characters of global long multiplets
$\text{Ch}_{h,\{h_i\}}^{(g)}(\t,\{z_j\})$ and of global short multiplets
$\chi_{h_1/2,\{h_i\}}^{(g)}(\t,\{z_j\})$,
\begin{equation}
\lim_{h \rightarrow \frac{h_1}{2}} \text{Ch}_{h,\{h_i\}}^{(g)}(\t,\{z_j\}) =
\chi_{\frac{h_1}{2},\{h_i\}}^{(g)}(\t,\{z_j\}) +
\chi_{\frac{h_1+1}{2},\{h_i + \delta_{i,1}\}}^{(g)}(\t,\{z_j\}).
\end{equation}
Note that the lowest exponent of $q$ in the second short character on the RHS
is larger by $\frac12$ than that in the first. This allows us to write the character of
a short multiplet in terms of that of long multiplets perturbatively and to all orders.
For simplicity, we omit the arguments $(\t,\{z_j\})$ of each character.
\begin{align}
\chi_{\frac{h_1}{2},\{h_i\}}^{(g)} &=
\text{Ch}_{h \rightarrow \frac{h_1}{2},\{h_i\}}^{(g)}
- \chi_{\frac{h_1+1}{2},\{h_i + \delta_{i,1}\}}^{(g)} \nonumber \\
&= \text{Ch}_{h \rightarrow \frac{h_1}{2},\{h_i\}}^{(g)}
- \text{Ch}_{h \rightarrow \frac{h_1+1}{2},\{h_i+ \delta_{i,1}\}}^{(g)}
+\chi_{\frac{h_1+2}{2},\{h_i+ 2 \delta_{i,1}\}}^{(g)} \nonumber \\
&= \sum_{n=0}^\infty (-1)^n \text{Ch}_{h \rightarrow \frac{h_1+n}{2},\{h_i+ n \delta_{i,1}\}}^{(g)}
\end{align}

We can easily incorporate the series of long multiplet characters into the
Weyl determinant, using the fact that determinant of a matrix is linear in
one particular row. Different long multiplet characters that appear in the
series all depend on determinants of the same matrix except for the first row.
Thus, we present the character formulae for global short multiplets as follows,
\begin{align}\label{gchio}
\chi_{\frac{h_1}{2},\{h_i\}}^{(g)}(\t,\{z_j\}) =
&q^{h-\frac{c}{24}}
\frac{ \displaystyle \lvert \frac{y_j^{h_i+r-i+\frac{1}{2}} }{1+\delta^{i,1}y_jq^\frac{1}{2}} - \frac{y_j^{-(h_i+r-i+\frac{1}{2})} }{1+\delta^{i,1}y_j^{-1}q^\frac{1}{2}} \rvert}{| y_j^{r-i+\frac{1}{2}} - y_j^{-(r-i+\frac{1}{2})} \rvert } \nonumber \\
&\times  \frac{ \displaystyle (1+q^{1/2}) \bigg( \prod_{i=1}^{r}\prod_{\e=\pm 1}
(1+y_i^{\e} q^{1/2} ) \bigg)}{1-q}
\end{align}
for odd $\CN$ with rank $r$, and
\begin{align}\label{gchie}
\chi_{\frac{h_1}{2},\{h_i\}}^{(g)}(\t,\{z_j\}) =
&q^{h-\frac{c}{24}}
\frac{ \displaystyle \sum_{\e=\pm 1}  \lvert \dfrac{ y_j^{h_i+r-i} }{1+\delta^{i,1}y_jq^\frac{1}{2}} +\e \dfrac{y_j^{-(h_i+r-i)} }{1+\delta^{i,1}y_j^{-1}q^\frac{1}{2}}  \rvert}{ \lvert y_j^{r-i} + y_j^{-(r-i)}  \rvert} \nonumber \\
& \times \frac{ \displaystyle  \bigg( \prod_{i=1}^{r}\prod_{\e= \pm 1}
(1+y_i^{\e} q^{1/2} ) \bigg)}{1-q}
\end{align}
for even $\CN$ with rank $r$. Note that due to the Kronecker delta $\delta_{i,1}$,
the only modification from (\ref{gcho}) and (\ref{gche}) is the first row ($i=1$)
of the matrix in the determinant. This form of short multiplet character
resembles the known formula for short super Virasoro multiplet of the
small $\CN=4$ algebra \cite{Eguchi:1987wf}, see also (\ref{N4char2}) in particular.

Let us work out a simple example of the character formula. Consider the short
multiplet $A[1;1]_\frac12$ of the large $\CN=4$ algebra, which has been
depicted in Fig. \ref{N4 sporadic}. Note that the primary $[1;1]$ is equivalent
to $[1\, 0]$ in the orthogonal basis. We first focus on the determinant part of
(\ref{gchie}), for which we expand the factor that includes Kronecker delta:
\begin{align}
&\frac{\displaystyle \sum_{\e = \pm 1}
\begin{vmatrix}
\dfrac{y_1^2}{1+y_1 q^\half} +\e \dfrac{y_1^{-2}}{1+y_1^{-1} q^\half} & \dfrac{y_2^2}{1+y_2 q^\half} +\e \dfrac{y_2^{-2}}{1+y_2^{-1} q^\half} \\
y_1^0 +\e y_1^{0} & y_2^0 +\e y_2^{0}
\end{vmatrix}
}{\displaystyle
\lvert y_j^{2-i} + y_j^{-(2-i)}  \rvert} \nonumber \\
=& \frac{\displaystyle
\begin{vmatrix}
(y_1^2+y_1^{-2})-q^\half (y_1^3+y_1^{-3}) + \cdots &
(y_2^2+y_2^{-2})-q^\half (y_2^3+y_2^{-3}) + \cdots  \\
y_1^0 + y_1^{0} & y_2^0 + y_2^{0}
\end{vmatrix}
}{\displaystyle
\lvert y_j^{2-i} + y_j^{-(2-i)}  \rvert} \nonumber \\
=& \frac{\displaystyle
\begin{vmatrix}
y_1^2+y_1^{-2} & y_2^2+y_2^{-2}  \\
y_1^0 + y_1^{0} & y_2^0 + y_2^{0}
\end{vmatrix}}{\displaystyle
\lvert y_j^{2-i} + y_j^{-(2-i)}  \rvert}
-q^\half  \frac{\displaystyle
\begin{vmatrix}
y_1^3+y_1^{-3} & y_2^2+y_2^{-2}  \\
y_1^0 + y_1^{0} & y_2^0 + y_2^{0}
\end{vmatrix}}{\displaystyle
\lvert y_j^{2-i} + y_j^{-(2-i)}  \rvert} + \cdots .
\end{align}
Identifying each term to Weyl character formula (\ref{Weylchar})
and inserting back into (\ref{gchie}), we have
\begin{align}\label{characterexample}
\Big( \frac{q^{\frac12-\frac{c}{24}} }{1-q} \Big)^{-1} \chi_{\frac12,\{1,0\}}^{(g)}(\t,\{z_j\}) =
&\bigg(\chi_e(\{1,0\}) - q^\half \chi_e(\{2,0\}) +q\chi_e(\{3,0\}) - \cdots  \bigg) \nonumber \\
& \times (1+y_1 q^{1/2} )(1+y_1^{-1} q^{1/2} )(1+y_2 q^{1/2} )(1+y_2^{-1} q^{1/2} ).
\end{align}

Let us examine the RHS order-by-order.
In the order $q^0$ we have the primary representation, namely $\chi_e(\{1,0\})$,
or $[1;1]_\half$ in terms of fundamental weights.
In the next order $q^\half$ are all states that can be obtained by operating
one of four supercharges on the primary, as would appear at the first level
of the long multiplet. However, the LHS is a short multiplet which has
the null states corresponding to $\chi_e(\{2,0\})$,
or $[2;2]_1$ in terms of fundamental weights. Thus it is subtracted.
Then in the next order $q^1$, from all states that can be obtained by operating
two of four supercharges on the primary, as would appear at the second level
of the long multiplet, those that can be obtained by operating one supercharge
on the first-level null states $\chi_e(\{2,0\})$ are subtracted because they are
the null states. However, those corresponding to $\chi_e(\{3,0\})$
do not exist in the long multiplet due to Fermi-Dirac statistics,
yet have been subtracted. Therefore, they are added back.
Proceeding similarly, and recalling the effect of Virasoro operator $1/(1-q)$
that produces conformal descendants, one can confirm that the character
formula (\ref{characterexample}) is compatible with Figure \ref{N4 sporadic}.

Recall from the end of section \ref{subsecrecomb} that there is an exceptional
case of recombination rule related to conserved currents,
\begin{equation}
L[0 \cdots 0]_h \, \xrightarrow{\quad  h \to 0 \quad} \, V[0 \cdots 0]_0 \oplus A[1 \cdots 0]_{\frac12}.
\end{equation}
When a long multiplet whose primary is an R-singlet approaches the unitarity
bound, the corresponding short multiplet $V[0]_0$ counts as a zero degree of
freedom, and an extra component $[0]_1$ appears in the other
short multiplet. The character formulae can shed light on this phenomenon.

Consider the vacuum multiplet $V[0;0]_0$ of the large $\CN=4$ algebra.
Its character, by (\ref{gchie}), turns out to be
\begin{equation}\label{vacuumrecomb}
\chi_{0,\{0,0\}}^{(g)}(\t,\{z_j\}) = q^{-\frac{c}{24}} \frac{1-q}{1-q}.
\end{equation}
The $(1-q)$ factor in the numerator cancels the same factor in the denominator,
nullifying the effect of Virasoro operator $L_{-1}$.
In other words, in this vacuum multiplet, not only are there no conformal
primaries other than the superconformal primary $[0;0]_0$,
but also there are no conformal descendants that are derived
from the conformal primary by $L_{-1} \sim \partial$.
This is the manifestation of the conservation law
\begin{equation}
\partial \mathcal{J}_z =0 \quad \text{or} \quad \bar{\partial} \mathcal{\bar{J}}_{\bar{z}}=0,
\end{equation}
where the holomorphic conservation law holds when the vacuum multiplet belongs
to the holomorphic sector of full superconformal multiplet, and vice versa.
A superconformal multiplet with the vacuum multiplet $V[0]_0$ in its
(anti-)holomorphic sector is identified as the (anti-)holomorphic conserved current,
as discussed in section \ref{subsecdeformation} and will be discussed further
in section \ref{subsecconcur}.

On the LHS of (\ref{vacuumrecomb}), however, the primary component $[0]_h$
is not a conserved current since its conformal weight is non-zero, and
therefore the conformal descendants $(L_{-1})^n \ket{[0]_h}$ exist.
These descendants are precisely the extra $[0]_1$ component,
now treated as a conformal primary $\ket{[0]_1}$ and its descendants,
that appears in the second short multiplet $A[1 \cdots 0]_{\frac12}$
but not in the long multiplet $L[0 \cdots 0]_h$ as a conformal primary.

\section{Applications}\label{applicationsection}

Continuing based on the results of section \ref{multipletsection},
we discuss various aspects of two-dimensional superconformal field theories,
including stress-energy tensor, conserved currents,
supersymmetric deformations, and supersymmetry enhancement.

\subsection{Stress-Energy Tensor and Conserved Currents}\label{subsecconcur}

\subsubsection{Stress-Energy Tensor}\label{subsecstresstensor}

Any two-dimensional conformal field theories contain an identity operator and
a stress-energy tensor. We present in this subsection that there always exist
superconformal multiplets that contain the corresponding states
for all $\CN$ including both small and large $\CN=4$
with any value of the parameter $\a$.

It is obvious that the vacuum superconformal multiplet
$V\overline{ V}[0\cdots0|0\cdots0]_{0,0}$ is present
for any superconformal algebra.

The holomorphic stress-energy tensor $T$ in two dimensions
is a global conformal primary of scaling dimension
two and spin two. \eqref{genericalgebra} also implies that
$T$ must be neutral under the R-symmetry, and is a top
component in its superconformal multiplet.
We can show that there is a unique multiplet that
has the holomorphic (anti-holomorphic) stress-energy tensor
$[0\cdots0]_2\otimes \overline{[0\cdots0]}_0$
($[0\cdots0]_0\otimes \overline{[0\cdots0]}_2$)
as a top component for each $\CN$ except $\CN=6$
(each $\bar{\CN}$ except $\bar{\CN}=6$). As will be discussed further
in section \ref{subsecN6stress}, the $\CN=6$ superconformal
algebra has two candidate multiplets that have the stress-energy tensor.

\tikzset{block/.style={draw,text width=3cm, align=center,  inner sep=0.5ex, anchor=east, minimum height=4em}}
\tikzset{check/.style={text width=2cm, align=center,  inner sep=1ex, anchor=east, minimum height=4em}}
\begin{figure}[t]
\centering

\begin{tikzpicture}[node distance=2cm,auto,>=latex']
\node [block] (node1) {$[ \wedge^{3} V]_\frac{1}{2}$ \\ $[0 \, 0 \, 1 \, 0 \cdots 0]_\frac{1}{2}$};
\node [check, above of = node1] (node2) {};
\node [check, below of = node1] (node3) {};

\node [check, right of=node2] (node31) {};
\node [block, right of=node31] (node12) {$[ \wedge^{4} V]_1$ \\ $[0 \, 0 \, 0 \, 1 \, 0 \cdots 0]_1$};
\node [check, right of=node12] (node32) {};
\node [check, right of=node32] (node13) {$\cdots$};

\node [check, right of=node3] (node41) {};
\node [block, right of=node41] (node22) {$[\wedge^{2} V]_1$ \\ $[0 \, 1 \, 0 \cdots 0]_1$ \\ $\quad$ \\R-symmetry \\ current};
\node [check, right of=node22] (node42) {};
\node [block, right of=node42] (node23) {$[ \wedge^{1} V]_\frac{3}{2}$ \\ $[1 \, 0 \cdots 0]_\frac{3}{2}$\\ $\quad$ \\ supercurrent};
\node [check, right of=node23] (node43) {};
\node [block, right of=node43] (node24) {$[ \wedge^{0} V]_2$ \\ $[0 \cdots 0]_2$ \\ $\quad$ \\ stress-energy \\ tensor};
\path [line] (node1) -- (node12);
\path [line] (node12) -- (node13);
\path [line] (node1) -- (node22);
\path [line] (node22) -- (node23);
\path [line] (node23) -- (node24);
\end{tikzpicture}

\caption{Generic stress tensor multiplet, non-vacuum sector only.}
\label{stress multiplet}
\end{figure}
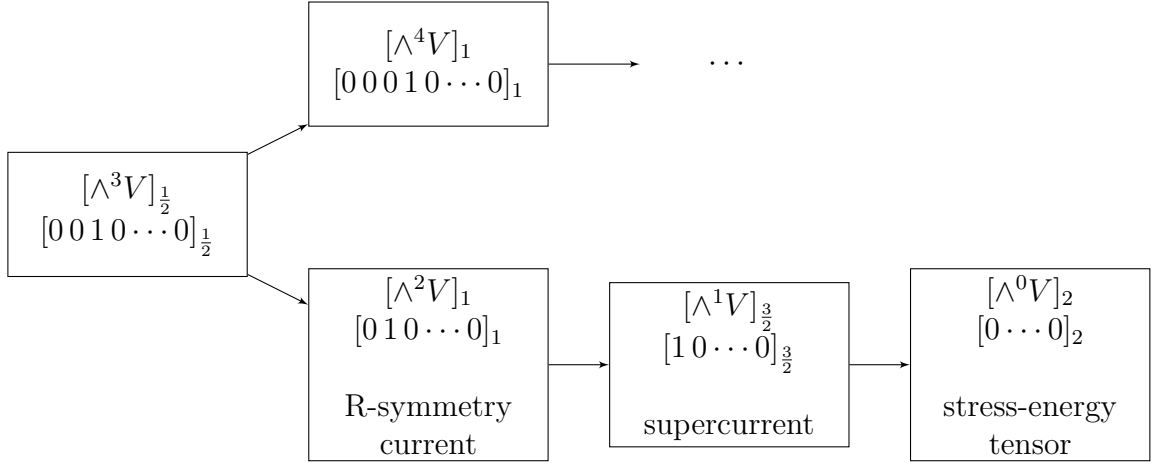
For all $\CN \geq 3$ except for the small $\CN=4$, the holomorphic stress-energy
tensor resides in a short multiplet $\CT\equiv A\overline{V}[0\,0\,1\,0\cdots\,0|0\cdots0]_{\frac12,0}$,
as depicted in Figure \ref{stress multiplet}. Note that
the primary of $\CT$ transforms in the $3^\text{rd}$ anti-symmetric
representation under the R-symmetry group $SO(\CN)$.
In the present work, such a short multiplet $\CT$ is referred to as a stress-energy tensor multiplet.

We also observe from Figure \ref{stress multiplet} that the stress-energy tensor multiplet $\CT$ has
other components such as $[\text{R-adjoint}]_1\otimes\overline{[0\cdots0]}_0$ and
$[\text{R-vector}]_\frac{3}{2}\otimes \overline{[0\cdots0]}_0$ at
level one and two. From \eqref{genericalgebra} that
$\{ G, G \} \sim L$ and $ \{ G, T \} \sim G $, one can identify the former as the R-symmetry currents
($T^{ab}$) while the latter as the supersymmetry currents ($G^a$).

The stress-energy tensor multiplet for small $\CN=4$ superconformal algebra
is a short multiplet $A\overline{V}[2|0]_{1,0}$. On the other hand, the stress-energy
tensor resides in a long multiplet $L\overline{V}[0|0]_{2-\frac{\CN}{2},0}$ for $\CN \leq 3$.

\subsubsection{Remark on the $\CN=6$ Stress-Energy Tensor and Anti-Symmetric Vector Products} \label{subsecN6stress}

The $\CN=6$ superconformal algebra has $SO(6)$ R-symmetry
where the third anti-symmetric tensor $\wedge^{3} V$
is no longer irreducible but decomposes into
two irreducible representations
 \begin{align}
  \wedge^{3} V = [0 \, 2 \, 0] \oplus [0 \, 0 \, 2],
\end{align}
self-dual and anti-self-dual three-forms. Accordingly,
the multiplet built on the primary $[\wedge^{3} V]_\frac12$
depicted in Figure \ref{N6 stress multiplet} can reduce to
two irreducible superconformal multiplets, each with
the primary $[0 \, 2 \, 0]_\frac{1}{2} $ and $[0 \, 0 \, 2]_\frac{1}{2}$.
The short multiplets $\CT=A\overline{V}[0 \, 2 \, 0|0\,0\,0]_{\frac{1}{2},0}$
and $\tilde \CT=A\overline{V}[0 \, 0 \, 2|0\,0\,0]_{\frac{1}{2},0}$
are conjugate to each other.
\tikzset{block/.style={draw,text width=3cm, align=center,  inner sep=1ex, anchor=east, minimum height=4em}}
\tikzset{check/.style={text width=2cm, align=center,  inner sep=1.5ex, anchor=east, minimum height=4em}}
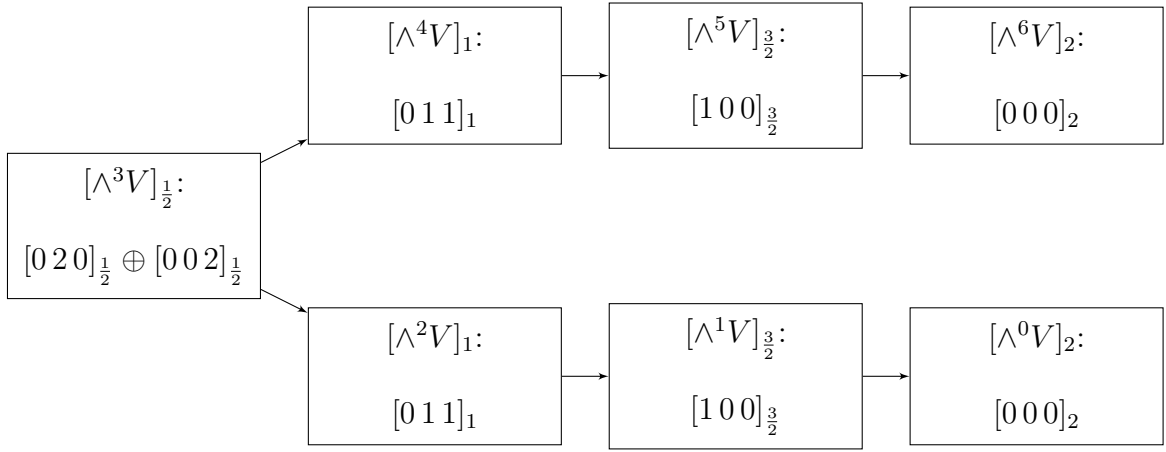
\begin{figure}[t]
\centering
\begin{tikzpicture}[node distance=2cm,auto,>=latex']
\node [block] (node1) {$[ \wedge^{3} V]_\frac{1}{2}$: \\ $\quad $ \\ $[0 \, 2 \, 0]_\frac{1}{2} \oplus [0 \, 0 \, 2]_\frac{1}{2}$};
\node [check, above of = node1] (node2) {};
\node [check, below of = node1] (node3) {};

\node [check, right of=node2] (node31) {};
\node [block, right of=node31] (node12) {$[ \wedge^{4} V]_1 $: \\ $\quad$ \\ $[0 \, 1 \, 1]_1$};
\node [check, right of=node12] (node32) {};
\node [block, right of=node32] (node13) {$[ \wedge^{5} V]_\frac{3}{2} $: \\ $\quad$ \\ $[1 \, 0 \, 0]_\frac{3}{2}$};
\node [check, right of=node13] (node33) {};
\node [block, right of=node33] (node14) {$[ \wedge^{6} V]_2$: \\ $\quad$ \\ $[0 \, 0 \, 0]_2$};

\node [check, right of=node3] (node41) {};
\node [block, right of=node41] (node22) {$[\wedge^{2} V]_1$: \\ $\quad$ \\ $[0 \, 1 \, 1]_1$};
\node [check, right of=node22] (node42) {};
\node [block, right of=node42] (node23) {$[ \wedge^{1} V]_\frac{3}{2}$: \\ $\quad$ \\ $[1 \, 0 \, 0]_\frac{3}{2}$};
\node [check, right of=node23] (node43) {};
\node [block, right of=node43] (node24) {$[ \wedge^{0} V]_2$: \\ $\quad$ \\ $[0 \, 0 \, 0]_2$};

\path [line] (node1) -- (node12);
\path [line] (node12) -- (node13);
\path [line] (node13) -- (node14);
\path [line] (node1) -- (node22);
\path [line] (node22) -- (node23);
\path [line] (node23) -- (node24);
\end{tikzpicture}

\caption{holomorphic $\CN=6$ stress tensor multiplet with left-moving sector only.
It is reducible into two parts conjugate to each other. We use
both the Dynkin label notation and the anti-symmetric
tensor product notation.}
\label{N6 stress multiplet}
\end{figure}

Both $\CT$ and $\tilde \CT$ contain the $[0\,0\,0]_2\otimes\overline{[0\,0\,0]}_0$
component. In terms of Dynkin labels, it is unclear which of the two
$[0 \, 0 \, 0]_2\otimes\overline{[0\,0\,0]}_0$ components
should be interpreted as the stress-energy tensor,
or even whether both could be the stress-energy tensor
or not. We can argue that only one of them should be
identified as a true stress-energy tensor.
Otherwise any $\CN\geq 7$ superconformal theories
would have two stress-energy tensors and thus become invalidated.
This is because the $\CN=7$ stress-energy tensor multiplet
$A\overline{V}[0\,0\,2|0\,0\,0]_{\half,0}$ can reduce to
such two $\CN=6$ multiplets $\CT$ and $\tilde \CT$ simultaneously.
See section \ref{subsecenhance} for the detail.

The anti-symmetric tensor product makes this point clearer. In this language, it is apparent
that one of the two $[0 \, 0 \, 0]_2$ in Figure \ref{N6 stress multiplet}
is a pseudoscalar with respect to the R-symmetry group while the
other is a genuine scalar. The former is produced from an axial vector at the previous level,
which in turn is produced from a 4-form at its previous level. However, for a genuine
stress-energy tensor, the algebra requires that R-vector and R-adjoint components
corresponding to the supercurrent and the R-current exist at the previous two levels.
This implies that, although the multiplet with an anti-self-dual 3-form primary appears to contain
a top component $[0 \, 0 \, 0]_2$, it cannot be identified as a stress-energy tensor, and
the self-dual counterpart can.

Note that this argument applies to every short multiplet $A[\wedge^{k} V]_\frac{1}{2}$
for every $\CN$: one of the top components is a scalar,
and the other is a pseudoscalar of the R-symmtry group.
Therefore, it matters to distinguish two short multiplets $A[\wedge^{k} V]_\frac{1}{2}$
and $A[\wedge^{\CN-k} V]_\frac{1}{2}$, although they appear to be identical in terms
of Dynkin labels. The case $\CN=6$ was special only because the stress-energy tensor
was involved.

An additional argument regarding supersymmetry enhancement and decomposition
that supports the result here will be presented in section \ref{subsecenhance}.

\subsubsection{Conserved currents}\label{subsecconcurr}

From Table \ref{N1table} through Table \ref{Notable}, one can see the
presence of top component $[0 \cdots 0]_1$ for all $\CN$.
It allows for a conserved current $[0 \cdots 0]_{1} \otimes \overline{[0 \cdots 0]}_{0}$
with spin 1 in any $(\CN,\bar{\CN})$ theory.
It is a Lorentz vector, has scaling dimension 1, and commutes with R-symmetry
and supersymmetry generators up to a total derivative. Thus, it qualifies as a flavor current.

Furthermore, a top component $[0 \cdots 0]_s$, thus a spin-$s$ conserved current
$[0 \cdots 0]_{s} \otimes \overline{[0 \cdots 0]}_{0}$ is allowed for every half value of $s$
starting from $s=1$.\footnote{From $s=1$ to $s=\frac{\CN}{2}$ they appear as top
components of some short multiplets. For $s > \frac{\CN}{2}$ they appear as generic
top components of long multiplets $L[0 \cdots 0]_{s-\frac{\CN}{2}}$. Note that in the
latter case $s$ can be any real number. See Tables \ref{Netable} and \ref{Notable}.}
When $s=\frac{3}{2}$, $[0 \cdots 0]_{\frac32} \otimes \overline{[0 \cdots 0]}_{0}$ corresponds to
an extra supercurrent, which will be discussed in section \ref{subsecenhance}.
The case $s=2$ corresponds to the stress-energy tensor which has been just discussed,
and $s \geq \frac{5}{2}$ corresponds to the higher-spin conserved currents.

In particular for each $\CN \geq 7$, a supersymmetric higher-spin conserved current
of spin $s=\frac{\CN}{2}-1$ appears in the stress tensor
multiplet, and thus is universal in all theories. In higher dimensions $d \geq 3$,
presence of the higher-spin currents indicates a locally free theory
\cite{Maldacena:2011jn,Boulanger:2013zza,Alba:2013yda,Alba:2015upa},
thus imposing an upper bound on the number of supersymmetries $\CN$
for interacting theories \cite{Cordova:2016emh}. In two dimensions
this is not necessarily true. One simple example is the three-state Pott's model
which is an interacting CFT with $W_3$-algebra. We close this subsection
with a remark that the above higher-spin current in the stress-energy tensor multiplet
may extend the superconformal algebra to a non-linear (super $\mathcal{W}$) algebra
\cite{Zamolodchikov:1985wn}.

\subsection{Deformations of CFT}\label{subsecdeform}
\subsubsection{Relevant and Marginal Deformations}

We turn into operators leading to relevant or marginal
deformations of a given superconformal theory that preserve
supersymmetry, as discussed briefly in section \ref{subsecdeformation}.
These operators have to be Lorentz scalars with scaling dimensions
$\D\leq 2$ and supersymmetric.
We thus look for top components from both left-moving and right-moving sectors
with $h_0 = \bar{h}_0 \leq 1$.
The deformation is marginal when the equality holds, and relevant when
the inequality is strict.

From Table \ref{N1table} through Table \ref{Notable}, we can conclude the followings.
\begin{enumerate}
\item \textbf{Marginal deformations.} For all $\CN$, including both small and large
$\CN=4$ with any value of the parameter $\a$, a top component $[0 \cdots 0]_1$
is allowed in some multiplet. For $\CN \geq 3$ it always appears as a sporadic
top component in a short multiplet with an R-vector primary.

Therefore, in all \emph{global} superconformal theories with any $(\CN,\bar{\CN})$
in two dimensions, a supersymmetric  marginal operator
$[0 \cdots 0]_1 \otimes \overline{[0 \cdots 0]}_1$, which is a Lorentz scalar
and an R-singlet, is allowed. For instance, a marginal operator
resides in an $\CN=5$ short multiplet $A\overline{A}[1\,0|1\,0]_{\half,\half}$.

\item \textbf{Relevant deformations.} For all $\CN \neq 4$ and for large $\CN=4$
with $\a=1$, a top component with $h=\frac{3}{4}$ that is an
R-spinor\footnote{For $\CN =1$ it is just the trivial representation [0]. For
$\CN =2$ it is the $[\frac{1}{2}] $ and $[-\frac{1}{2}]$, which are conjugate to
each other.} is allowed in some multiplet. Particularly for $\CN$ even, there
are two R-spinors conjugate to each other, and correspondingly there are two allowed
R-spinor top components with $h=\frac{3}{4}$. For $\CN \geq 3$ they always appear
as a generic top component at level one in a short multiplet with an R-spinor primary.
For instance, a component $[0\,0\,1]_{3/4}$ resides in an $\CN=6$ short multiplet
$A[0\,1\,0]_{1/4}$ at level one.

Therefore, a supersymmetric relevant deformation with $\D = \frac{3}{2}$ is allowed
for all \emph{global} superconformal theories with any $(\CN,\bar{\CN})$,
except that if any of these is 4 it has to be the large $\CN=4$ with $\a =1$.
This relevant deformation however breaks the R-symmetry because
it transforms as spinors under both SO$(\CN)$ and SO($\bar{\CN}$).

In case of the large $\CN=4$ with $\a \neq 1$, relevant deformations with
scaling dimensions $\D = 1+\frac{\a}{1+\a}$ and $\D = 1+\frac{1}{1+\a}$ are
similarly allowed, provided that both sectors have large $\CN=4$
symmetry and share the common value of $\a$. In particular, note that
the small $\CN=4$ superconformal algebra (, i.e., $\a \to \infty$) does
not admit a relevant deformation.

\end{enumerate}

In particular, the marginal deformation $[0;0]_1 \otimes \overline{[0;0]}_1$
is guaranteed to exist in a large $(4,4)$ superconformal theory.
It is known that the stress-energy tensor is actually a quasi-primary with respect
to the Virasoro symmetry: it is a super Virasoro descendant of the vacuum.
Therefore, whenever the stress tensor multiplet $A\overline{V}[1;1|0;0]_{\half,0}$
exists, there also exists in the same super Virasoro multiplet
a global multiplet $A\overline{A}[1;1|1;1]_{\half,\half}$, where the marginal
deformation resides. Therefore, a large $(4,4)$ superconformal field theory
always contains a marginal deformation, and thus exhibits a non-trivial moduli space.
This marginal deformation is in fact exactly marginal, and corresponds to
moduli of the type IIB string theory on
$AdS_3 \times S^3 \times S^3 \times S^1$, see \cite{Elitzur:1998mm,Gukov:2004ym}.

This argument can be applied to higher $\CN$,
but it results in the existence of an irrelevant deformation,
which is subject to less interest.

A relevant deformation that resides in the stress tensor multiplet
is referred to as a \emph{universal mass} \cite{Cordova:2016xhm}.
It is a deformation that is guaranteed to exist because it appears in the
stress tensor multiplet, which breaks the conformal symmetry
and often the R-symmetry as well. It results in a deformed super-Poincar\'e
algebra with central or non-central charge extension.
Study of the universal mass in higher dimensions have led to many interesting
results, see \cite{Cordova:2016xhm} and references therein.

However, the universal mass does not exist in two dimensions. Every relevant
deformation in two dimensions resides in a global superconformal
multiplet whose primary has conformal weights $h=\bar{h}=\frac14$.
It is obvious that this global multiplet itself is not a stress tensor multiplet
nor a flavor current multiplet. One can further argue that this multiplet cannot
belong to the same super Virasoro multiplet as the global stress tensor multiplet.
This is because the stress tensor multiplet is the lowest super Virasoro descendant of the vacuum,
and the primary of $h=\bar{h}=\frac14$ cannot become
another descendant of the vacuum.
This implies that a non-central charge extension
of Poincar\'e supersymmetry cannot be smoothly connected
to a superconformal symmetry via relevant deformations.

Lists of allowed top components, conserved currents and deformations discussed
in sections \ref{subsecconcur} and \ref{subsecdeform} can be found in Tables
\ref{concurtable1} and \ref{concurtable2}.
\begin{table}[t] \centering
\begin{tabular}{| c | c | c |} \hline
Case & Allowed top component & Primary of the multiplet \\ \hline \hline
\multirow{4}{*}{All $\CN$} & $[0 \cdots 0]_0$ & $[0 \cdots 0]_0$ \\
 & $[0 \cdots 0]_1$ & $[1 \, 0 \cdots 0]_\frac{1}{2}$ \\
 & $[0 \cdots 0]_2$ & $[0 \, 0 \, 1 \, 0 \cdots 0]_\frac{1}{2}$\\
 & $[0 \cdots 0]_\frac{Z \geq 5}{2}$ & $[0 \cdots 0 \, 1 \, 0 \cdots 0]_\frac{1}{2}$\\
\hline
All $\CN \neq 4$ and & [spinor]$_\frac{3}{4}$& [conjugate spinor]$_\frac{1}{4}$ \\
large $\CN = 4$ with $\a=1$ & $[0 \cdots 0]_\frac{3}{2}$ & $[0 \, 1 \, 0 \cdots 0]_\frac{1}{2}$\\ \hline
 \end{tabular}
\caption{\label{concurtable1} List of allowed top components and multiplets in which they reside.}
\end{table}
\begin{table}[t] \centering
\begin{tabular}{| c | c | c |} \hline
Supersymmetric operators & Top component & Multiplet in which it resides \\ \hline \hline
Marginal deformation & $[0 \cdots 0]_{1} \otimes \overline{[0 \cdots 0]}_{1} $ & $[1 \, 0 \cdots 0]_\frac{1}{2} \otimes \overline{[1 \, 0 \cdots 0]}_\frac{1}{2}$  \\
Relevant deformation & [spinor]$_\frac{3}{4} \otimes \overline{[\text{spinor}]}_\frac{3}{4}$ & [dual spinor]$_\frac{1}{4} \otimes \overline{[\text{dual spinor}]}_\frac{1}{4}$ \\ \hline
Unit operator & $[0 \cdots 0]_0 \otimes \overline{[0 \cdots 0]}_0$ & $[0 \cdots 0]_0 \otimes \overline{[0 \cdots 0]}_0$ \\
Flavor current (left) & $[0 \cdots 0]_1 \otimes \overline{[0 \cdots 0]}_0$ & $[1 \, 0 \cdots 0]_\frac{1}{2} \otimes \overline{[0 \cdots 0]}_0$  \\
Stress-energy tensor (left) & $[0 \cdots 0]_2 \otimes \overline{[0 \cdots 0]}_0$ & $[0 \, 0 \, 1 \, 0 \cdots 0]_\frac{1}{2} \otimes \overline{[0 \cdots 0]}_0 $  \\
Higher-spin currents (left) & $[0 \cdots 0]_\frac{Z=3,5,6,\cdots}{2} \otimes \overline{[0 \cdots 0]}_0$ & $[0 \cdots 0 \, 1 \, 0 \cdots 0]_\frac{1}{2} \otimes \overline{[0 \cdots 0]}_0 $   \\ \hline
 \end{tabular}
\caption{\label{concurtable2} List of marginal or relevant deformations and conserved
currents, and multiplets in which they reside. For conserved currents including stress
tensor, corresponding versions with left and right exchanged are also required.}
\end{table}

\subsubsection{Recombination Rules Revisited}

In section \ref{subsecrecomb} we have discussed recombination rules:
how long multiplets decompose into short multiplets as they hit the unitarity bound.
There, we considered only one sector. That is, a long multiplet in the
left-moving sector was decomposed into short multiplets while the one in the
right-moving sector remained unchanged.

Having discussed conserved currents and deformations of CFT,
it is fruitful to consider the case where multiplets in the left-moving
and the right-moving sectors approach the unitarity bound simultaneously.
In particular, we are interested in recombination rules where marginal deformations
or flavor currents appear. We write some of the recombination rules for
multiplets that have non-generic R-symmetry Dynkin labels but for any value of
$(\CN, \bar{\CN})$ below,

\begin{align}
L\overline{L}[0 \cdots 0 | 0 \cdots 0]_{h,\bar{h}}
\, \xrightarrow{\quad  h, \bar{h} \to 0 \quad} \, &
\underbrace{V\overline{V}[0 \cdots 0 | 0 \cdots 0]_{0,0}}_{\text{unit operator}} \oplus
\underbrace{A\overline{A}[1 \, 0 \cdots 0 | 1 \, 0 \cdots 0]_{\half,\half}}_{\text{marginal deformation}} \nonumber \\
\oplus & \underbrace{A\overline{V}[1 \, 0 \cdots 0 | 0 \cdots 0]_{\half,0} \oplus V\overline{A}[0 \cdots 0 | 1 \, 0 \cdots 0]_{0,\half}}_{\text{flavor currents}} , \label{recomb1} \\
L\overline{L}[1 \, 0 \cdots 0 | 0 \cdots 0]_{h,\bar{h}}
\, \xrightarrow{\quad  h \to \frac{1}{2}, \bar{h} \to 0 \quad} \, &
\underbrace{A\overline{V}[1 \, 0 \cdots 0 | 0 \cdots 0]_{\half,0}}_{\text{flavor current (left)}} \oplus
\underbrace{A\overline{A}[1 \, 0 \cdots 0 | 1 \, 0 \cdots 0]_{\half,\half}}_{\text{marginal deformation}} \oplus \cdots, \label{recomb2} \\
L\overline{L}[1 \, 0 \cdots 0 | 1 \, 0 \cdots 0]_{h,\bar{h}}
\, \xrightarrow{\quad  h, \bar{h} \to \frac{1}{2} \quad} \, &  \underbrace{A\overline{A}[1 \, 0 \cdots 0 | 1 \, 0 \cdots 0]_{\half,\half}}_{\text{marginal deformation}} \oplus \cdots .\label{recomb3}
\end{align}

When a marginal deformation cannot remain marginal beyond the leading order,
its scaling dimension should receive quantum corrections, and
the corresponding short multiplet, combined with other short multiplets,
should be up-lifted to a long multiplet.
In other words, a short multiplet that contains an exactly marginal operator must
not participate in any of the recombinations (\ref{recomb1})-(\ref{recomb3}). It
leads to a constraint that all marginal deformations become exactly marginal only if
they do not break any flavor symmetry. Otherwise, one of
\eqref{recomb1} and \eqref{recomb2} has to happen.

\subsection{Supersymmetry Enhancement}\label{subsecenhance}

It is useful to understand how an $\CN$-superconformal
multiplet can decompose into various multiplets of fewer
superconformal symmetries $\CN'<\CN$.
Note that, in the language of $\CN'$ superconformal algebra,
$\CN$ supercharges decompose into $\CN'$ supercharges and
extra $(\CN-\CN')$ fermionic conserved charges. The R-symmetry algebra
decomposes into the R$'$-symmetry algebra corresponding to the $\CN'$-supersymmetry,
flavor symmetry algebra commuting with the R$'$-symmetry algebra, and the remaining off-diagonal
generators charged under both R$'$-symmetry and flavor symmetry.

Let us in particular consider the (holomorphic) stress-energy tensor multiplet
$\CT^{(\CN+1)}$ of the $\CN+1$ superconformal algebra.
$\CT^{(\CN+1)}$ can split into various $\CN$-multiplets that
must include the following multiplets of $\CN$ superconformal algebra:
\begin{enumerate}
\item a stress-energy tensor multiplet $\CT^{(\CN)}$ that
has the holomorphic stress-energy tensor,
$\CN$ supercurrents and R-currents.

\item a short multiplet that has an extra R-neutral supercurrent as a top component.
In order that it be part of the enhanced $\CN+1$ supercurrents,
the $\CN$ off-diagonal R-currents $[1 \cdots 0]_1 \otimes \overline{[0 \cdots 0]}_0$
should be contained in the same short multiplet.

\end{enumerate}

Conversely, if a theory with $\CN$-supersymmetry contains all the multiplets
enumerated above with mentioned properties,
the theory is \emph{enhanced} to $\CN+1$.

As an illustration, let us consider the stress tensor multiplet
of large $\CN=4$ algebra with an arbitrary value of $\a$, $A\overline{V}[1;1|0;0]_{\frac12,0}$.
The multiplet and its decomposition to the $\CN=3$ algebra
are described in Figure \ref{N4 sporadic} and \ref{N4 sporadic decomposed},
respectively. This decomposition consists of two $\CN=3$ multiplets, $L\overline{V}[0|0]_{\frac{1}{2},0}$
which is the $\CN=3$ stress tensor multiplet, and $A\overline{V}[2|0]_{\frac{1}{2},0}$ which is the extra
supercurrent multiplet. The extra supercurrent multiplet $A\overline{V}[2|0]_{\frac{1}{2},0}$
has a top component $[0]_\frac{3}{2}\otimes\overline{[0]}_0$ which can be identified as
the extra supercurrent, and $[2]_1\otimes\overline{[0]}_0$ at the first level which corresponds to the
off-diagonal R-currents, as expected.
Therefore, the $\CN=3$ global superconformal theory in two dimensions can have an
enhanced large $\CN=4$ theory if and only if there exists the extra supercurrent multiplet
$A\overline{V}[2|0]_{\frac{1}{2},0}$.

This argument directly generalizes to generic values of $\CN$, as is clear from the fact that
\begin{equation}\label{ruleofdecom}
\wedge^k V \text{ in $SO(\CN+1$)}
\rightarrow
\wedge^k V \oplus \wedge^{k-1} V \text{ in $SO(\CN$)}.
\end{equation}
For a generic $\CN$, the stress tensor multiplet of $\CN+1$ superconformal algebra is
decomposed into the stress tensor multiplet of $\CN$ superconformal algebra and an
extra supercurrent multiplet thereof:
\begin{align}\label{genericenhancement}
\CT^{(\CN+1)} = A\overline{V}[\wedge^3 V | 0]_{\half,0} ~\leftrightarrow ~
\CT^{(\CN)}  = & ~A\overline{V}[\wedge^3 V | 0]_{\half,0} ~ \oplus \,
 ~
\hspace*{-0.3cm}\underbrace{A\overline{V}[\wedge^2 V | 0]_{\half,0} }_{\text{extra supercurrent multiplet}},
\end{align}
where the number of supersymmetries $\bar{\CN}$ in the right-moving sector
is arbitrary and irrelevant.

Conversely, a global superconformal theory with a generic number $\CN$ of
supersymmetries in two
dimensions is enhanced to an $\CN+1$ theory if and only if it contains the extra
supercurrent multiplet $A\overline{V}[\wedge^2 V | 0]_{\half,0}$.

Note that for large $\CN=4$ or $\CN=6$, $A\overline{V}[\wedge^2 V | 0]_{\half,0}$ or
$A\overline{V}[\wedge^3 V | 0]_{\half,0}$ on the RHS of (\ref{genericenhancement})
is reducible to two irreducible parts. Although only one irreducible part
corresponds to a genuine extra supercurrent multiplet or stress tensor multiplet
(see section \ref{subsecN6stress}),
both parts are required in order to enhance the theory into $\CN+1$.

There are several types of non-generic cases of supersymmetry enhancement.
Although what happens in each case is highly analogous to the generic case,
we make remarks on the differences.

\subsubsection{$\CN \leq 2$ to $\CN+1$}
For $\CN \leq 2$ where the smaller R-symmetry is abelian, the representation
$\wedge^3 V$ does not have a sensible interpretation. However, similar relations
to (\ref{genericenhancement}) hold:
\begin{align}\label{N3enhancement}
\CT^{(3)} = L\overline{V}[0 | 0 ]_{\half,0} \quad &\leftrightarrow \quad
\CT^{(2)} = L\overline{V}[0 | 0 ]_{1,0} \, \oplus
\underbrace{L\overline{V}[0 | 0 ]_{\half,0}}_{\text{e.s.m. for }\CN=2}, \\
\label{N2enhancement}
\CT^{(2)} = L\overline{V}[0 | 0 ]_{1,0} \quad &\leftrightarrow \quad
\CT^{(1)} = L\overline{V}[0 | 0 ]_{\frac32,0} \, \oplus
\underbrace{L\overline{V}[0 | 0 ]_{1,0}}_{\text{e.s.m. for }\CN=1},
\end{align} where e.s.m. abbreviates the extra supercurrent multiplet.

\subsubsection{$\CN=2$ to small $\CN=4$}

$\CN=4$ superconformal algebra that appears in the generic enhancement rule
(\ref{genericenhancement}) is the large one with $SO(4)$ R-symmetry group.
Meanwhile, the small $\CN=4$ with $SU(2)$ R-symmetry group can be decomposed
into or enhanced from the $\CN=2$ theory, considering that $SO(2) \subset SU(2)$.
That is,
\begin{align}\label{N4smallenhancement}
\CT^{(4s)} = A\overline{V}[2 | 0 ]_{1,0} \quad \leftrightarrow \quad
&\CT^{(2)} = L\overline{V}[0 | 0 ]_{1,0}
\oplus \underbrace{A\overline{V}[2 | 0 ]_{1,0} \oplus A\overline{V}[-2 | 0 ]_{1,0} }_{\text{e.s.m. for }\CN=2}.
\end{align}

Therefore, the $\CN=2$ global superconformal theory in two dimensions is
enhanced into a small $\CN=4$ theory if and only if \emph{both} extra supercurrent
multiplets $A\overline{V}[2 | 0 ]_{1,0}$ and $A\overline{V}[-2 | 0 ]_{1,0}$ exist.
Note that two extra supercurrents are
required since there are two more supercurrents in $\CN=4$ than in $\CN=2$.

As an application of \eqref{N4smallenhancement}, let us discuss
new anomalies in $(\CN,\bar{\CN})=(2,2)$ superconformal theories
resolving a long-standing puzzle that, unlike generic $(2,2)$
superconformal theories, the conformal manifold
of a $(4,4)$ superconformal theory does not factorize into
a product of K\"ahler manifolds \cite{Gomis:2016sab}. The authors
of \cite{Gomis:2016sab} have shown that whenever the operator
product expansion (OPE) between a chiral multiplet $A\overline{A}[1|1]_{\frac12,\frac12}$
and a twisted chiral multiplet $A\overline{A}[1|-1]_{\frac12,\frac12}$ develops a pole,
\begin{align}\label{ddd}
  \CO(x_1) \tilde{\CO}(x_2) \sim \frac{1}{x_1^{--} - x_2^{--}} J_{++}(x_2) + \text{regular terms},
\end{align}
the aforementioned anomaly occurs and the factorization fails.
Here $\CO$ and $\tilde{\CO}$ are the primaries of chiral
and twisted chiral multiplets while $J_{++}$ is a conserved current of
spin one and $U(1)$ R-charge two.

One can argue that any $(2,2)$ superconformal theories
have enhanced $(4,4)$ superconformal symmetry if and only if
the OPE between $\CO$ and $\tilde \CO$ has a pole. We present
a sketch of the proof below.

Let us start with a $(2,2)$ superconformal theory
that has a chiral multiplet and a twisted chiral
multiplet. Suppose that the OPE \eqref{ddd}
has a pole, which suggests that the given $(2,2)$
superconformal theory has an extra supercurrent multiplet $A\overline{V}[2|0]_{1,0}$
and its complex conjugate $A\overline{V}[-2|0]_{1,0}$.
This is because $J_{\pm \pm}$ is contained in
the short multiplet $A\overline{V}[\pm2|0]_{1,0}$ as a component.
The existence of such extra supercurrent multiplets
leads to the enhancement from $(2,2)$ to $(4,4)$, as explained above.

On the other hand, let us assume that the $(2,2)$ superconformal
theory has enhanced $(4,4)$ superconformal symmetry.
Both chiral and twisted chiral multiplets as well as
their conjugates are then combined in a $(4,4)$
short multiplet $A\overline{A}[1|1]_{\frac12, \frac12}$,
\begin{equation}\label{N4toN2}
\underbrace{A\overline{A}[1 | 1]_{\half,\half}}_{\text{in } \CN=(4,4)} \quad \leftrightarrow \quad
\underbrace{\bigoplus_{\pm} A\overline{A}[\pm 1 | \pm 1]_{\half,\half}}_{\text{in } \CN=(2,2)}\ .
\end{equation}
Its OPE with itself must give rise to the small (4,4) super Virasoro
vacuum multiplet with an identity and the stress-energy tensor
multiplet  $\CT^{(4s)} = A\overline{V}[2 | 0]_{1,0} $ included,
as described schematically below:
\begin{align} \label{N4OPE}
(4,4)\text{-theory : } &A\overline{A}[1 | 1]_{\half,\half} \times A\overline{A}[1 | 1]_{\half,\half}
\quad \sim \quad
V\overline{V}[0|0]_{0,0} \oplus \CT^{(4s)} \oplus \cdots.
\end{align}
Decomposing back to the $(2,2)$ theory, the RHS of (\ref{N4OPE}) must include
the extra supercurrent multiplets
$A\overline{V}[\pm 2 | 0 ]_{1,0}$. Meanwhile, among various
OPEs that appear in the LHS of (\ref{N4OPE}), it is the OPE
between the chiral multiplet $A\overline{A}[1 | 1]_{\half,\half}$ and
the twisted chiral multiplet $A\overline{A}[1 | -1]_{\half,\half}$
that accounts for the extra supercurrent multiplet,
\begin{align} \label{N2OPE}
(2,2)\text{-theory : } & A\overline{A}[1 | 1]_{\half,\half}\times A\overline{A}[1 | -1]_{\half,\half}\supset A\overline{V}[2 | 0]_{1,0}.
\end{align}
This completes the proof.

\subsubsection{$\CN=5$ to $\CN=6$}\label{5to6enhance}

\tikzstyle{line} = [draw, -latex']

\tikzset{block/.style={draw,text width=2.4cm, align=center,  inner sep=0.5ex, anchor=east, minimum height=2.5em}}
\tikzset{node/.style={text width=4cm, align=center,  inner sep=0ex, anchor=east, minimum height=4em}}

\begin{figure}[t]
\centering
\begin{tikzpicture}[node distance=1.8cm,auto,>=latex']
\node [block] (node1) {$[0 \, 2 \, 0]_\frac{1}{2}$};
\node [node, right of=node1] (node6) {};
\node [block, right of=node6] (node2) {$[0 \, 1 \, 1]_1$};
\node [node, right of=node2] (node7) {};
\node [block, right of=node7] (node3) {$[1 \, 0 \, 0]_\frac{3}{2}$};
\node [node, right of=node3] (node8) {};
\node [block, right of=node8] (node4) {$[0 \, 0 \, 0]_2$};
\node [node, right of=node4] (node5) {$\qquad$ : $\CT^{(6)}$};

\node [block, below of=node1] (node11) {$[0 \, 2]_\frac{1}{2}$};
\node [block, below of=node2] (node12) {$[1 \, 0]_1 \oplus [0 \, 2]_1$};
\node [block, below of=node3] (node13) {$[0 \, 0]_\frac{3}{2} \oplus [1 \, 0]_\frac{3}{2}$};
\node [block, below of=node4] (node14) {$[0 \, 0]_2$};
\node [node, below of=node5] (node15) {$\qquad$ : $\CT^{(5)}$};

\path [line] (node1) -- (node2);
\path [line] (node2) -- (node3);
\path [line] (node3) -- (node4);
\path [line] (node11) -- (node12);
\path [line] (node12) -- (node13);
\path [line] (node13) -- (node14);

\end{tikzpicture}

\caption{The $\CN=6$ stress tensor multiplet decomposed into $\CN=5$.}
\label{N5 enhance}
\end{figure}
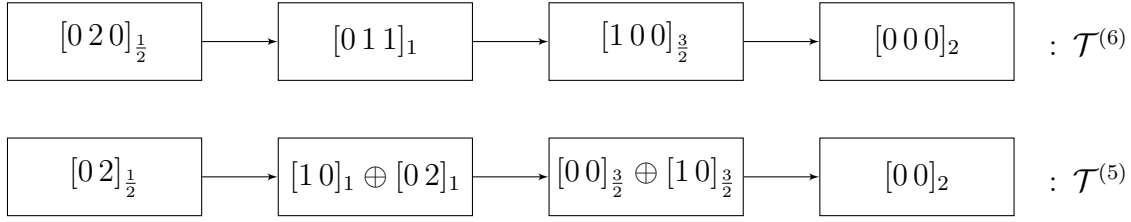

The enhancement of $\CN=5$ to $\CN=6$ is the most interesting.

Consider the genuine irreducible $\CN=6$ stress tensor multiplet denoted
by $\CT^{(6)}$. Its decomposition into conformal multiplets is depicted
in the upper row of Figure \ref{N5 enhance}. Note that the primary is
self-dual part of the representation $\wedge^3 V = [0 \, 2 \, 0] \oplus [0 \, 0 \, 2]$.

Decomposition of this multiplet into $SO(5)_R$ representations according to the
general rule (\ref{ruleofdecom}) is depicted in the lower row of Figure
\ref{N5 enhance}.
This is precisely the stress tensor multiplet $\CT^{(5)}$ of $\CN=5$:
$A[0 \, 2]_\frac{1}{2}$ with a $[0 \, 0]_2$ at the third level.

Therefore, the irreducible $\CN=6$ stress tensor multiplet is identical to the
$\CN=5$ stress tensor multiplet. In other words, an $\CN=5$ global superconformal
theory is \emph{automatically enhanced} into an $\CN=6$ theory.

\section{Relation to Super Virasoro Algebra}\label{virasorosection}

\subsection{Implications to the Super Virasoro Algebra}\label{subsecimplyalgebra}

Among deformations and conserved currents discussed in the last section,
some are guaranteed to exist because they belong to the same multiplet
as the stress tensor, which must exist in any superconformal field theories.
In this subsection, we will systematically investigate such
components and their implications to the super Virasoro algebra, which is not
fully understood for generic values of $\CN$.

Let us start with the case $\CN=3$, which is the simplest among interesting cases.
For concreteness, consider an $(\CN=3, \bar{\CN})$ theory with any $\bar{\CN}$ for
the right-moving sector. See the uppermost part of Figure \ref{stressmultiplets},
where structure of the stress-tensor multiplet $\CT^{(3)}$ is depicted. Note that the
product with $\overline{V}[0]_0$ in the right-moving sector is implied for all multiplets.
\tikzset{block/.style={draw,text width=1.5cm, align=center,  inner sep=1ex, anchor=east, minimum height=4em}}
\tikzset{check/.style={text width=2cm, align=center,  inner sep=1ex, anchor=east, minimum height=4em}}
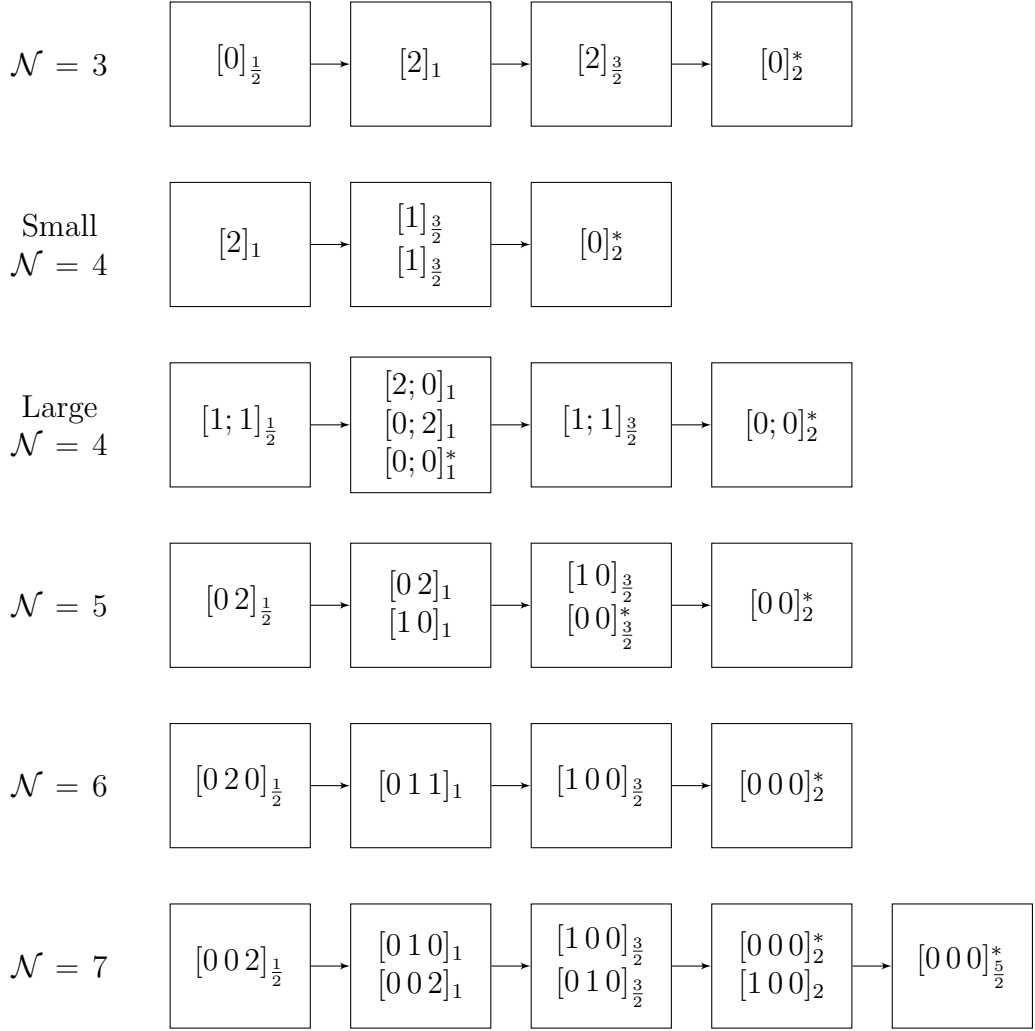
\begin{figure}[t!]
\centering
\begin{tikzpicture}[node distance=2.4cm,auto,>=latex']
\node [check] (node30) {$\CN=3$};
\node [block, right of=node30] (node31) {$[0]_\half$};
\node [block, right of=node31] (node32) {$[2]_1$};
\node [block, right of=node32] (node33) {$[2]_\frac{3}{2}$};
\node [block, right of=node33] (node34) {$[0]^*_2$};

\path [line] (node31) -- (node32);
\path [line] (node32) -- (node33);
\path [line] (node33) -- (node34);

\node [check, below of=node30] (node10) {Small $\CN=4$};
\node [block, right of=node10] (node11) {$[2]_1$};
\node [block, right of=node11] (node12) {$[1]_\frac32$ \\ $[1]_\frac32$ };
\node [block, right of=node12] (node13) {$[0]^*_2$};

\path [line] (node11) -- (node12);
\path [line] (node12) -- (node13);

\node [check, below of=node10] (node40) {Large $\CN=4$};
\node [block, right of=node40] (node41) {$[1;1]_\frac{1}{2}$};
\node [block, right of=node41] (node42) {$[2;0]_1$ \\ $[0;2]_1$ \\ $[0;0]^*_1$};
\node [block, right of=node42] (node43) {$[1;1]_\frac{3}{2}$};
\node [block, right of=node43] (node44) {$[0;0]^*_2$};

\path [line] (node41) -- (node42);
\path [line] (node42) -- (node43);
\path [line] (node43) -- (node44);

\node [check, below of=node40] (node50) {$\CN=5$};
\node [block, right of=node50] (node51) {$[0 \, 2]_\frac{1}{2}$};
\node [block, right of=node51] (node52) {$[0 \, 2]_1$ \\ $[1 \, 0]_1$};
\node [block, right of=node52] (node53) {$[1 \, 0]_\frac{3}{2}$ \\ $[0 \, 0]^*_\frac{3}{2}$};
\node [block, right of=node53] (node54) {$[0 \, 0]^*_2$};

\path [line] (node51) -- (node52);
\path [line] (node52) -- (node53);
\path [line] (node53) -- (node54);

\node [check, below of=node50] (node60) {$\CN=6$};
\node [block, right of=node60] (node61) {$[0 \, 2 \, 0]_\frac{1}{2}$};
\node [block, right of=node61] (node62) {$[0 \, 1 \, 1]_1$};
\node [block, right of=node62] (node63) {$[1 \, 0 \, 0]_\frac{3}{2}$};
\node [block, right of=node63] (node64) {$[0 \, 0 \, 0]^*_2$};

\path [line] (node61) -- (node62);
\path [line] (node62) -- (node63);
\path [line] (node63) -- (node64);

\node [check, below of=node60] (node70) {$\CN=7$};
\node [block, right of=node70] (node71) {$[0 \, 0 \, 2]_\frac{1}{2}$};
\node [block, right of=node71] (node72) {$[0 \, 1 \, 0]_1$ \\ $[0 \, 0 \, 2]_1$};
\node [block, right of=node72] (node73) {$[1 \, 0 \, 0]_\frac{3}{2}$ \\ $[0 \, 1 \, 0]_\frac{3}{2}$};
\node [block, right of=node73] (node74) {$[0 \, 0 \, 0]^*_2$ \\ $[1 \, 0 \, 0]_2$};
\node [block, right of=node74] (node75) {$[0 \, 0 \, 0]^*_\frac{5}{2}$};

\path [line] (node71) -- (node72);
\path [line] (node72) -- (node73);
\path [line] (node73) -- (node74);
\path [line] (node74) -- (node75);

\end{tikzpicture}
\caption{Stress tensor multiplets for $3 \leq \CN \leq 7$, where $\CN$ denotes number
of supercharges in the left-moving sector. Product with $V[0]_0$ in the right-moving
sector is implied. Top components are marked with an asterisk.}
\label{stressmultiplets}
\end{figure}

In addition to the components $[2]_1 \otimes \overline{[0]}_0$, $[2]_\frac32 \otimes
\overline{[0]}_0$, and $[0]_2 \otimes \overline{[0]}_0$ that we have already
identified as holomorphic R-currents $T^{[ab]}_0$, supercurrents $G^a_r$,
and stress-energy tensor $L_0$, respectively, we find another component
$[0]_\half \otimes \overline{[0]}_0$ in the multiplet. This implies that
a spin-$\half$ conserved current $[0]_{\half} \otimes \overline{[0]}_{0}$ exists
in any $\CN=3$ superconformal theories.

Furthermore, the multiplet structure contains additional information about the current.
Not only does the current possess conformal weight $h= \half$ and transform as an R-singlet,
but its anti-commutation relation with the supercurrents $G^a_r$ must also yield
the R-currents $T^{[ab]}_m$. Denoting the current as $\G_r$ ($r \in \mathbb{Z} + \frac12$)
and matching dimensions and R-indices, one can predict that for all modes
$r,s \in \mathbb{Z} + \frac12$ and $n \in \mathbb{Z} $,
\begin{subequations}
\begin{align}
[L_n,\G_r] &= (-\frac{n}{2}-r) \G_{n+r}, \\
[\G_r, T^{[ab]}_n] &= 0, \\
\{ \G_r, G^a_s \} &\sim \e^{abc} T^{[bc]}_{r+s},
\end{align}
\end{subequations}
must be part of the $\CN=3$ super Virasoro algebra. This is precisely what was found
in \cite{Ademollo:1975an}.

Let us proceed to large $\CN=4$. Small $\CN=4$ is not interesting in this respect,
as the stress tensor multiplet therein does not contain any extra component other than
$\{ L,G,T \}$, see the second part of Figure \ref{stressmultiplets}. For the rest of this
subsection, we no longer mention explicitly the right-moving sector, but assume that
tensor product with the vacuum multiplet $\overline{V}[0]_0$ in the right-moving sector with
any number of supercharges $\bar{\CN}$ is always implied.

In addition to the components $[2;0]_1$, $[1;1]_\frac{3}{2}$, and $[0;0]_2$ that we have
already identified as R-currents $T^{\pm i}_0$, supercurrents $G^a_r$, and
stress-energy tensor $L_0$, we find a sporadic top component
$[0;0]_1$ and a primary component $[1;1]_\frac{1}{2}$. Therefore, conserved
currents $[0;0]_{1}$ and $[1;1]_{\half}$ are guaranteed to exist in
any large $\CN=4$ superconformal theories.

Following \cite{Sevrin:1988ew}, we denote the respective currents by $U_0$ and $Q^a_r$,
where $r=\frac{1}{2}$ and $-\frac{1}{2}$ correspond to each other's conjugate.
We can see, not to mention their transformation properties under the R-symmetry
and their scaling dimensions 1 and $\half$, that $U_0$ commutes with supercurrents
$G^a_r$, and that $\{ G^a_r,Q^b_s \}$ leads to $T^{\pm i}_0$ and $U_0$.
These currents with the respective properties are precisely what are present in the
full $\CN =4$ super Virasoro algebra \cite{Sevrin:1988ew}.

Some of the relevant (anti-)commutation relations are presented below,
\begin{subequations}\label{N4supervirasoro}
\begin{align}
[T^{+i}_n, G^a_r ] &= \eta^{+i}_{ab} (G^b_{n+r} - 2 \frac{1}{1+\a} n Q^b_{n+r}), \\
[T^{-i}_n, G^a_r ] &= \eta^{-i}_{ab} (G^b_{n+r} + 2 \frac{\a}{1+\a} n Q^b_{n+r}), \\
\{ Q^{a}_r, G^b_s \} &= 2(\eta^{+i}_{ab} T^{+i}_{r+s} - \eta^{-i}_{ab} T^{-i}_{r+s}) +\d^{ab} U_{r+s}, \\
[U_n, Q^a_r] &= 0, \qquad [U_n, G^a_r] = n Q^a_{n+r}, \\
\{ Q^a_r , Q^b_s \} &= -\d^{ab}\d_{r+s,0} \frac{c(1+\a)^2}{12 \a}, \\
\{ U_m , U_n \} &= -m \d_{m+n,0} \frac{c(1+\a)^2}{12 \a}.
\end{align}
\end{subequations}
As in the global subalgebra (\ref{N4largealgebra}), $\a$ parametrizes the
$\mathfrak{su}(2)$ levels:
\begin{equation}\label{Kaclevels}
k_+ = c(1+\a)/(6\a), \qquad k_-=c(1+\a)/6,
\end{equation}
where $c$ is the central charge. Note that $c$ does not enter the
global subalgebra while $\a$ does.

Therefore, although the global subalgebra is complete with $\{ L,G,T \}$ alone,
we can find by physical reasoning what additional operators must exist,
which actually appear in the Virasoro algebra. Admittedly, this approach may
not exhaust all operators that appear in the Virasoro algebra.

Let us turn to the next example: $\CN=5$. We can see that, in addition to the familiar
operators $T^{[ab]}_0$, $G^a_r$, and $L_0$, there are conserved currents of dimensions
$\frac{1}{2},1,\frac{3}{2}$ that are 3-form, 4-form, and 5-form, respectively, under the
R-symmetry group. The 5-form is also supersymmetric.
Note that for odd $SO(\CN)$, $k$-form and $(\CN-k)$-form are identical.
Thus, an additional R-singlet current with dimension $\frac{3}{2}$ that is supersymmetric,
an R-vector current with dimension 1, and an R-adjoint current with dimension $\half$
are guaranteed to exist. The first two are identified as an extra supercurrent and
extra R-currents that enhance the supersymmetry into $\CN=6$.
This has been the topic of section \ref{5to6enhance}.

Due to peculiarity discussed in section \ref{subsecN6stress} and \ref{5to6enhance},
the $\CN=6$ stress
tensor multiplet contains no non-trivial components except for the primary,
which is a self-dual 3-form under the R-symmetry with dimension $\frac{1}{2}$.
Thus, implications to $\CN=6$ super Virasoro algebra are more or less contained
in the last paragraph.

Proceeding further, we can conclude that for all $\CN \geq 7$, conserved currents
$U^{[a_1a_2 \cdots a_k]}_{m(r)} \, (k=3,4,\cdots,\CN)$ that are $k$-forms under the
R-symmetry with dimensions $\frac{k}{2}-1$ are guaranteed to exist. The mode
number is $m \in \mathbb{Z}$ for $k$ even, thus bosonic currents,
and $r \in \mathbb{Z}+\frac{1}{2}$ for $k$ odd, thus fermionic currents.
Extending the mode numbers, commutation relations with the
supercurrent is schematically written as
\begin{equation}
\{ G^{a_0}_{r}, U^{[a_1 a_2 \cdots a_k]}_{s} \} \sim U^{[a_0 a_1 \cdots a_k]}_{r+s} + \d^{a_0, a_i} U^{[a_1 a_2 \cdots a_k] \backslash a_i}_{r+s} + \cdots.
\end{equation}

Note that the conserved currents discussed in this subsection are to be distinguished
from those discussed in section \ref{subsecconcurr}, which were supersymmetric
and allowed for every half-integer value of spin, but not guaranteed to exist.
Only the $\CN$-form coincides.

\subsection{Relation to Super Virasoro Characters in $\CN=2$ and Small $\CN=4$}\label{subsecsvc}

Recall that the superconformal multiplets analyzed in section \ref{multipletsection}
are representations of global subalgebras of the larger super Virasoro algebras.
In this sense, super Virasoro multiplets can be decomposed into the global
superconformal multiplets. In this subsection, we discuss how the
super Virasoro multiplets are decomposed, and compare recombination rules
of super Virasoro multiplets and global superconformal multiplets
for $\CN=2$ and small $\CN=4$, for which the super Virasoro multiplets
have been thoroughly studied.

\subsubsection{$\CN=2$}

Super Virasoro multiplets in $\CN=2$ have been classified in \cite{Boucher:1986bh}.
A multiplet is long, or \emph{massive},
if its primary $[j]_h$ lies \emph{above} the unitarity bound,
and it is short, or \emph{massless},
if the primary lies \emph{on} the unitarity bound.
The unitarity bound is described by a set of line segments (see \cite{Boucher:1986bh,Bae:2018qym})
\begin{equation}
2h-2rj+(\frac{c}{3}-1)(r^2-\frac{1}{4})=0 \quad (r \in \frac{1}{2}+\mathbb{Z})
\end{equation} defined in the range
\begin{equation}
(\frac{c}{3}-1)(|r|-\frac{1}{2}) \leq \text{sign}(r) \cdot j \leq (\frac{c}{3}-1)(|r|+\frac{1}{2})+1.
\end{equation}
The line segment labelled by $r$ describes a set of primaries $[j]_h$
that would be annihilated by the supercharge $G_{-|r|}^{\text{sign}(r)}$.
Note that the global unitarity bound (\ref{unicon2}) corresponds to the
line segments $r=\pm \frac{1}{2}$ extended to infinity.

It is known that the $\CN=2$ super Virasoro algebra has an outer automorphism
referred to as spectral flow \cite{Schwimmer:1986mf}.
Although the \emph{extended} $\CN=2$ algebra that incorporates
the spectral flow by one unit has many applications (see e.g. \cite{Keller:2012mr}),
we do not impose such an invariance here.

Character of an $\CN=2$ super Viasoro multiplet with primary $[j]_h$,
massive or massless, is written as (see \cite{Dobrev:1986hq,Matsuo:1986cj,Kiritsis:1986rv})
\begin{align}
\text{Massive: } \text{Ch}_{h,j}(\tau,z)&=
q^{h-\frac{c}{24}}y^j \prod_{n=1}^\infty \frac{(1+q^{n-\frac{1}{2}}y) (1+q^{n-\frac{1}{2}}y^{-1})}{(1-q^n)^2}, \\
\text{Massless: } \chi_{h(j,r),j}(\tau,z)&=
q^{h-\frac{c}{24}}y^j \prod_{n=1}^\infty \frac{(1+q^{n-\frac{1}{2}}y) (1+q^{n-\frac{1}{2}}y^{-1})}{(1-q^n)^2} \frac{1}{1+q^{|r|}y^{\text{sign}(r)}},
\end{align}
where $q=e^{2\pi i \t}$ and $y=e^{2 \pi i z}$ as in section \ref{subsecchar}.
Note that $h(j,r)$ in the massless character ensures that
$h$ be on the unitarity bound,
where $r$ labels which line segment $(h,j)$ locates on.
These multiplets satisfy a recombination rule as $h$ in the massive multiplet
approaches the unitarity bound given $j$:
\begin{equation} \label{svrecombN2}
\text{Ch}_{h,j}(\tau,z) \, \xrightarrow{\quad h \rightarrow h(j,r) \quad} \,
\chi_{h(j,r),j}(\tau,z) + \chi_{h(j,r)+|r|,j+\text{sign}(r)}(\tau,z)
\end{equation}

Structure of the character can be analyzed as follows.
Starting from the primary which contributes $q^{h-\frac{c}{24}}y^j$,
two factors in the numerator take
negative-mode fermionic supercharges $G^\pm_{-n+\frac{1}{2}}$ into effect
while those in the denominator
take negative-mode bosonic operators $L_{-n}, T_{-n}$ into effect.
The extra factor for the massless case simply cancels the
operation by $G^{\text{sign}(r)}_{-\frac{1}{2}}$, which produces the null states.

Meanwhile, we can write characters of global $\CN=2$ superconformal multiplets
enlisted in Table \ref{N2table} using more general formula (\ref{gche}) and (\ref{gchie}),
\begin{align}
\text{Massive: } \text{Ch}_{h,j}^{(g)}(\tau,z)&=q^{h-\frac{c}{24}}y^j \frac{(1+q^{\frac{1}{2}}y) (1+q^{\frac{1}{2}}y^{-1})}{(1-q)}, \\
\text{Massless: } \chi_{\frac{|j|}{2},j}^{(g)}(\tau,z)&=q^{h-\frac{c}{24}}y^j \frac{(1+q^{\frac{1}{2}}y) (1+q^{\frac{1}{2}}y^{-1})}{(1-q)} \frac{1}{1+q^{\frac{1}{2}}y^{\text{sign}(j)}}.
\end{align} These multiplets also satisfy a recombination rule that has been already written down in (\ref{N2recomb}) using different notation:
\begin{equation} \label{glrecombN2}
\text{Ch}_{h,j}^{(g)}(\tau,z) \, \xrightarrow{\quad h \rightarrow \frac{|j|}{2} \quad} \, \chi_{\frac{|j|}{2},j}^{(g)}(\tau,z) + \chi_{\frac{|j|+1}{2},j+\text{sign}(j)}^{(g)}(\tau,z)
\end{equation}

For simplicity and without loss of generality, let us assume $r>0$.
Decomposition of super Virasoro multiplets into global multiplets is now straightforward.
We first write for massive multiplets and massless multiplets with $r \geq \frac{3}{2}$:
\begin{align}
\text{Ch}_{h,j}(\tau,z)&= \text{Ch}_{h,j}^{(g)}(\tau,z) \underbrace{\frac{1}{1-q} \prod_{n=2}^\infty \frac{(1+q^{n-\frac{1}{2}}y) (1+q^{n-\frac{1}{2}}y^{-1})}{(1-q^n)^2}}_{(*)} , \label{svdecomN2.1}\\
\chi_{h(j,r \geq \frac{3}{2}),j}(\tau,z)&= \text{Ch}_{h,j}^{(g)}(\tau,z) \underbrace{\frac{1}{(1-q)(1+q^ry)} \prod_{n=2}^\infty \frac{(1+q^{n-\frac{1}{2}}y) (1+q^{n-\frac{1}{2}}y^{-1})}{(1-q^n)^2}}_{(**)}. \label{svdecomN2.2}
\end{align}

An important point to note from (\ref{svdecomN2.1}) and (\ref{svdecomN2.2})
is that only long global multiplets appear on the RHS.
This is because
i) the unitarity bound for $r \geq \frac{3}{2}$ is higher than the
global bound $h=\frac{|j|}{2}$, so the global bound is never saturated
by primaries of the super Virasoro multiplets,
and ii) every monomial $q^a y^b$ that appears in $(*)$ or $(**)$, except for 1,
satisfies $a > \frac{|b|}{2}$, so the corresponding global multiplet
\begin{equation}
\text{Ch}_{h,j}^{(g)}(\tau,z) \times q^a y^b = \text{Ch}_{h+a,j+b}^{(g)}(\tau,z)
\end{equation} remains long: $h+a > \frac{|j+b|}{2},$ provided that $h > \frac{|j|}{2}$.

However, the massless character with $r=\frac{1}{2}$ requires a special treatment
because it no longer satisfies i) in the last paragraph.
Therefore, we need to single out a global short multiplet $\chi_{h=\frac{j}{2},j}^{(g)}(\tau,z)$:
\begin{equation}
\chi_{h=\frac{j}{2},j}(\tau,z)= \chi^{(g)}_{\frac{j}{2},j} (\tau,z) +
\text{Ch}_{\frac{j}{2},j}^{(g)}(\tau,z)  \underbrace{\frac{1}{1+q^\frac{1}{2} y} \bigg( \frac{1}{1-q} \prod_{n=2}^\infty \frac{(1+q^{n-\frac{1}{2}}y) (1+q^{n-\frac{1}{2}}y^{-1})}{(1-q^n)^2} -1 \bigg)}_{(\#)} . \label{svdecomN2.3}
\end{equation}
Every monomial $q^a y^b$ in $(\#)$ satisfies $a > \frac{|b|}{2}$,
and appears with a positive coefficient.
Therefore, a massless super Virasoro multiplet on the segment $r=\pm \half$
is decomposed into exactly one short global multiplet with the same primary,
and infinitely many long global multiplets.\footnote{Note that the
global long multiplet $\text{Ch}_{\frac{j}{2},j}^{(g)}(\tau,z)$ only works as a proxy for
producing other long multiplets by multiplication with $q^ay^b$'s.
It itself does not make sense because the unitarity bound has been reached.}

The decomposition of super Virasoro multiplets can be applied to
their recombination rule (\ref{svrecombN2}).
Again, cases $r \geq \frac32$ and $r=\half$ need to be treated separately.

First, for $r \geq \frac32$, super Virasoro multiplet \emph{recombination},
or \emph{branching} if we are looking at it in the opposite direction,
occurs at $h$ higher than the global unitarity bound,
so the same constituent long global multiplets are simply regrouped
to form different sets of super Virasoro multiplets:
\begin{align}\label{svglrecombN2.1}
\text{Ch}_{h,j}(\tau,z) =& \,
\text{Ch}_{h,j}^{(g)}(\tau,z) \frac{1}{1-q}
\prod_{n=2}^\infty \frac{(1+q^{n-\frac{1}{2}}y) (1+q^{n-\frac{1}{2}}y^{-1})}{(1-q^n)^2} \nonumber \\
=& \, (\text{Ch}_{h,j}^{(g)}(\tau,z) + \text{Ch}_{h+r,j+1}^{(g)}(\tau,z)  )
\frac{1}{(1-q)(1+q^ry)}
\prod_{n=2}^\infty \frac{(1+q^{n-\frac{1}{2}}y) (1+q^{n-\frac{1}{2}}y^{-1})}{(1-q^n)^2} \nonumber \\
\xrightarrow{ h \rightarrow h(j,r) } \, &
(\text{Ch}_{h(j,r),j}^{(g)}(\tau,z) + \text{Ch}_{h(j,r)+r,j+1}^{(g)}(\tau,z) )
\frac{1}{(1-q)(1+q^ry)}
\prod_{n=2}^\infty \frac{(1+q^{n-\frac{1}{2}}y) (1+q^{n-\frac{1}{2}}y^{-1})}{(1-q^n)^2} \nonumber \\
=& \, \chi_{h(j,r),j}(\tau,z) + \chi_{h(j,r)+r,j+1}(\tau,z) .
\end{align}

On the other hand, for $r=\frac{1}{2}$ where super Virasoro
and global unitarity bounds coincide, the super Virasoro multiplet recombination
occurs simultaneously as \emph{one} of its constituent global multiplet
also experiences the recombination.
\begin{align}\label{svglrecombN2.2}
\text{Ch}_{h,j}(\tau,z) =& \,
\text{Ch}_{h,j}^{(g)}(\tau,z) \frac{1}{1-q}
\prod_{n=2}^\infty \frac{(1+q^{n-\frac{1}{2}}y) (1+q^{n-\frac{1}{2}}y^{-1})}{(1-q^n)^2} \nonumber \\
=& \, \text{Ch}_{h,j}^{(g)}(\tau,z)+ \text{Ch}_{h,j}^{(g)}(\tau,z)
\bigg( \frac{1}{1-q} \prod_{n=2}^\infty \frac{(1+q^{n-\half}y) (1+q^{n-\half}y^{-1})}{(1-q^n)^2}  -1 \bigg)\nonumber \\
\xrightarrow{ h \rightarrow \frac{j}{2} } \, &
\Big( \chi_{\frac{j}{2},j}^{(g)}(\tau,z) + \chi_{\frac{j+1}{2},j+1}^{(g)}(\tau,z) \Big) +
\text{Ch}_{\frac{j}{2},j}^{(g)}(\tau,z)
\bigg( \frac{1}{1-q} \prod_{n=2}^\infty \frac{(1+q^{n-\half}y) (1+q^{n-\half}y^{-1})}{(1-q^n)^2}  -1 \bigg)\nonumber \\
=& \chi_{\frac{j}{2},j}^{(g)}(\tau,z) +
\text{Ch}_{\frac{j}{2},j}^{(g)}(\tau,z) \frac{1}{1+q^\frac{1}{2} y}
\bigg( \frac{1}{1-q} \prod_{n=2}^\infty \frac{(1+q^{n-\half}y) (1+q^{n-\half}y^{-1})}{(1-q^n)^2}  -1 \bigg)\nonumber  \\
&+ \chi_{\frac{j+1}{2},j+1}^{(g)}(\tau,z) +\text{Ch}_{\frac{j+1}{2},j+1}^{(g)}(\tau,z)
\frac{1}{1+q^\half y}
\bigg( \frac{1}{1-q} \prod_{n=2}^\infty \frac{(1+q^{n-\half}y) (1+q^{n-\half}y^{-1})}{(1-q^n)^2}  -1 \bigg) \nonumber \\
=& \chi_{\frac{j}{2},j}(\tau,z) + \chi_{\frac{j+1}{2},j+1}(\tau,z)
\end{align}

\subsubsection{Small $\CN=4$}

The small $\CN=4$ super Virasoro multiplets are classified similarly
to their global counterparts: a multiplet with primary $[R]_h$ is bounded by
$h \geq \frac{R}{2}$, and is massless, or short, if the bound is saturated
and massive, or long, otherwise.
Their characters are similar to the case $\CN=2$ except that now
each R-representation may contain many states, graded by $z$.
The central charge needs to be discretized by $c=6k$ $(k=1,2,\cdots )$.

Using the orthogonal basis, we define $l=\frac{R}{2}$ so that the
orthogonal weight $l$ can take half-integer values
and represents conventional $SU(2)$ spin $T^3_0$.
The unitarity bound becomes $h \geq l$ in the NS sector we are working on.
Then, the characters are given by\footnote{Another fugacity for the $U(1)$ charge
that distinguishes two copies of $SU(2) \subset SO(4)$ may be introduced,
as has been in \cite{Eguchi:1987wf}. However, we simply set the fugacity to unity
because it plays no role in our argument.} (see \cite{Eguchi:1987sm,Eguchi:1987wf})
\begin{align}
\text{Ch}_{h,l} (\tau,z) =&
q^{h-\frac{k}{4}}
\bigg( \prod_{n=1}^\infty
\frac{(1+y^{\frac{1}{2}}q^{n-\frac{1}{2}})^2(1+y^{-\frac{1}{2}}q^{n-\frac{1}{2}})^2}{(1-yq^n)(1-q^n)^2(1-y^{-1}q^{n})}
\bigg) \nonumber \\
& \times \sum_{m=-\infty}^{\infty} \frac{y^{(k+1)m+l}-y^{-(k+1)m-l-1} }{1-y^{-1}}
q^{(k+1)m^2+(2l+1)m}, \label{N4char1}
\end{align}
where $2l+1=1,2,\cdots,k$, and
\begin{align}
\chi_{h=l,l} (\tau,z) =&
q^{l-\frac{k}{4}}
\bigg( \prod_{n=1}^\infty
\frac{(1+y^{\frac{1}{2}}q^{n-\frac{1}{2}})^2(1+y^{-\frac{1}{2}}q^{n-\frac{1}{2}})^2}{(1-yq^n)(1-q^n)^2(1-y^{-1}q^{n})}
\bigg) \frac{1}{1-y^{-1}} \nonumber \\
& \times \sum_{m=-\infty}^{\infty}
\bigg( \frac{y^{(k+1)m+l}}{(1+y^{\frac{1}{2}} q^{m+\frac{1}{2}})^2} -
\frac{y^{-(k+1)m-l-1}}{(1+y^{-\frac{1}{2}} q^{m+\frac{1}{2}})^2} \bigg)
q^{(k+1)m^2+(2l+1)m}, \label{N4char2}
\end{align}
where $2l+1=1,2,\cdots,k+1$. The factor $\frac{1}{1-y^{-1}}$ is often
absorbed into the product in the literature, but we choose to
write it separately to manifest the appearance of Weyl character formula.

Meanwhile, characters of the global multiplets read
\begin{align}
\text{Ch}_{h,l}^{(g)} (\tau,z) =&
q^{h-\frac{k}{4}}
\frac{(1+y^{\frac{1}{2}}q^{\frac{1}{2}})^2(1+y^{-\frac{1}{2}}q^{\frac{1}{2}})^2}{1-q}
\frac{y^l - y^{-l-1}}{1-y^{-1}}, \label{N4char3} \\
\chi_{h=l,l}^{(g)} (\tau,z) =&
q^{l-\frac{k}{4}}
\frac{(1+y^{\frac{1}{2}}q^{\frac{1}{2}})^2(1+y^{-\frac{1}{2}}q^{\frac{1}{2}})^2}{1-q}
\frac{1}{1-y^{-1}}
\bigg( \frac{y^l}{(1+y^{\frac{1}{2}}q^{\frac{1}{2}})^2 } -
\frac{y^{-l-1}}{(1+y^{-\frac{1}{2}}q^{\frac{1}{2}})^2 }\bigg).
\label{N4char4}
\end{align}
From (\ref{N4char1})-(\ref{N4char4}) we can infer the recombination rule
of (\ref{N4recomb}), common to the super Virasoro and global multiplets.
\begin{align}
\text{Ch}_{h,l} \, & \xrightarrow{ h \to l } \,
\chi_{l,l} +2 \chi_{l+\half,l+\half} +\chi_{l+1,l+1}, \label{N4recombsv} \\
\text{Ch}_{h,l}^{(g)} \, & \xrightarrow{ h \to l } \,
\chi_{l,l}^{(g)} +2 \chi_{l+\frac{1}{2},l+\frac{1}{2}}^{(g)}+ \chi_{l+1,l+1}^{(g)}. \label{N4recombgl}
\end{align}

Note that this formulae are not directly derived from (\ref{gche}) and (\ref{gchie}),
because in the small $\CN=4$ algebra the R-symmetry is $SU(2)$ rather than $SO(4)$,
and the supercharges do not form a vector representation thereof.
However, essential properties appear in common. In particular, note how the
$(1+y^{\pm \half} q^\half)^2$ factors nullify the action of two supercharges simultaneously.

Decomposition of super Virasoro multiplets into global multiplets is done
in a similar manner to $\CN=2$.
Massive super Virasoro multiplets are decomposed into
long global multiplets only, while massless super Virasoro multiplets are
decomposed into short and long global multiplets.

One thing we would like to check is when a massive super Virasoro multiplet
hits the unitarity bound to branch into massless multiplets (\ref{N4recombsv}),
if its global constituents also branch into the global constituents of the
massless multiplets via (\ref{N4recombgl}).
For this purpose, we only need to consider global constituents of
massive multiplets that also saturate the global unitarity bound
as the massive multiplet saturates its super Virasoro unitarity bound,
and also short global multiplets in the decomposition of
massless super Virasoro multiplets.
It seems reasonable to believe that the other parts,
which remain manifestly long throughout the recombination/branching,
would simply regroup among themselves to belong to appropriate super Virasoro multiplets.

Consider the infinite sum in (\ref{N4char1}) and (\ref{N4char2}).
As we are only interested in global constituents that
hit the unitarity bound as $h \to l$ ($h=l$ is automatic for (\ref{N4char2})),
we look for monomials $q^{a-\frac{k}{4}} y^b$ such that $a \leq b$.
Apart from the product term that only contains monomials
$q^a y^b$ with $a \geq b$, we require
\begin{align}
0 &\geq (k+1)m^2+(2l+1)m+l - (k+1)m - l \nonumber \\
&= (k+1) m \big(m-(1-\frac{2l+1}{k+1}) \big). \label{ineqform}
\end{align}

However, since $0< \frac{1}{k+1} \leq \frac{2l+1}{k+1} \leq 1$,
the only way to satisfy (\ref{ineqform}) is by $m=0$,
which saturates the inequality.
Therefore, $m \neq 0$ in the infinite sum always leads to global multiplets
that are manifestly long even when the super Virasoro multiplet hits the
unitarity bound, and $m=0$ may
contribute to global multiplets that hit their unitarity bound
simultaneously with the super Virasoro multiplet.

Finally, we can write down the decomposition of small $\CN=4$ multiplets as
\begin{align}
\text{Ch}_{h,l}(\tau,z) =&
\text{Ch}_{h,l}^{(g)}(\tau,z)
\bigg( \prod_{n=2}^\infty
\frac{(1+y^{\frac{1}{2}}q^{n-\frac{1}{2}})^2(1+y^{-\frac{1}{2}}q^{n-\frac{1}{2}})^2}{(1-yq^{n-1})(1-q^n)(1-q^{n-1})(1-y^{-1}q^{n-1})}
\bigg) \nonumber \\
&+(\text{Ch}_{h',l'}^{(g)}(\tau,z) \text{ such that }h'-l' > h-l),
\label{svtogl1} \\
\chi_{h=l,l}(\tau,z) =&
\chi_{h=l,l}^{(g)}(\tau,z)
\bigg( \prod_{n=2}^\infty
\frac{(1+y^{\frac{1}{2}}q^{n-\frac{1}{2}})^2(1+y^{-\frac{1}{2}}q^{n-\frac{1}{2}})^2}{(1-yq^{n-1})(1-q^n)(1-q^{n-1})(1-y^{-1}q^{n-1})}
\bigg) \nonumber \\
&+(\text{Ch}_{h',l'}^{(g)}(\tau,z) \text{ such that }h' > l'). \label{svtogl2}
\end{align}
Then, from the fact that a common multiplicative factor appears on the
right-hand-sides of (\ref{svtogl1}) and (\ref{svtogl2}),
it can be inferred that the global multiplet decomposition of (\ref{N4recombsv})
is indeed compatible with (\ref{N4recombgl}).

\subsubsection{Large $\CN = 4$}

For completeness, we conclude by relating characters of the large
$\CN=4$ global multiplets to their analogues in super Virasoro algebra
that are studied in \cite{Petersen:1989zz,Petersen:1989pp}.
In particular, we consider representations of $\tilde{A}_\g$,
which is a subalgebra of the full large $\CN=4$ superconformal algebra $A_\g$.
It is obtained from the latter by decoupling four dimension-$\half$
operators and one dimension-$1$ current, denoted by $Q^a_r$ and
$U_m$ in (\ref{N4supervirasoro}), see \cite{Knizhnik:1986wc,Goddard:1988wv}.
It is straightforward to obtain characters of representations of $A_\g$
from that of $\tilde{A}_\g$, see \cite{Petersen:1989zz}.

We first present the character formulae for the large $\CN=4$ global
subalgebra, massive or massless, which are special cases of (\ref{gche}),
(\ref{Weylchar}), and (\ref{gchie}).
\begin{align}\label{N4gche}
\text{Ch}_{h,\{h_1, h_2\}}^{(g)}(\t,\{z_1,z_2\}) &=
q^{h-\frac{c}{24}} \cdot
\frac{ \displaystyle  \bigg( \prod_{i=1}^{r}\prod_{\e=\pm 1}
(1+y_i^{\e} q^{\half} ) \bigg)}{1-q} \nonumber \\
&\times \frac{\displaystyle \sum_{\e = \pm 1}
\begin{vmatrix}
y_1^{h_1+1} + \e y_1^{-h_1-1} & y_2^{h_1+1} + \e y_2^{-h_1-1} \\
y_1^{h_2} + \e y_1^{-h_2} & y_2^{h_2} + \e y_2^{-h_2}
\end{vmatrix}}
{\begin{vmatrix}
y_1+ y_1^{-1} & y_2 + y_2^{-1} \\ 2 & 2
\end{vmatrix}},
\end{align}
\begin{align}\label{N4gchie}
\chi_{h=\frac{h_1}{2},\{h_1, h_2\}}^{(g)}(\t,\{z_1,z_2\}) &=
q^{\frac{h_1}{2}-\frac{c}{24}} \cdot
\frac{ \displaystyle  \bigg( \prod_{i=1}^{r}\prod_{\e=\pm 1}
(1+y_i^{\e} q^{\half} ) \bigg)}{1-q} \nonumber \\
&\times \frac{ \displaystyle \sum_{\e = \pm 1}
\begin{vmatrix} \displaystyle
\frac{y_1^{h_1+1}}{1+y_1 q^\half} + \e \frac{y_1^{-h_1-1}}{1+y_1^{-1} q^\half} & \displaystyle
\frac{y_2^{h_1+1}}{1+y_2 q^\half} + \e \frac{y_2^{-h_1-1}}{1+y_2^{-1} q^\half} \\
y_1^{h_2} + \e y_1^{-h_2} & y_2^{h_2} + \e y_2^{-h_2}
\end{vmatrix}}
{\begin{vmatrix}
y_1+ y_1^{-1} & y_2 + y_2^{-1} \\ 2 & 2
\end{vmatrix}}.
\end{align}

We then present the character formulae for the 
massive and massless representations of $\tilde{A}_\g$ \cite{Petersen:1989zz,Petersen:1989pp}
in our convention  below.  
In particular, we move from the $SU(2)$ isospins, 
related to the fundamental Dynkin labels by $l^{+}=\frac{R_1}{2}$
and $l^{-}=\frac{R_2}{2}$, with fugacities $\{ z_+, z_- \} $
into the orthogonal basis $h_1 = l^{+}+l^{-} $, $h_2 = l^{+} - l^{-}$
with fugacities $\{ y_1 = z_+ z_-  , y_2=z_+ z_-^{-1} \} $.
\begin{align}\label{N4che}
\text{Ch}^{\tilde{A}_\g}_{h, \{ h_1,h_2 \} }(\t, &\{ z_1,z_2 \} ) \nonumber \\
= q^{h-\frac{c}{24}} \bigg( \prod_{n=1}^{\infty} &
\frac{\displaystyle \prod_{i=1}^{2}\prod_{\e=\pm 1} (1+y_i^{\e} q^{n-\half} )}
{\displaystyle (1-q^n)^3 \cdot \prod_{\e_1, \e_2 = \pm 1}(1-y_1^{\e_1}y_2^{\e_2}q^n)} \bigg)
\times \sum_{a,b=-\infty}^{\infty}  \bigg( q^{k_+ a^2 + (h_1+h_2+1)a +k_- b^2 + (h_1-h_2+1)b} \nonumber \\
\times & \frac{\displaystyle \sum_{\e=\pm 1}
\begin{vmatrix}
y_1^{k_+ a + k_- b + h_1 +1}+\e y_1^{-k_+ a - k_- b - h_1 -1} &
y_2^{k_+ a + k_- b + h_1 +1}+\e y_2^{-k_+ a - k_- b - h_1 -1}  \\
y_1^{k_+ a - k_- b + h_2 }+\e y_1^{-k_+ a + k_- b - h_2} &
y_2^{k_+ a - k_- b + h_2 }+\e y_2^{-k_+ a + k_- b - h_2}
\end{vmatrix} }{\begin{vmatrix} y_1 + y_1^{-1} & y_2 + y_2^{-1} \\ 2 & 2 \end{vmatrix}} \bigg),
\end{align}
\begin{align}\label{N4chie}
\chi^{\tilde{A}_\g}_{h=\frac{h_1}{2}, \{ h_1,h_2 \} }(\t, &\{ z_1,z_2 \} ) \nonumber \\
= q^{\frac{h_1}{2}-\frac{c}{24}} \bigg( \prod_{n=1}^{\infty} &
\frac{\displaystyle \prod_{i=1}^{2}\prod_{\e=\pm 1} (1+y_i^{\e} q^{n-\half} )}
{\displaystyle (1-q^n)^3 \cdot \prod_{\e_1, \e_2 = \pm 1}(1-y_1^{\e_1}y_2^{\e_2}q^n)} \bigg)
\times \sum_{a,b=-\infty}^{\infty}  \bigg( q^{k_+ a^2 + (h_1+h_2+1)a +k_- b^2 + (h_1-h_2+1)b} \nonumber \\
\times & \frac{\displaystyle \sum_{\e=\pm 1}
\begin{vmatrix} \displaystyle
\frac{y_1^{k_+ a + k_- b + h_1 +1}}{1+y_1 q^{a+b+\frac12}}
+\e \frac{y_1^{-k_+ a - k_- b - h_1 -1}}{1+y_1^{-1} q^{a+b+\frac12}} & \displaystyle
\frac{y_2^{k_+ a + k_- b + h_1 +1}}{1+y_2 q^{a+b+\frac12}}
+\e \frac{y_2^{-k_+ a - k_- b - h_1 -1}}{1+y_2^{-1} q^{a+b+\frac12}}  \\
y_1^{k_+ a - k_- b + h_2 }+\e y_1^{-k_+ a + k_- b - h_2} &
y_2^{k_+ a - k_- b + h_2 }+\e y_2^{-k_+ a + k_- b - h_2}
\end{vmatrix} }{\begin{vmatrix} y_1 + y_1^{-1} & y_2 + y_2^{-1} \\ 2 & 2 \end{vmatrix}} \bigg),
\end{align}
where $k_{\pm}$ are the $\mathfrak{su}(2)$ levels defined in (\ref{Kaclevels}).
Note the recombination rules,
\begin{align}
\text{Ch}^{\tilde{A}_\g}_{h,\{h_1, h_2 \} } \, & \xrightarrow{ h \to \frac{h_1}{2} } \,
\chi^{\tilde{A}_\g}_{\frac{h_1}{2},\{h_1, h_2 \} } + \chi^{\tilde{A}_\g}_{\frac{h_1+1}{2},\{h_1+1, h_2 \} }, \label{largeN4recombsv} \\
\text{Ch}^{(g)}_{h,\{h_1, h_2 \} } \, & \xrightarrow{ h \to \frac{h_1}{2} } \,
\chi^{(g)}_{\frac{h_1}{2},\{h_1, h_2 \} } + \chi^{(g)}_{\frac{h_1+1}{2},\{h_1+1, h_2 \} }. \label{largeN4recombgl}
\end{align}

Comparing (\ref{N4gche}) through (\ref{N4chie}), it is clear that the $\tilde{A}_\g$
characters can be written in terms of the global multiplet characters.
Specifically, the summands over $a$ and $b$ in (\ref{N4che}) and (\ref{N4chie})
are precisely the second lines of (\ref{N4gche}) and (\ref{N4gchie}) for $a=b=0$.
More generally, the summand in (\ref{N4che}) is the second line of
(\ref{N4gche}) with
\begin{align}
h &\to h+k^+ a^2 + (h_1+h_2+1)a +k^- b^2 + (h_1-h_2+1)b, \nonumber \\
h_1 &\to h_1 + k^+ a + k^- b, \qquad \text{and} \qquad h_2 \to h_2 + k^+ a - k^- b.
\end{align}
The summand in (\ref{N4chie}) is more involved, but after expanding the
denominators in the first row of the determinant, it can be expressed as a
sum of a massless global character and an infinite series of massive global
characters.

However, it is not very illuminating to write the $\tilde{A}_\g$
characters entirely in terms of the global characters.
Instead, we write analogues of (\ref{svtogl1}) and (\ref{svtogl2})
to illuminate how the saturation of unitarity bound occurs in terms of
the global constituents.

To do so, the following inequality is crucial:
\begin{align}
&k_+ a^2 + (h_1+h_2+1)a +k_- b^2 + (h_1-h_2+1)b - \frac12 k_+ a - \frac12 k_- b \nonumber \\
= & k_+ a(a + \frac{h_1+h_2+1 - \frac12 k_+}{k_+})+k_- b(b + \frac{h_1-h_2+1 - \frac12 k_-}{k_-}) \nonumber \\ \geq  & 0 \quad \text{(for } a,b \in \mathbb{Z} ).
\end{align}
Equality holds if and only if $a=b=0$.
It is valid because the spectral flow constrains isospins of $\tilde{A}_\g$
multiplets by
\begin{equation}\begin{cases}
1 \leq h_1 \pm h_2+1=2l^{\pm} +1 \leq k^{\pm}, & \text{(massive)} \\
1 \leq h_1 \pm h_2+1=2l^{\pm} +1 \leq k^{\pm}+1. & \text{(massless)}
\end{cases}\end{equation}

The inequality implies that only the contribution from $a=b=0$ in (\ref{N4che})
may saturate the unitarity bound $h \geq \frac{h_1}{2}$ as the LHS does,
and that only the contribution from $a=b=0$ in (\ref{N4chie}) may be
massless global multiplets. Therefore, we can generalize (\ref{svtogl1}) and
(\ref{svtogl2}) into the large $\CN=4$ characters as follows,
\begin{align}
\text{Ch}^{\tilde{A}_\g}_{h,\{ h_1,h_2 \} }=&
\text{Ch}^{(g),\CN=4}_{h,\{ h_1,h_2 \} }
\bigg( \prod_{n=2}^\infty
\frac{\displaystyle \prod_{i=1}^{2}\prod_{\e=\pm 1} (1+y_i^{\e} q^{n-\half} )}
{\displaystyle (1-q^n) (1-q^{n-1})^2 \cdot \prod_{\e_1, \e_2 = \pm 1}(1-y_1^{\e_1}y_2^{\e_2}q^{n-1})}
\bigg) \nonumber \\
&+(\text{Ch}^{(g),\CN=4}_{h',\{ h'_1,h'_2 \} } \text{ such that }h'-\frac{h'_1}{2} > h-\frac{h_1}{2}),
\label{svtogl3} \\
\chi^{\tilde{A}_\g}_{\frac{h_1}{2},\{ h_1,h_2 \} } =&
\chi^{(g),\CN=4}_{\frac{h_1}{2},\{ h_1,h_2 \} }
\bigg( \prod_{n=2}^\infty
\frac{\displaystyle \prod_{i=1}^{2}\prod_{\e=\pm 1} (1+y_i^{\e} q^{n-\half} )}
{\displaystyle (1-q^n) (1-q^{n-1})^2 \cdot \prod_{\e_1, \e_2 = \pm 1}(1-y_1^{\e_1}y_2^{\e_2}q^{n-1})}
\bigg) \nonumber \\
&+(\text{Ch}^{(g),\CN=4}_{h',\{ h'_1,h'_2 \} } \text{ such that }h' > \frac{h'_1}{2}). \label{svtogl4}
\end{align}
Again, from the fact that a common multiplicative factor appears on the
right-hand-sides of (\ref{svtogl3}) and (\ref{svtogl4}),
it can be inferred that the global multiplet decomposition of (\ref{largeN4recombsv})
is indeed compatible with (\ref{largeN4recombgl}).

\section*{Acknowledgments}
We would like to thank Hyekyung Choi, Seok Kim, Kimyeong Lee, Sameer Murthy, Jaewon Song, and David Tong
for useful discussions.
The research of SJ.L. is supported in part by the National Research Foundation of Korea (NRF) Grant NRF-$2017$R$1$C$1$B$1011440$ and NRF Grants No. $2006$-$0093850$.
SJ.L. is grateful to the Aspen Center for Physics (supported by National Science Foundation grant PHY-1607611)
and to the $2019$ Pollica summer workshop (supported in part by the Simons Foundation and by the INFN)
for their kind hospitality during some stages of this work.

\newpage

\appendix
\section{Racah-Speiser Algorithm}\label{appendix}

In this appendix, we review the Racah-Speiser Algorithm that has been used
throughout this article. Similarly to \cite{Cordova:2016emh}, our focus will primarily be
on how to use the algorithm rather than its construction, and we refer the
readers to \cite{Fuchs:1997jv} for the latter. Moreover, we are particularly
interested in its application to special orthogonal groups $SO(\CN)$.

Racah-Speiser Algorithm, in short, is an efficient algorithm that yields
a tensor product of two irreducible representations of a Lie group
as a direct sum of irreducible representations, while avoiding overly detailed
relations between the weights best described by Clebsch-Gordan coefficients.

Consider two irreducible representations of a Lie group $G_r$ with rank $r$,
denoted by their highest weights $\l_1$ and $\l_2$. Our objective is to find
irreducible representations $\Lambda^a$ with corresponding multiplicities $m^i$, such that
\begin{equation}\label{RSalg}
\l_1 \otimes \l_2 = \bigoplus_a m^a \Lambda^a,
\end{equation}
where $m^a \Lambda^a$ indicates that $\Lambda^a$ appears $m^a $ times in the sum.

Let us denote by $\{ \m_2^a \} \, (a=1,\cdots,\text{dim}\l_2 )$
the complete set of weights of the irreducible representation $\l_2$.
Then, consider the set of weights $\{ \l_1 + \m_2^a \} \, (a=1,\cdots,\text{dim}\l_2 )$.
Each weight belongs to one of three categories, and contributes to the
RHS of (\ref{RSalg}), after appropriate treatments:

\begin{itemize}

\item Some of the weights live in the Weyl chamber, that is, all of their Dynkin labels
are non-negative. These weights do not require any special treatment, and each of
them contributes to the RHS of (\ref{RSalg}) as a highest weight of an irreducible
representation $\Lambda^a$ with multiplicity 1.

\item Some of the weights do not live in the Weyl chamber, however, can be
brought into one by a series of \emph{reflections}. Here, the \emph{reflection}
is defined as a shift by the Weyl vector $\r$, followed by a Weyl reflection and
then a negative shift by the same Weyl vector $\r$. Using fundamental
Dynkin labels, the $i^{th}$ Weyl reflection $\s^i$ of a weight $[R_1 \cdots R_r]$
can be written as
\begin{equation}\label{weylreflection}
\s^i([R_1 \cdots R_r]) = [R_1 \cdots R_r] - R_i[A_{i1} \cdots A_{ir}],
\end{equation}
where $A_{ij}$ denotes the Cartan matrix of $\mathfrak{g}_r$.
Incorporating the shifts by the Weyl vector $\r=[1 \cdots 1]$,
the $i^{th}$ reflection is summarized as
\begin{equation}\label{reflection}
r^i([R_1 \cdots R_r]) = [R_1 \cdots R_r] - (R_i+1) [A_{i1} \cdots A_{ir}].
\end{equation}
Note that there are $r$ different reflections, where $r$ is the rank of the Lie group.
Number of reflections required to bring a weight into the Weyl chamber is
defined modulo 2. Then, each of the weights that belongs to this category contributes
to the RHS of (\ref{RSalg}) after being brought into the Weyl chamber via $W$
reflections, with multiplicity $(-1)^W$.

\item Finally, some multiplets are reflected into itself. By inspection of (\ref{reflection}),
one can see that a weight belongs to this category if and only if at least one of its
fundamental Dynkin labels equals $-1$. Considering the $-1$ factor that entails
each reflection, these weights naturally have vanishing contribution to
the RHS of (\ref{RSalg}).

\end{itemize}

To summarize, each weight in the set $\{ \l_1 + \m_2^a \} $
whose Dynkin labels do not include $-1$, contributes to the RHS of (\ref{RSalg})
by an irreducible representation whose highest weight equals itself, or a reflection of itself.
If an odd number of reflection is required in the process, it contributes negatively,
cancelling a positive but identical contribution from another weight. At the end, the
cancellation is always complete and there is no remaining contribution with a
negative multiplicity. Note that exchanging the role of two representations
$l_1$ and $l_2$ gives the same result.

Throughout this article, the Racah-Speiser Algorithm is always performed in a
special orthogonal group $SO(\CN)$, with one of the two multiplicands being the
vector representation that represents the supercharges. In such cases, the minimum
Dynkin label that can appear in any of the weights in the set $\{ \l_1+\m_2^a\}$ is $-2$,
and the classification and treatment of the weights become extremely simple.
That is, whenever there is a Dynkin label equal to $-1$ we dispose the weight, and
else if the $k^{th}$ label is $-2$, we add the $k^{th}$ row of the Cartan matrix to the
weight in accordance with (\ref{reflection}) to bring the $k^{th}$ label to 0 then to
cancel with an identical weight. Of course, if the reflection produces a $-1$ for
another label, we dispose the weight.

Let us take the product $[0 \, 1 \, 0] \otimes [1 \, 0 \, 0]$ in $SO(7)$ as an example.
The weight system of the highest weight representation $[1 \, 0 \, 0]$ is given by
\begin{equation}
[1 \, 0 \, 0], \quad [-1 \, 1 \, 0], \quad [0 \, -1 \, 2], \quad [0 \, 0 \, 0],
\quad [0 \, 1 \, -2], \quad [1 \, -1 \, 0], \quad [-1 \, 0 \, 0],
\end{equation}
and thus the set of weights $\{ \l_1 + \m_2^a \}$ is
\begin{equation}
[1 \, 1 \, 0], \quad [-1 \, 2 \, 0], \quad [0 \, 0 \, 2], \quad [0 \, 1 \, 0],
\quad [0 \, 2 \, -2], \quad [1 \, 0 \, 0], \quad [-1 \, 1 \, 0].
\end{equation}
The second and seventh weights are disposed because of the Dynkin label $-1$.
To bring the fifth weight $[0 \, 2 \, -2]$ into the Weyl chamber, we add the third row of
the Cartan matrix $[0 \, -1 \, 2]$ to get $[0 \, 1 \, 0]$, which cancels the fourth weight.
Therefore,
\begin{equation}
[0 \, 1 \, 0] \otimes [1 \, 0 \, 0] = [1 \, 1 \, 0] \oplus [0 \, 0 \, 2] \oplus [1 \, 0 \, 0].
\end{equation}
When $[0 \, 1 \, 0]$ is the primary of a multiplet, the product with vector $[1 \, 0 \, 0]$
represents states at the first level of the multiplet. The first piece $[1 \, 1 \, 0]$
represents null states when the unitarity bound is saturated, then we can identify
the remaining parts as
\begin{equation}
\wedge^2V \otimes \wedge^1V = \wedge^3 V \oplus \wedge^1 V,
\end{equation}
in agreement with (\ref{wedgeproducts}) and the discussion followed.



\providecommand{\href}[2]{#2}\begingroup\raggedright\begin{thebibliography}{}

\end{thebibliography}\endgroup


\begin{thebibliography}{10}

\bibitem{Haag:1974qh}
R.~Haag, J.~T. Lopuszanski and M.~Sohnius, \emph{{All Possible Generators of
  Supersymmetries of the s Matrix}},
  \href{https://doi.org/10.1016/0550-3213(75)90279-5}{\emph{Nucl. Phys.}
  {\bfseries B88} (1975) 257}.

\bibitem{Virasoro:1969zu}
M.~S. Virasoro, \emph{{Subsidiary conditions and ghosts in dual resonance
  models}}, \href{https://doi.org/10.1103/PhysRevD.1.2933}{\emph{Phys. Rev.}
  {\bfseries D1} (1970) 2933}.

\bibitem{Kac_1968}
V.~G. Kac, \emph{Simple irreducible graded lie algebras of finite growth},
  \href{https://doi.org/10.1070/im1968v002n06abeh000729}{\emph{Mathematics of
  the {USSR}-Izvestiya} {\bfseries 2} (1968) 1271}.

\bibitem{Moody:1968zz}
R.~V. Moody, \emph{{A new class of Lie algebras}},
  \href{https://doi.org/10.1016/0021-8693(68)90096-3}{\emph{J. Algebra}
  {\bfseries 10} (1968) 211}.

\bibitem{Zamolodchikov:1985wn}
A.~B. Zamolodchikov, \emph{{Infinite Additional Symmetries in Two-Dimensional
  Conformal Quantum Field Theory}},
  \href{https://doi.org/10.1007/BF01036128}{\emph{Theor. Math. Phys.}
  {\bfseries 65} (1985) 1205}.

\bibitem{DiFrancesco:1997nk}
P.~Di~Francesco, P.~Mathieu and D.~Senechal, \emph{{Conformal Field Theory}},
  Graduate Texts in Contemporary Physics. Springer-Verlag, New York, 1997,
  \href{https://doi.org/10.1007/978-1-4612-2256-9}{10.1007/978-1-4612-2256-9}.

\bibitem{Blumenhagen:2009zz}
R.~Blumenhagen and E.~Plauschinn, \emph{{Introduction to conformal field
  theory}}, \href{https://doi.org/10.1007/978-3-642-00450-6}{\emph{Lect. Notes
  Phys.} {\bfseries 779} (2009) 1}.

\bibitem{Ademollo:1975an}
M.~Ademollo et~al., \emph{{Supersymmetric Strings and Color Confinement}},
  \href{https://doi.org/10.1016/0370-2693(76)90061-7}{\emph{Phys. Lett.}
  {\bfseries 62B} (1976) 105}.

\bibitem{Ademollo:1976pp}
M.~Ademollo et~al., \emph{{Dual String with U(1) Color Symmetry}},
  \href{https://doi.org/10.1016/0550-3213(76)90483-1}{\emph{Nucl. Phys.}
  {\bfseries B111} (1976) 77}.

\bibitem{Schoutens:1987uu}
K.~Schoutens, \emph{{A Nonlinear Representation of the $d=2$ SO(4) Extended
  Superconformal Algebra}},
  \href{https://doi.org/10.1016/0370-2693(87)90772-6}{\emph{Phys. Lett.}
  {\bfseries B194} (1987) 75}.

\bibitem{Schoutens:1988ig}
K.~Schoutens, \emph{{O(n) Extended Superconformal Field Theory in Superspace}},
  \href{https://doi.org/10.1016/0550-3213(88)90539-1}{\emph{Nucl. Phys.}
  {\bfseries B295} (1988) 634}.

\bibitem{Sevrin:1988ew}
A.~Sevrin, W.~Troost and A.~Van~Proeyen, \emph{{Superconformal Algebras in
  Two-Dimensions with N=4}},
  \href{https://doi.org/10.1016/0370-2693(88)90645-4}{\emph{Phys. Lett.}
  {\bfseries B208} (1988) 447}.

\bibitem{Eguchi:1987sm}
T.~Eguchi and A.~Taormina, \emph{{Unitary Representations of $N=4$
  Superconformal Algebra}},
  \href{https://doi.org/10.1016/0370-2693(87)91679-0}{\emph{Phys. Lett.}
  {\bfseries B196} (1987) 75}.

\bibitem{Eguchi:1987wf}
T.~Eguchi and A.~Taormina, \emph{{Character Formulas for the $N=4$
  Superconformal Algebra}},
  \href{https://doi.org/10.1016/0370-2693(88)90778-2}{\emph{Phys. Lett.}
  {\bfseries B200} (1988) 315}.

\bibitem{Odake:1988bh}
S.~Odake, \emph{{Extension of $N=2$ Superconformal Algebra and Calabi-yau
  Compactification}},
  \href{https://doi.org/10.1142/S021773238900068X}{\emph{Mod. Phys. Lett.}
  {\bfseries A4} (1989) 557}.

\bibitem{Odake:1989dm}
S.~Odake, \emph{{Character Formulas of an Extended Superconformal Algebra
  Relevant to String Compactification}},
  \href{https://doi.org/10.1142/S0217751X90000428}{\emph{Int. J. Mod. Phys.}
  {\bfseries A5} (1990) 897}.

\bibitem{Englert:1987fm}
F.~Englert, A.~Sevrin, W.~Troost, A.~Van~Proeyen and P.~Spindel, \emph{{Loop
  Algebras and Superalgebras Based on S(7)}},
  \href{https://doi.org/10.1063/1.528065}{\emph{J. Math. Phys.} {\bfseries 29}
  (1988) 281}.

\bibitem{Gastmans:1987up}
R.~Gastmans, A.~Sevrin, W.~Troost and A.~Van~Proeyen, \emph{{Infinite
  Dimensional Extended Superconformal Lie Algebras}},
  \href{https://doi.org/10.1142/S0217751X87000077}{\emph{Int. J. Mod. Phys.}
  {\bfseries A2} (1987) 195}.

\bibitem{Sevrin:1988ps}
A.~Sevrin, W.~Troost, A.~Van~Proeyen and P.~Spindel, \emph{{Extended
  Supersymmetric Sigma Models on Group Manifolds. 2. Current Algebras}},
  \href{https://doi.org/10.1016/0550-3213(88)90070-3}{\emph{Nucl. Phys.}
  {\bfseries B311} (1988) 465}.

\bibitem{Schoutens:1988tg}
K.~Schoutens, \emph{{Representation Theory for a Class of SO($N$) Extended
  Superconformal Operator Algebras}},
  \href{https://doi.org/10.1016/0550-3213(89)90163-6}{\emph{Nucl. Phys.}
  {\bfseries B314} (1989) 519}.

\bibitem{Gates:1998ss}
S.~J. Gates, Jr. and L.~Rana, \emph{{Superspace geometrical representations of
  extended super Virasoro algebras}},
  \href{https://doi.org/10.1016/S0370-2693(98)00937-X}{\emph{Phys. Lett.}
  {\bfseries B438} (1998) 80}
  [\href{https://arxiv.org/abs/hep-th/9806038}{{\ttfamily hep-th/9806038}}].

\bibitem{deBoer:1998kjm}
J.~de~Boer, \emph{{Six-dimensional supergravity on S**3 x AdS(3) and 2-D
  conformal field theory}},
  \href{https://doi.org/10.1016/S0550-3213(99)00160-1}{\emph{Nucl. Phys.}
  {\bfseries B548} (1999) 139}
  [\href{https://arxiv.org/abs/hep-th/9806104}{{\ttfamily hep-th/9806104}}].

\bibitem{Ito:1998vd}
K.~Ito, \emph{{Extended superconformal algebras on AdS(3)}},
  \href{https://doi.org/10.1016/S0370-2693(99)00070-2}{\emph{Phys. Lett.}
  {\bfseries B449} (1999) 48}
  [\href{https://arxiv.org/abs/hep-th/9811002}{{\ttfamily hep-th/9811002}}].

\bibitem{Henneaux:1999ib}
M.~Henneaux, L.~Maoz and A.~Schwimmer, \emph{{Asymptotic dynamics and
  asymptotic symmetries of three-dimensional extended AdS supergravity}},
  \href{https://doi.org/10.1006/aphy.2000.5994}{\emph{Annals Phys.} {\bfseries
  282} (2000) 31} [\href{https://arxiv.org/abs/hep-th/9910013}{{\ttfamily
  hep-th/9910013}}].

\bibitem{Kraus:2007vu}
P.~Kraus, F.~Larsen and A.~Shah, \emph{{Fundamental Strings, Holography, and
  Nonlinear Superconformal Algebras}},
  \href{https://doi.org/10.1088/1126-6708/2007/11/028}{\emph{JHEP} {\bfseries
  11} (2007) 028} [\href{https://arxiv.org/abs/0708.1001}{{\ttfamily
  0708.1001}}].

\bibitem{Cordova:2016emh}
C.~Cordova, T.~T. Dumitrescu and K.~Intriligator, \emph{{Multiplets of
  Superconformal Symmetry in Diverse Dimensions}},
  \href{https://doi.org/10.1007/JHEP03(2019)163}{\emph{JHEP} {\bfseries 03}
  (2019) 163} [\href{https://arxiv.org/abs/1612.00809}{{\ttfamily
  1612.00809}}].

\bibitem{Cordova:2016xhm}
C.~Cordova, T.~T. Dumitrescu and K.~Intriligator, \emph{{Deformations of
  Superconformal Theories}},
  \href{https://doi.org/10.1007/JHEP11(2016)135}{\emph{JHEP} {\bfseries 11}
  (2016) 135} [\href{https://arxiv.org/abs/1602.01217}{{\ttfamily
  1602.01217}}].

\bibitem{Gukov:2004ym}
S.~Gukov, E.~Martinec, G.~W. Moore and A.~Strominger, \emph{{The Search for a
  holographic dual to AdS(3) x S**3 x S**3 x S**1}},
  \href{https://doi.org/10.4310/ATMP.2005.v9.n3.a3,
  10.1142/9789812775344_0035}{\emph{Adv. Theor. Math. Phys.} {\bfseries 9}
  (2005) 435} [\href{https://arxiv.org/abs/hep-th/0403090}{{\ttfamily
  hep-th/0403090}}].

\bibitem{Knizhnik:1986wc}
V.~G. Knizhnik, \emph{{Superconformal Algebras in Two-dimensions}},
  \href{https://doi.org/10.1007/BF01028940}{\emph{Theor. Math. Phys.}
  {\bfseries 66} (1986) 68}.

\bibitem{Bershadsky:1986ms}
M.~A. Bershadsky, \emph{{Superconformal Algebras in Two-dimensions With
  Arbitrary $N$}},
  \href{https://doi.org/10.1016/0370-2693(86)91100-7}{\emph{Phys. Lett.}
  {\bfseries B174} (1986) 285}.

\bibitem{Minwalla:1997ka}
S.~Minwalla, \emph{{Restrictions imposed by superconformal invariance on
  quantum field theories}},
  \href{https://doi.org/10.4310/ATMP.1998.v2.n4.a4}{\emph{Adv. Theor. Math.
  Phys.} {\bfseries 2} (1998) 783}
  [\href{https://arxiv.org/abs/hep-th/9712074}{{\ttfamily hep-th/9712074}}].

\bibitem{Dolan:2002zh}
F.~A. Dolan and H.~Osborn, \emph{{On short and semi-short representations for
  four-dimensional superconformal symmetry}},
  \href{https://doi.org/10.1016/S0003-4916(03)00074-5}{\emph{Annals Phys.}
  {\bfseries 307} (2003) 41}
  [\href{https://arxiv.org/abs/hep-th/0209056}{{\ttfamily hep-th/0209056}}].

\bibitem{Boucher:1986bh}
W.~Boucher, D.~Friedan and A.~Kent, \emph{{Determinant Formulae and Unitarity
  for the N=2 Superconformal Algebras in Two-Dimensions or Exact Results on
  String Compactification}},
  \href{https://doi.org/10.1016/0370-2693(86)90260-1}{\emph{Phys. Lett.}
  {\bfseries B172} (1986) 316}.

\bibitem{Fulton:1991}
W.~Fulton and H.~Joe, \emph{{Representation Theory: A First Course}}.
  Springer-Verlag, 1991.

\bibitem{Maldacena:2011jn}
J.~Maldacena and A.~Zhiboedov, \emph{{Constraining Conformal Field Theories
  with A Higher Spin Symmetry}},
  \href{https://doi.org/10.1088/1751-8113/46/21/214011}{\emph{J. Phys.}
  {\bfseries A46} (2013) 214011}
  [\href{https://arxiv.org/abs/1112.1016}{{\ttfamily 1112.1016}}].

\bibitem{Boulanger:2013zza}
N.~Boulanger, D.~Ponomarev, E.~D. Skvortsov and M.~Taronna, \emph{{On the
  uniqueness of higher-spin symmetries in AdS and CFT}},
  \href{https://doi.org/10.1142/S0217751X13501625}{\emph{Int. J. Mod. Phys.}
  {\bfseries A28} (2013) 1350162}
  [\href{https://arxiv.org/abs/1305.5180}{{\ttfamily 1305.5180}}].

\bibitem{Alba:2013yda}
V.~Alba and K.~Diab, \emph{{Constraining conformal field theories with a higher
  spin symmetry in d=4}},  \href{https://arxiv.org/abs/1307.8092}{{\ttfamily
  1307.8092}}.

\bibitem{Alba:2015upa}
V.~Alba and K.~Diab, \emph{{Constraining conformal field theories with a higher
  spin symmetry in $d>3$ dimensions}},
  \href{https://doi.org/10.1007/JHEP03(2016)044}{\emph{JHEP} {\bfseries 03}
  (2016) 044} [\href{https://arxiv.org/abs/1510.02535}{{\ttfamily
  1510.02535}}].

\bibitem{Elitzur:1998mm}
S.~Elitzur, O.~Feinerman, A.~Giveon and D.~Tsabar, \emph{{String theory on
  AdS(3) x S**3 x S**3 x S**1}},
  \href{https://doi.org/10.1016/S0370-2693(99)00101-X}{\emph{Phys. Lett.}
  {\bfseries B449} (1999) 180}
  [\href{https://arxiv.org/abs/hep-th/9811245}{{\ttfamily hep-th/9811245}}].

\bibitem{Gomis:2016sab}
J.~Gomis, Z.~Komargodski, H.~Ooguri, N.~Seiberg and Y.~Wang, \emph{{Shortening
  Anomalies in Supersymmetric Theories}},
  \href{https://doi.org/10.1007/JHEP01(2017)067}{\emph{JHEP} {\bfseries 01}
  (2017) 067} [\href{https://arxiv.org/abs/1611.03101}{{\ttfamily
  1611.03101}}].

\bibitem{Bae:2018qym}
J.-B. Bae, S.~Lee and J.~Song, \emph{{Modular Constraints on Superconformal
  Field Theories}}, \href{https://doi.org/10.1007/JHEP01(2019)209}{\emph{JHEP}
  {\bfseries 01} (2019) 209}
  [\href{https://arxiv.org/abs/1811.00976}{{\ttfamily 1811.00976}}].

\bibitem{Schwimmer:1986mf}
A.~Schwimmer and N.~Seiberg, \emph{{Comments on the N=2, N=3, N=4
  Superconformal Algebras in Two-Dimensions}},
  \href{https://doi.org/10.1016/0370-2693(87)90566-1}{\emph{Phys. Lett.}
  {\bfseries B184} (1987) 191}.

\bibitem{Keller:2012mr}
C.~A. Keller and H.~Ooguri, \emph{{Modular Constraints on Calabi-Yau
  Compactifications}},
  \href{https://doi.org/10.1007/s00220-013-1797-8}{\emph{Commun. Math. Phys.}
  {\bfseries 324} (2013) 107}
  [\href{https://arxiv.org/abs/1209.4649}{{\ttfamily 1209.4649}}].

\bibitem{Dobrev:1986hq}
V.~K. Dobrev, \emph{{Characters of the Unitarizable Highest Weight Modules Over
  the $N=2$ Superconformal Algebras}},
  \href{https://doi.org/10.1016/0370-2693(87)90510-7}{\emph{Phys. Lett.}
  {\bfseries B186} (1987) 43}.

\bibitem{Matsuo:1986cj}
Y.~Matsuo, \emph{{Character Formula of C $<$ 1 Unitary Representation of $N=2$
  Superconformal Algebra}},
  \href{https://doi.org/10.1143/PTP.77.793}{\emph{Prog. Theor. Phys.}
  {\bfseries 77} (1987) 793}.

\bibitem{Kiritsis:1986rv}
E.~Kiritsis, \emph{{Character Formulae and the Structure of the Representations
  of the $N=1$, $N=2$ Superconformal Algebras}},
  \href{https://doi.org/10.1142/S0217751X88000795}{\emph{Int. J. Mod. Phys.}
  {\bfseries A3} (1988) 1871}.

\bibitem{Petersen:1989zz}
J.~L. Petersen and A.~Taormina, \emph{{Characters of the $N=4$ Superconformal
  Algebra With Two Central Extensions}},
  \href{https://doi.org/10.1016/0550-3213(90)90084-Q}{\emph{Nucl. Phys.}
  {\bfseries B331} (1990) 556}.

\bibitem{Petersen:1989pp}
J.~L. Petersen and A.~Taormina, \emph{{Characters of the $N=4$ Superconformal
  Algebra With Two Central Extensions: 2. Massless Representations}},
  \href{https://doi.org/10.1016/0550-3213(90)90141-Y}{\emph{Nucl. Phys.}
  {\bfseries B333} (1990) 833}.

\bibitem{Goddard:1988wv}
P.~Goddard and A.~Schwimmer, \emph{{Factoring Out Free Fermions and
  Superconformal Algebras}},
  \href{https://doi.org/10.1016/0370-2693(88)91470-0}{\emph{Phys. Lett.}
  {\bfseries B214} (1988) 209}.

\bibitem{Fuchs:1997jv}
J.~Fuchs and C.~Schweigert, \emph{{Symmetries, Lie algebras and
  representations: A graduate course for physicists}}. Cambridge University
  Press, 2003.

\end{thebibliography}

\bibliographystyle{JHEP}
\providecommand{\href}[2]{#2}\begingroup\raggedright\endgroup

\end{document}